\documentclass[11pt,a4paper]{article}
\pdfoutput = 1
\usepackage{jcappub}
\usepackage[utf8]{inputenc}
\usepackage{amsmath,amssymb}
\usepackage{graphicx}
\usepackage{hyperref}
\usepackage{natbib}

\subheader{\footnotesize
        \vspace*{-3.7em}
        \begin{flushright}
                CERN-TH-2020-068
        \end{flushright}
}

\newcommand{\edit}[1]{{#1}}

\newcommand{\bp}{\textbf{p}}
\newcommand{\bq}{\textbf{q}}
\newcommand{\bx}{\textbf{x}}
\newcommand{\bk}{\textbf{k}}
\newcommand{\bJ}{\textbf{J}}
\newcommand{\bv}{\textbf{v}}
\newcommand{\bPsi}{\boldsymbol{\Psi}}
\newcommand{\bs}{\textbf{s}}
\newcommand{\bu}{\textbf{u}}

\newcommand{\half}{\frac{1}{2}}
\newcommand{\third}{\frac{1}{3}}

\newcommand{\dDelta}{\dot{\Delta}}

\def \dtq{\int d^3 \bq \ }

\newcommand{\avg}[1]{\ensuremath{\left\langle #1 \right\rangle}}

\def\Gpc{\, h^{-1} \, {\rm Gpc}}

\def\kMpc{\, h \, {\rm Mpc}^{-1}}

\def\hq{\hat{q}} 
\def\hn{\hat{n}}
\def\hk{\hat{k}} 
\def\PL{P_{\rm lin}}

\author[a]{Shi-Fan Chen}
\author[b]{Zvonimir Vlah}
\author[a]{Martin White}

\affiliation[a]{Department of Physics, University of California,
Berkeley, CA 94720}
\affiliation[b]{Theory Department, CERN, CH-1211 Geneve 23, Switzerland}

\emailAdd{shifan\_chen@berkeley.edu}
\emailAdd{zvonimir.vlah@cern.ch}
\emailAdd{mwhite@berkeley.edu}

\title{Consistent Modeling of Velocity Statistics and Redshift-Space Distortions in One-Loop Perturbation Theory}


\abstract{The peculiar velocities of biased tracers of the cosmic density field contain important information about the growth of large scale structure and generate anisotropy in the observed clustering of galaxies.  Using N-body data, we show that velocity expansions for halo redshift-space power spectra are converged at the percent-level at perturbative scales for most line-of-sight angles $\mu$ when the first three pairwise velocity moments are included, and that the third moment is well-approximated by a counterterm-like contribution. We compute these pairwise-velocity statistics in Fourier space using both Eulerian and Lagrangian one-loop perturbation theory using a cubic bias scheme and a complete set of counterterms and stochastic contributions.  We compare the models and show that our models fit both real-space velocity statistics and redshift-space power spectra for both halos and a mock sample of galaxies at sub-percent level on perturbative scales using consistent sets of parameters, making them appealing choices for the upcoming era of spectroscopic, peculiar-velocity and kSZ surveys. }

\begin{document}
\maketitle
\flushbottom


\section{Introduction}

The large-scale structure (LSS) of the Universe contains a trove of information relevant to astrophysics, cosmology and fundamental physics, including the initial conditions from the early universe and constraints on cosmological parameters and gravity \cite{Wei13,PDG18,Ame18}. As cosmological distances are typically inferred through redshifts, a common theme in LSS observations is the necessity to operate in redshift space, where the peculiar velocities of observed targets lead to structure beyond what exists in real space \citep{Kai87,Ham92}. These so-called redshift-space distortions (RSD) present both a modeling challenge and additional information by encoding information about cosmic velocities in observed densities, for example allowing us to measure \edit{the derivative of the linear growth factor $f D = \text{d} D/\text{d}\ln a$, where $f(a)$ and $D(a)$ are the linear-theory growth rate and growth factor} (see e.g.\ refs.~\cite{Wei13,PDG18} for recent reviews).
Current and upcoming spectroscopic surveys such as DESI \cite{DESI} and EUCLID \cite{EUCLID18} will test these measurements at unprecedented precision. At the same time, the rise of next-generation ground-based CMB experiments \cite{SimonsObs,CMBS4} as well as renewed interest in low-redshift peculiar velocity surveys \cite{Howlett17,Kim19,Graziani20} in recent years makes it likely that direct measurements of the peculiar velocity statistics underlying redshift space distortions will become available in the near future, offering complementary probes for theories of structure formation. These developments make it timely to revisit our understanding of velocities in large scale structure and their link to redshift space distortions. 

The evolution of the LSS at high redshifts and large scales is well modeled by linear perturbation theory \cite{Pee80,Pea99,Dod03}, and the reach of the perturbation theory can be extended to intermediate scales by including higher order terms in the equations of motion \cite{Ber02}.
In this paper we shall consider 1-loop perturbation theory in both the Eulerian (EPT; \cite{Juszkiewicz81,Vishniac83,Goroff86,Makino92,Jain94,Ber02,BNSZ12,CHS12,McDRoy09,Perko16,Des16,MerPaj14}) and Lagrangian 
(LPT; \cite{Buc89,Mou91,Hiv95,TayHam96,Mat08a,Mat08b,CLPT,Whi14,ZheFri14,Mat15,VlaSelBal15,VlaWhiAvi15}) formulations, and their extensions as an effective field theories
\cite{PorSenZal14,VlaWhiAvi15,McQWhi16}.  EPT has been extensively employed in the analysis of large-scale structure surveys, with the most recent incarnation being refs.~\cite{DAmico19,Ivanov19,Colas19}.
LPT provides a natural means of modeling biased tracers in redshift space \cite{Mat08a,Mat08b}, including resummation of the advection terms which is important for modeling features in the clustering signal, and deals directly with the displacement vectors of the cosmic fluid, making it an ideal framework within which to understand their derivatives, i.e.\ cosmological velocities.

The goal of this paper is to develop a consistent Fourier-space model of both peculiar-velocity and redshift-space statistics. Our strategy is twofold: first, since the redshift-space power spectrum of galaxies can be understood in terms of series expansions of their velocity statistics, we explore the convergence of these expansions to understand their requirements and limitations. Our analysis of these expansions for halo power spectra uses nonlinear velocity spectra measured directly from simulations, which include nonlinear bias and fingers-of-god \cite{Jackson72}, and is a continuation of that in ref.~\cite{Vlah16}, who explored these convergence properties within the Zeldovich approximation, and refs.~\cite{Okumura12,Okumura12II}, who explored them in the context of matter and halo power spectra. 
Similar expansions using velocity statistics from N-body data have also been studied in configuration space for the Gaussian and Edgeworth streaming models \cite{ReiWhi11,Wan14,VlaCasWhi16,Uhlemann15}. Second, we use one-loop perturbation theory with effective corrections for small scale effects to model the requisite velocity statistics. Our work builds naturally on previous work in configuration space combining velocity statistics and the correlation function in LPT, particularly within the context of the Gaussian streaming model \cite{Pee80,Fis95,ReiWhi11,Rei12,Wan14,VlaCasWhi16}, though modeling these statistics in Fourier space enables us to more effectively extend the reach of perturbation theory.  We compare and contrast the behavior of these velocity statistics in both EPT and LPT.

This work is organized as follows. We begin in Section~\ref{sec:nbody} by describing the N-body simulations that we use throughout the paper. In Section~\ref{sec:vel_exp} we briefly review two methods of expanding velocity statistics in the redshift-space power spectrum (the moment expansion approach and the Fourier streaming model) and study their convergence at the level of velocity statistics measured from N-body simulations. We describe the modeling of these velocity statistics in perturbation theory in Section~\ref{sec:spectra} providing a comparison of and translation between the two approaches.  Finally, in Section~\ref{sec:rsd_pk} the velocity expansions and PT modeling of velocities are combined to yield a consistent model for the power spectrum within one-loop perturbation theory.  We conclude with a discussion of our results in Section~\ref{sec:conclusions}. In Appendices, we compare our work to existing models (\ref{app:others}, \ref{app:gsm}), discuss differences between power spectrum wedges and multipoles (\ref{app:wedges}) and provide details of our numerical calculations (\ref{app:fftpt},\ref{app:Hankel},\ref{app:bessel},\ref{app:python}).

\section{N-Body Simulations}

\label{sec:nbody}

In this paper we will use N-body data for two purposes: (1) to test the convergence of various velocity-based expansions for redshift space distortions using exact velocity statistics extracted from simulations and (2) to investigate the extent to which these velocity statistics can be modeled within 1-loop perturbation theory and combined to model the redshift-space power spectrum for biased tracers. To this end we make use of the halo catalogs\footnote{The data are available at http://www.hep.anl.gov/cosmology/mock.html.  Of the 5 realizations, the data for the first were corrupted so we used only the last 4.} from the simulations described in ref.~\cite{Sunayama16}.  These were the same simulations used in ref.~\cite{VlaCasWhi16}, to which the reader is referred for further discussion.  Briefly, there were 4 realizations of a $\Lambda$CDM ($\Omega_m=0.2648$, $\Omega_bh^2=0.02258$, $h=0.71$, $n_s=0.963$,  $\sigma_8=0.8$) cosmology simulated with $4096^3$ particles in a $4\,h^{-1}$Gpc box.  We measured the halo power spectrum in two mass bins ($12.5<{\rm lg} M<13.0$ and $13.0<{\rm lg}M<13.5$; all masses in $h^{-1}M_\odot$) at $z=0.8$ and $0.55$, in both real and redshift space.  We compute the power spectra in bins of width $0.0031\,h\,{\rm Mpc}^{-1}$, which is small enough that effects due to binning are $\mathcal{O}(0.1\%)$ for the theories we wish to test.  We additionally computed the Fourier-space pairwise velocity statistics up to fourth order in real space. The aforementioned quantities were all computed using the publically available {\tt nbodykit} software \cite{nbdkit}. The number densities and rough estimates for the linear biases of the halo samples we consider are given in Table~\ref{tab:nbody}.

The total volume simulated, $256\,h^{-3}{\rm Gpc}^3$, is equivalent to $>40$ and $>25$ full-sky surveys for redshift slices $0.5<z<0.6$ and $0.75<z<0.85$, respectively.  The statistical errors from the simulations should thus be much smaller than those of any future survey confined to a narrow redshift slice and are dominated by systematic errors in the algorithms or physics missing from the simulations themselves.  In fact, the simulations were run with ``derated'' time steps and halo masses were adjusted to match the halo abundance of a simulation with finer time steps \cite{Sunayama16}.  As detailed in ref.~\cite{VlaCasWhi16}, tests of halo catalogs produced with and without derated time steps lead us to assign a systematic error of several percent to the clustering statistics measured in these simulations.  Of direct relevance to redshift-space statistics, by comparing the mean-infall velocity and pairwise velocity dispersion on very large scales with linear theory predictions we see evidence that the velocities are underpredicted by about 1-2\% by $z = 0.55$. In particular we note that agreement with theory can be improved on all scales if we increase N-body velocities by such a constant factor. To keep the measured redshift-space power spectrum and velocity statistics consistent, we do not apply this correction. Rather, we choose to focus our analysis primarily on the redshift bin $z = 0.8,$ relevant in the near term for spectroscopic surveys such as DESI \cite{DESI} and where the accumulated effects of this systematic are less severe, noting that a few percent error is well within the error budget for simulations of this form.

\begin{table}
\begin{center}
\begin{tabular}{cccc}
${\rm lg}M$ & Redshift & $\bar{n}$ & $b$ \\ \hline
$12.5-13.0$ &   0.55   & 0.61      & 1.45 \\
$13.0-13.5$ &   0.55   & 0.19      & 1.93 \\
$12.5-13.0$ &   0.8    & 0.53      & 1.72\\
$13.0-13.5$ &   0.8    & 0.15      & 2.32 \\
`Galaxies'  &   0.8    & 0.80      & 1.97
\end{tabular}
\caption{Number densities and bias values for the samples we use.  Halo masses are $\log_{10}$ of the mass in $h^{-1}M_\odot$, number densities are times $10^{-3}\,h^3\,{\rm Mpc}^{-3}$.  The last row, labeled `Galaxies', refers to the mock galaxy sample drawn from the halo occupation distribution described in the text.}
\label{tab:nbody}
\end{center}
\end{table}

Finally we construct a mock galaxy sample at $z\simeq 0.8$ using a simple HOD applied to the dark matter halo catalogs.  Since it is not our goal to match any particular sample, but rather to investigate how well our model performs on a sample covering a wide range of halo masses and with satellite galaxies, we simply populate all halos above $M_{\rm cut}=10^{12.5}\,h^{-1}M_\odot$ with a ``central'' galaxy taken to be comoving with the halo and at the halo center.  We also draw a Poisson number of satellites with
\begin{equation}
  \langle N_{\rm sat} \rangle = \Theta\left(M-M_{\rm cut}\right)\left( \frac{M}{ M_1}\right)
  \qquad , \quad
  M_1=10^{14}\,h^{-1}M_\odot
\label{eqn:hod}
\end{equation}
and arrange them following a spherically symmetric NFW profile \cite{NFW} scaled by the halo concentration and virial radius.   In addition to the halo velocity, the satellites have a random, line-of-sight velocity drawn from a Gaussian with width equal to the halo velocity dispersion.  This sample has complex, scale-dependent bias and finger-of-god velocity dispersion on small scales providing a test of the ability of our model to fit observed galaxy samples which exhibit both properties.

\section{Redshift Space Distortions: Velocity Expansions and Convergence}
\label{sec:vel_exp}
\subsection{Formalism}

In large-scale surveys, line-of-sight positions are typically inferred by measuring redshifts. Since redshifts are affected by the peculiar motions of the observed objects, these inferred redshift-space positions $\bs$ will be shifted from the ``true'' positions $\bx$ of these objects according to $\bs = \bx + \hn (\hn \cdot \bv) / \mathcal{H}$, where $\hn$ is the unit vector along the line-of-sight and $\mathcal{H}=aH$ is the conformal Hubble parameter \cite{Pea99,Dod03}.
Overdensities in redshift space are thus related to their real space counterparts via number conservation as
\begin{align}
    1 + \delta_s(\bs, \tau) &= \int\ d^3\bx\ \big(1 + \delta_g(\bx,\tau) \big)\ \delta_D(\bs - \bx - \bu) \nonumber \\
    (2\pi)^3 \delta_D(\bk) + \delta_s(\bk) &= \int\ d^3\bx \ \big(1 + \delta_g(\bx,\tau) \big)\ e^{i\bk \cdot(\bx + \bu(\bx))},
\end{align}
where we have defined the shorthand $\bu = \hn (\hn \cdot \bv) / \mathcal{H}$. From the above, the redshift space power spectrum can be written as a special case of the (Fourier transformed) velocity moment-generating function \cite{VlaWhi19}
\begin{equation}
    \tilde{M}(\bJ,\bk) = \frac{k^3}{2\pi^2} \int\ d^3r\ e^{i\bk \cdot \textbf{r}}\ \avg{(1+\delta_g(\bx_1)) (1+\delta_g(\bx_2)) e^{i\bJ\cdot \Delta\bu}}_{\bx_1 - \bx_2 = \textbf{r}},
    \label{eqn:gen_func}
\end{equation}
where we have defined the pairwise velocity $\Delta\bu = \bu_1 - \bu_2$ and the $k^3/(2\pi^2)$ in inserted for convenience.  Specifically, we have
\begin{equation}
    \frac{k^3}{2\pi^2} P_s(\bk) = \tilde{M}(\bJ=\bk,\bk) = \frac{k^3}{2\pi^2} \int\ d^3r\ e^{i\bk \cdot \textbf{r}}\ \avg{(1+\delta_g(\bx_1)) (1+\delta_g(\bx_2)) e^{i\bk\cdot \Delta\bu}}_{\bx_1 - \bx_2 = \textbf{r}}.
    \label{eqn:rsd_power_spectrum}
\end{equation}
Note that the moment generating function with $\bJ = 0$ is directly proportional to the real space power spectrum, i.e.\  $\tilde{M}_0 = k^3 P(k) / (2\pi^2) = \Delta^2(k)$, where $\Delta^2(k)$ is the power per log interval in wavenumber in real space.

There exist many approaches to model the redshift space power spectrum (see e.g.\ refs.\ \cite{White15,GF19,VlaWhi19} for recent reviews). Roughly speaking, these techniques can be understood as different series expansions of the exponential in Equation~\ref{eqn:rsd_power_spectrum} (see e.g.\ the discussion in ref.~\cite{VlaWhi19}; a related discussion on the correlation function and velocity expansions in configuration space can be found in ref.~\cite{Cuesta20}). Our main objective here is to explore the effectiveness of two Fourier-space based approaches: the moment expansion (ME), or ``distribution function approach'' \cite{SelMcD11}, and the recently proposed Fourier Streaming Model (FSM) \cite{VlaWhi19}.

In the moment expansion approach the redshift-space power spectrum is derived by expanding the exponential in Equation~\ref{eqn:rsd_power_spectrum} such that
\begin{equation}
    \frac{k^3}{2\pi^2} P_s(\bk) = \tilde{M}(\bJ = \bk) = \frac{k^3}{2\pi^2} \sum_{n=0}^\infty \frac{i^n}{n!} k_{i_1} \cdots k_{i_n} \tilde{\Xi}^{(n)}_{i_1 \cdots
    i_n}(\bk)
    \label{eqn:moment_expansion}
\end{equation}
where the density-weighted pairwise velocity moments are defined to be the Fourier transforms of $\Xi_{i_1 \cdots i_n}^{(n)} = \avg{(1 + \delta_1)(1+\delta_2) \Delta\bu_{i_1}\cdots\Delta\bu_{i_n}}$.  For example, the first and second moments are the mean pairwise velocity between halos separated by distance $\textbf{r}$, $\Xi^{(1)}_{i} = v_{12,i}(\textbf{r})$, and the pairwise velocity dispersion, $\Xi^{(2)}_{ij} = \sigma_{12,ij}(\textbf{r})$\footnote{Since redshift-space distortions depend only on line-of-sight velocities the only nonzero contributions in Equation~\ref{eqn:moment_expansion} are those due to $k_{\hn} = k \mu$, where $\mu$ is the cosine of the angle between the line-of-sight (LOS) and wave vector, which in turn multiplies only velocity statistics projected along the LOS $\hn$. However, models of large-scale structure naturally predict not only the LOS component but the full tensorial quantity
\begin{equation}
    \Xi^{\prime (n)}_{i_1 ... i_n} = \mathcal{H}^{-n} \avg{(1 + \delta_1)(1+\delta_2) \Delta \bv_{i_1} \cdots \Delta \bv_{i_n}},
\end{equation}
where $\Delta \bv = \bv_1 - \bv_2$, along with its Fourier transform $\tilde{\Xi}^{\prime}$, such that the statistics of $\bu$ are given by the e.g.\  $\tilde{\Xi}^{(1)}_i = \tilde{\Xi}^{\prime (1)}_{\hn} \hn_i$. However, due to the symmetric structure of these velocity moments, the tensor components of $\Xi^\prime$ can be mapped 1-1 to the multipole moments of $\Xi$, and for this reason we will refer to them interchangeably throughout the text.}.

In the Fourier Streaming Model, the redshift-space power spectrum is evaluated by applying the cumulant theorem to the logarithm
\begin{equation}
    \ln \big[1 + \Delta(k) \big] = \ln \big[1 + \tilde{M}(\bJ=0,\bk) \big] + i J_i \tilde{C}^{(1)}_i(\bk) - \frac{1}{2} J_i J_j \tilde{C}^{(2)}_{ij} + ...
\end{equation}
The first few cumulants are related to the Fourier pairwise velocity moments by
\begin{align}
    \tilde{C}^{(1)}_i(\bk) &= \frac{k^3}{2\pi^2} \frac{\tilde{\Xi}_i(\bk)}{1 + \Delta^2} \nonumber \\
    \tilde{C}^{(2)}_{ij}(\bk) &= \frac{k^3}{2\pi^2} \frac{\tilde{\Xi}_{ij}(\bk)}{1 + \Delta^2} - \tilde{C}^{(1)}_i \tilde{C}^{(1)}_j \nonumber \\
    \tilde{C}^{(3)}_{ijk}(\bk) &= \frac{k^3}{2\pi^2} \frac{\tilde{\Xi}_{ijk}(\bk)}{1 + \Delta^2} - \tilde{C}^{(2)}_{\{ij} \tilde{C}^{(1)}_{k\}} - \tilde{C}^{(1)}_i \tilde{C}^{(1)}_j \tilde{C}^{(1)}_k\nonumber \\
    \tilde{C}^{(4)}_{ijkl}(\bk) &= \frac{k^3}{2\pi^2} \frac{\tilde{\Xi}_{ijkl}(\bk)}{1 + \Delta^2} - \tilde{C}^{(3)}_{\{ijk} \tilde{C}^{(1)}_{l\}} - \tilde{C}^{(2)}_{\{ij} \tilde{C}^{(2)}_{kl\}} - \tilde{C}^{(1)}_i \tilde{C}^{(1)}_j \tilde{C}^{(1)}_k \tilde{C}^{(1)}_l,
\end{align}
The redshift-space power spectrum is then
\begin{equation}
    1 + \frac{k^3}{2\pi^2} P_s(\bk) = \big(1 + \Delta^2(k) \big) \exp\left[ \sum_{n=1}^{\infty} \frac{i^n}{n!} k_{i_1} ... k_{i_n} \tilde{C}^{(n)}_{i_1 ... i_n}(\bk)  \right].
\end{equation}
At any order the nonlinearity of the exponential in the FSM will produce a resummation of select terms when compared to the moment expansion. Indeed, ref.~\cite{VlaWhi19} found distinct differences in the rate of convergence for the case of Zeldovich matter dynamics. However, the two expansions are necessarily equivalent order-by-order in the Taylor-series expanded pairwise velocities, and on scales where $\Delta^2 \lesssim 1$, they will tend to behave similarly. Evaluating whether the differences between the two expansions are significant for halos and galaxies with nonlinear bias and dynamics will be one of the goals of the following sections.

\subsection{Comparison of methods using simulated data}

The Fourier-space velocity expansions described in the previous subsection can be tested by comparing the redshift-space power spectra measured in N-body simulations to velocity power spectra measured from the same simulations. Our aim in this subsection is to use this comparison to test the convergence of each expansion at $n^{\rm th}$ order in both the moment expansion and Fourier streaming approaches. Since the velocity expansions are effectively expansions in both $k$ and $\mu$ we will focus on their convergence in terms of power spectrum wedges, sufficiently finely binned such that their values are equivalent to $P(k,\mu_i)$ where $\mu_i$ is the central value of each angular bin, but comment on the extension to power spectrum multipoles where appropriate.

\begin{figure}
    \centering
    \includegraphics[width=\textwidth]{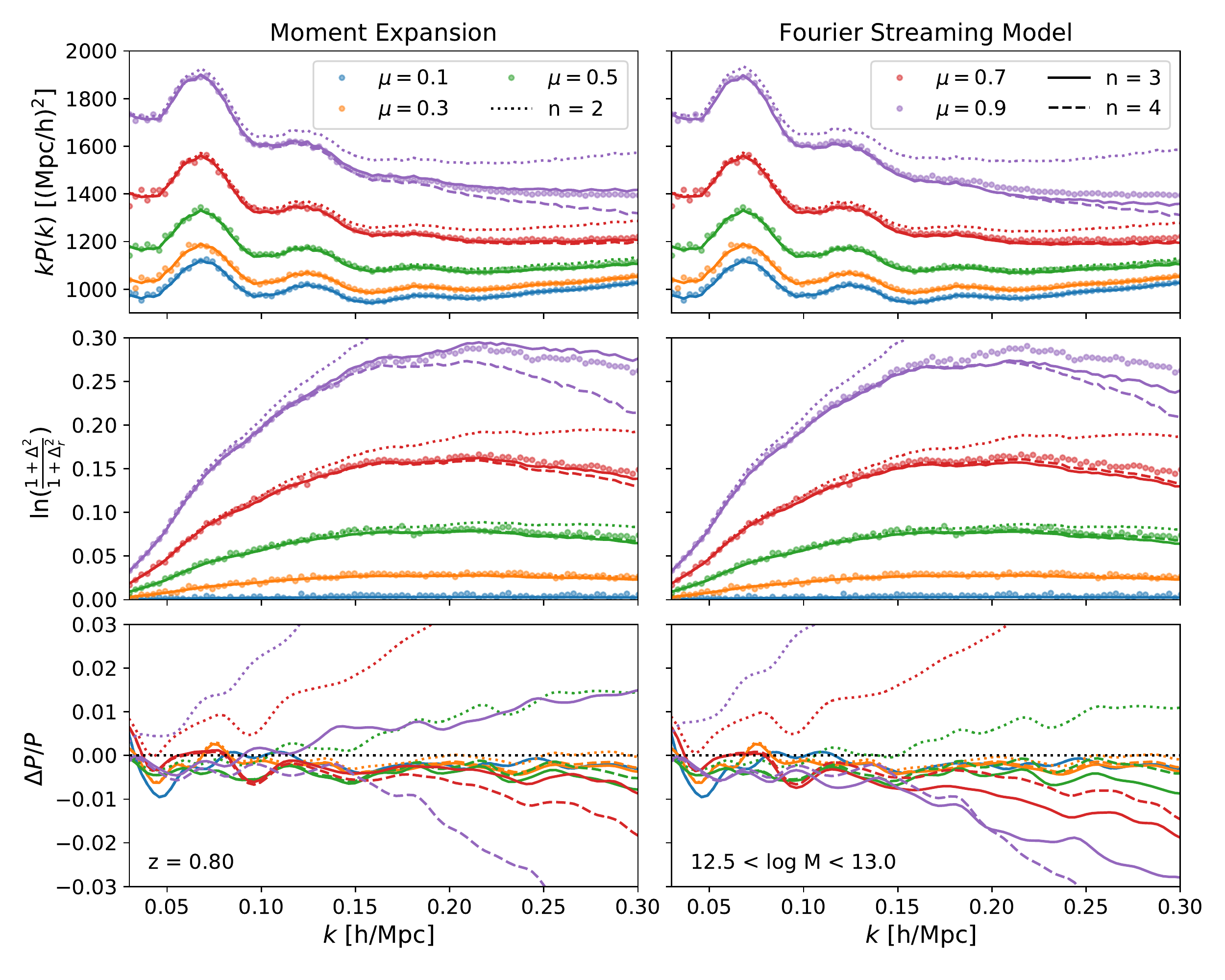}
    \caption{Convergence for the moment expansion (left) and Fourier streaming model (right) at each order in velocity statistics -- using inputs extracted from simulation data -- for halos of mass $12.5 < \log M < 13.0$ (in $h^{-1}M_\odot$) and $z = 0.8$. The top, middle, and bottom columns show five wedges $P(k,\mu)$ represented as $k P(k)$, the log ratio of $1 + \Delta$ in real and redshift space, and the error of each method (smoothed for presentation) and order compared to N-body data. While going from $n=2$ to $n=3$ dramatically improves agreement at essentially all scales, especially for large $\mu$, going to $n=4$ mostly only improves the asymptotic convergence at low $k$ and $\mu$ at the mostly subpercent level without significant improvement at higher $k$ and $\mu$.}
    \label{fig:data_expansion}
\end{figure}

\begin{figure}
    \centering
    \includegraphics[width=\textwidth]{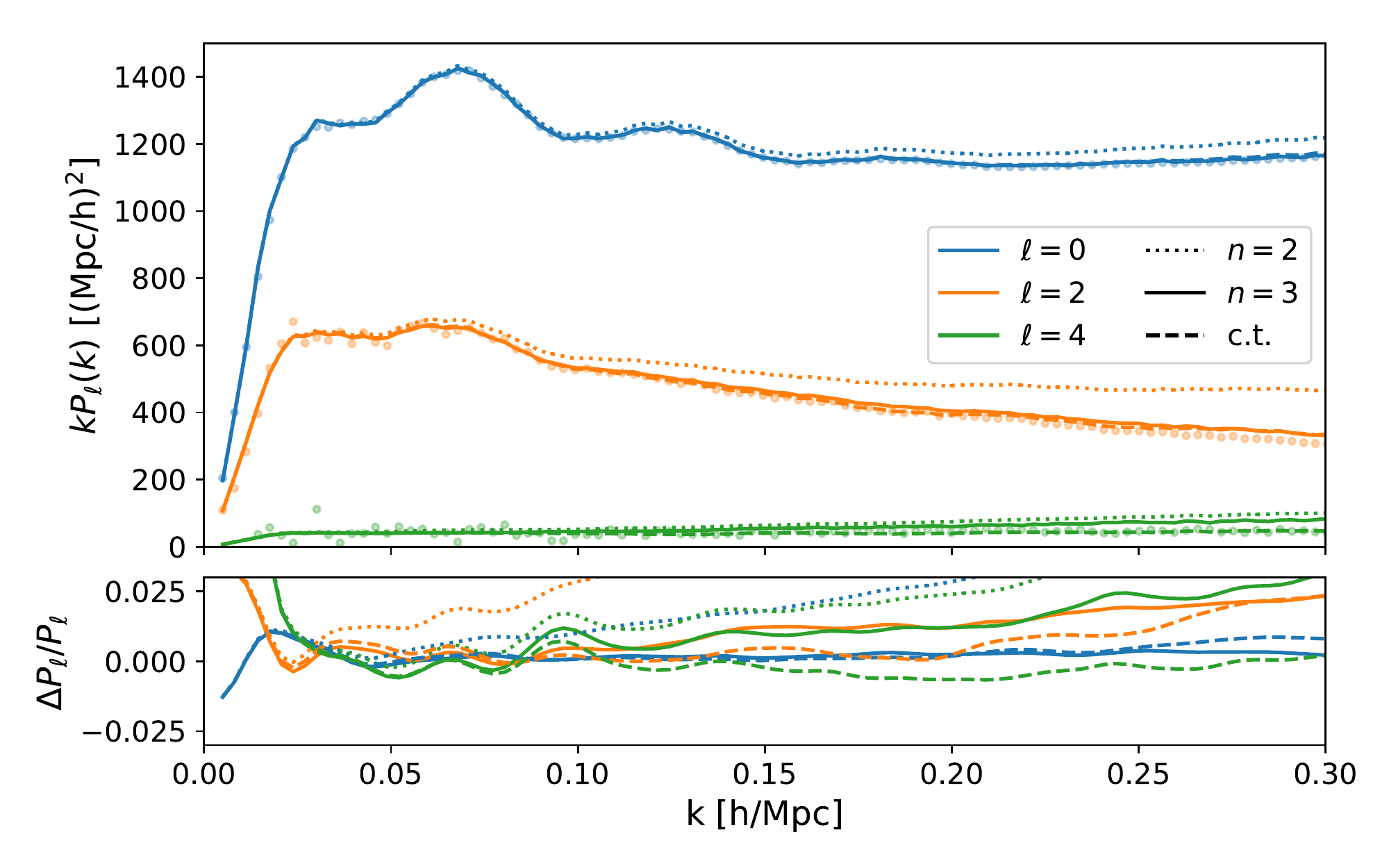}
    \caption{Convergence of the moment expansion at $z = 0.8$ for the first three multipoles of the redshift space power spectrum. The top panel shows $k P_\ell$ while the bottom panel shows the fractional error in each expansion, smoothed to highlight systematic trends. Similarly to the wedges, going from $n = 2$ to $n = 3$ presents substantial improvements in all three multipoles, with the agreeement in the quadrupole going from worse than 50 percent for $n = 2$ to a few percent at perturbative scales ($k < 0.25 \kMpc$).  In interpreting these differences it is important to bear in mind that for any observation the error on the quadrupole and hexadecapole are dominated by the monopole contribution and are therefore fractionally much larger than for the monopole.
    }
    \label{fig:data_plms}
\end{figure}

Figure~\ref{fig:data_expansion} shows the convergence of the moment expansion and Fourier streaming model for halos of mass $12.5 < \log M < 13.0$ in units of $h^{-1} M_\odot$ at $z = 0.8$ at orders $n = 2$, 3, 4 in each method using velocity spectra $\tilde{\Xi}^{(n)}(\bk)$ from simulations.  The dots show power spectrum wedges (arranged by color in $\mu$) extracted from simulations, while the curves show predictions for each model when keeping velocity statistics up to $n^{th}$ order. The top two rows show the wedges expressed as $k P(k,\mu)$ and the ratio $\ln( [1+\Delta^2_s]/[1+\Delta^2_r])$, while the bottom row shows the fractional difference between the data and models. The ME and FSM behave very similarly, except at high $k$ and $\mu$ where they diverge.  This can be understood from the fact that the redshift-to-real-space logarithm shown in the middle row is significantly below unity for most of the angles and scales shown, except for the $\mu = 0.9$ wedge where it reaches 30\% and where the ME seems to have somewhat better convergence properties at high $k$. In both models, going from $n = 2$ to $n = 3$ dramatically improves the broadband shape predictions at $k > 0.05\kMpc$, especially in the highest $\mu$ bins where the improvement can be in the tens of percents. As a further test, we compute the multipoles predicted by the moment expansion at $n = 2$  and $3$ and compare them to the data in the right panel of Figure~\ref{fig:data_plms}. Once again, while staying at $n = 2$ grossly mis-estimates the power spectrum quadrupole, going to $n = 3$ yields excellent agreement on these scales. A similar improvement when incorporating third-order velocity statistics extracted from simulations was seen by refs.~\cite{Uhlemann15,Cuesta20} in configuration space in the context of correlation function multipoles (see Appendix \ref{app:gsm} for further discussion of configuration space). Interestingly, the fractional error on the quadrupole in both cases grows slightly faster than the the fractional error in the highest $\mu$ bin in Figure~\ref{fig:data_expansion} (rather than the fractional error of some intermediate wedge), while the fractional error on the hexadecapole far exceeds that of any wedge. We comment on these counter-intuitively large errors for multipoles and implications for data analyses in Appendix~\ref{app:wedges}.

Going to $n = 4$ improves the behavior at low $k$ and $\mu$, but it does not improve -- indeed somewhat worsens -- the recovery of the broadband shape over the  scales smaller than $k\sim0.15\kMpc$. This suggests that the reach of both the ME and FSM are limited to perturbative scales, $k |\Delta \textbf{u}| \lesssim 1$, by the magnitude of the halo velocities and $n=3$ almost saturates this reach.  Indeed, at the scale where the virial velocities of halos become important one might expect that all velocity moments and cumulants contribute significantly to the redshift-space power, slowing the convergence of the velocity expansions. The fact that the inclusion of higher velocity moments does not obviously improve convergence suggests that extending treatments of RSD beyond industry-standard 1-loop order for extended reach in $k$ might give meager returns beyond those generated from overfitting with more parameters. We have chosen to focus on this mass bin and redshift for ease of presentation but note that the other samples discussed in Section~\ref{sec:nbody} exhibit qualitatively similar behavior; however, we caution that halos at even higher redshifts --- relevant to futuristic galaxy surveys \cite{Ferraro19,Schlegel19,MSE19,Ellis19} or 21-cm surveys \cite{Slosar19} for example --- might behave differently due both to the diminishing magnitude of large-scale velocities and differences in virial motions at high redshifts.

The above results suggest that in order to reproduce the broadband shape of $P(k,\mu)$ at the percent level on perturbative scales (k $\sim 0.25 \kMpc$) it should be sufficient to model velocity statistics up to third order. However, as we have already discussed we can expect that the higher velocity statistics will be dominated by stochastic contributions, i.e. the small scale virial motions of galaxies or halos. In this limit, neglecting the connected contributions to the correlator (see refs.~\cite{Vla12,Vla13} for similar decomposition), we have
\begin{equation}
    \Xi^{(3)}_{ijk}(\textbf{r}) = \avg{ (1 + \delta_1) (1 + \delta_2) \Delta \bu_i \Delta \bu_j \Delta \bu_k} \approx  \avg{\Delta \bu_{\{i} \Delta \bu_j}  \Xi^{(1)}_{k\}}(\textbf{r})  \approx \sigma_v^2 \delta_{\{ij}  \Xi^{(1)}_{k\}}(\textbf{r}) \nonumber
\end{equation}
where the curly brackets indicate a sum over symmetric combinations of $i, j, k$. At leading order in the moment expansion this is equivalent to a counterterm-like contribution
\begin{equation}
    P_{s}(\bk) \ni \frac{1}{2} k_\parallel^3 \sigma_v^2\  \tilde{\Xi}^{(1)}_\parallel(\bk) \approx \frac{1}{2} \sigma_v^2 k^2 \mu^4 P_{\rm L}(k),
\end{equation}
where $P_{\rm L}$ stands for the linear theory prediction with appropriate factors of bias. The predictions for using the moment expansion at $n = 2$ combined with this contribution are shown in dashed lines in Figure~\ref{fig:data_plms}. In addition to providing excellent agreement in the monopole and quadrupole, the counterterm also gives a good fit to the hexadecapole. This supports the  assumption we made above of keeping only the disconnected piece of the $n=3$ velocity moment, indicating that due to the relativly large contribution of the small-scale part of the velocity dispersion, $\sigma_v^2$, this term dominates over the connected contributions on the scales of interest.  We anticipate that this conclusion would only be strengthened by considering small-scale virial motions of satellite galaxies. This suggests that we focus our modeling efforts on the first two velocity moments, and in the next two sections we shall discuss the modeling of these moments in 1-loop perturbation theory.

\begin{figure}
    \centering
    \includegraphics[width=\textwidth]{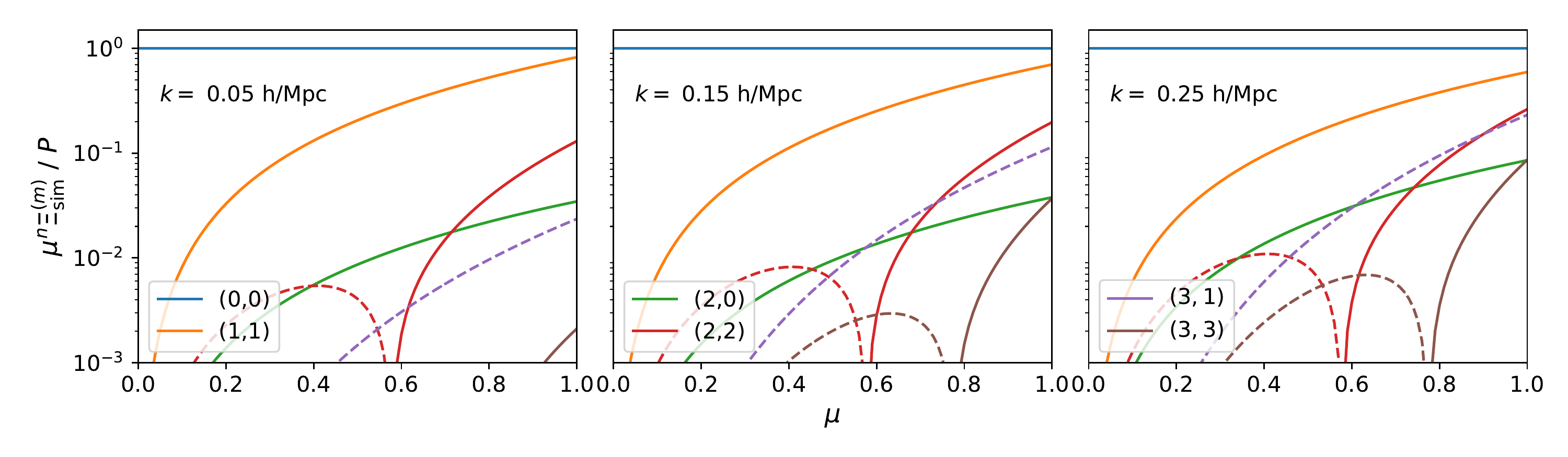}
    \caption{Angular contributions $(n,m)$ to the redshift-space power spectrum from the $m^{th}$ multipole of the $n^{th}$ velocity moment at three wavenumbers $k = 0.05$, $0.15$, $0.25 \kMpc$ as a fraction of the real-space power spectrum. The anisotropic signal is dominated by the first moment at all scales. For higher multipole moments, for example the quadrupole of the second moment, the absolute magnitude of the contribution to $P_s(k,\mu)$ is small at intermediate $\mu$ due to the occurence of zeros in $\mathcal{L}_\ell$.
    }
    \label{fig:components_mu}
\end{figure}

Finally, it is instructive to consider the relative roles played by the multipole moments of the velocity moments in the redshift-space power spectrum. By symmetry we can write each line-of-sight velocity moment as
\begin{equation}
    \tilde{\Xi}^{(n)}_{\rm LOS}(\bk) = \sum_{\ell = 0}^n \tilde{\Xi}^{(n)}_\ell(k) \mathcal{L}_\ell(\mu),
\end{equation}
where $\mathcal{L}_\ell(\mu)$ are Legendre polynomials of the line-of-sight angle; \edit{since each moment $\tilde{\Xi}^{(n)}$ gets multiplied by $(k\mu)^n$ in the moment expansion, the components $\tilde{\Xi}^{(n)}_\ell$ contribute with the angular structure $\mu^n \mathcal{L}_\ell(\mu)$}. As an example, in Figure~\ref{fig:components_mu} we have plotted the thus-enumerated contributions to $P_s(k,\mu)$ at three representative wavenumbers as a fraction of the real-space power spectrum at that wavenumber. At all of these scales, which cover the reach of perturbation theory at low redshifts, the anisotropic signal is dominated by the first moment, which contributes proportionally to $\mu \mathcal{L}_1$, with the relative importance of higher moments roughly increasing with LOS angle $\mu$. Moreover, the root structure of Legendre polynomials with $\ell > 0$ plays an interesting role in the relative prominence of each contribution--for example, while the quadrupole moment of $\Xi^{(2)}$ is typically larger in absolute magnitude than the monopole, its relative importance at intermediate $\mu$ can be comparatively suppressed due to proximity to the root of $\mathcal{L}_2(\mu)$ at $\mu = 1/\sqrt{3}$, and similarly for the octopole moment of $\Xi^{(3)}$. On the other hand, beyond these intermediate $\mu$ we expect the contamination of the cosmological signal by small scale (FoG) effects, as well as the importance higher velocity moments, to be increasingly large. Indeed, as we will see for realistic (galaxy) samples the monopole of $\Xi^{(2)}$ will tend to contain a large, constant small-scale contribution, further increasing its relative importance over the quadrupole. Roughly speaking, then, the contributions to the redshift-space power spectrum rank in importane as $\tilde{\Xi}^{(0)}_0, \tilde{\Xi}^{(1)}_1, \tilde{\Xi}^{(2)}_0, \tilde{\Xi}^{(2)}_2, \tilde{\Xi}^{(3)}_1$, and so on. 

\section{Pairwise Velocity Spectra in Perturbation Theory}
\label{sec:spectra}

In this section we present formulae for the real-space pairwise velocity spectra required for both the ME and FSM in Lagrangian and Eulerian perturbation theory. These quantities live naturally in configuration space, where they can be directly interpreted as density-weighted pairwise velocities, while in Fourier space they must be broken down into components to be measured.  While we shall primarily employ the velocity spectra for computation of the redshift-space power spectrum, we emphasize that pairwise velocity statistics are well-defined, Galilean invariant quantities and have the potential to be measured (in redshift space) by future kSZ and peculiar velocity surveys \cite{Sug16,Howlett17,Kim19}.  They are therefore interesting in their own right.  Our results for the zeroth, first and second moments of the pairwise velocity in LPT are the Fourier-space analogues of the results presented in ref.~\cite{VlaCasWhi16}, though we differ slightly in the treatment of counterterms in the velocity dispersion, include stochastic contributions to both densities and velocities and a superset of the density-bias expressions given in ref.~\cite{VlaWhi19}. We organize the expressions so that they can be efficiently evaluated numerically by converting the angular integrals into sums over spherical Bessel functions, then treating the resulting tower of Hankel transforms via the FFTLog algorithm \cite{Ham00,VlaSelBal15,VlaCasWhi16}. The explicit form of these Hankel transforms is given in Appendix \ref{app:Hankel}. Throughout this section and the next we will compare our theoretical predictions to velocity statistics of the same halos studied in Section~\ref{sec:vel_exp} (i.e. $12.5 < \log M < 13.0$ at $z = 0.8$). Results for the other mass bins and redshifts are qualitatively similar, though the potential for even higher systematics in the N-body data at lower $z$ are an important caveat.  We shall consider our mock galaxy catalogs when we combine the ingredients into the redshift-space power spectrum.

\subsection{Background}

\subsubsection{Lagrangian and Eulerian Perturbation Theory}
\label{sssec:PT}

The two conventional frameworks within which to perturbatively model cosmological structure formation are Eulerian and Lagrangian perturbation theory (see the references in the introduction). Lagrangian perturbation theory models cosmological structure formation by tracking the trajectories $\bx(\bq,t) = \bq + \bPsi(\bq,t)$ of infinitesimal fluid elements originating at Lagrangian positions $\bq$. These fluid elements cluster under the influence of gravity and their displacements obey the equation of motion $\ddot{\bPsi} + \mathcal{H}\dot{\bPsi} = - \nabla \Phi(\bx)$ --- where the dotted derivatives are with respect to conformal time $\tau$, $\mathcal{H} = a H$ is the conformal Hubble parameter and $\Phi$ is the gravitational potential --- which we solve for order-by-order in terms of the initial density contrast $\delta_0$ as $\bPsi = \bPsi^{(1)} + \bPsi^{(2)} + \cdots$, where
\begin{equation}
    \bPsi^{(n)}_i(\bq) = \frac{i^n}{n!} \int_{\bk,\ \bp_1 ... \bp_n} e^{i\bk \cdot \bq}\ \delta^D_{\bk - \bp}\  L^{(n)}_i(\bp_1,...,\bp_n)\ \tilde{\delta}_0(\bp_1) ... \tilde{\delta}_0(\bp_n),
\end{equation}
were we use the shorthands $\bp = \sum_i \bp_i$, $\delta^D_{\bk-\bp}=(2\pi)^3\delta^{(D)}(\bk-\bp)$ and $\int_\bp = \int  d^3\bp/(2\pi)^3$. Expressions for the n$^{\rm th}$ order kernels can be found in, for example, ref.~\cite{Mat08a}. By contrast, Eulerian perturbation theory (EPT, often also called standard perturbation theory: SPT), solves perturbatively for the density and velocity at the observed, Eulerian position $\bx$ (see e.g.\ ref.~\cite{Ber02}), i.e. 
\begin{align}
\delta(\bk) &= \sum_n \int_{\bp_1\ldots \bp_n} \delta^D_{\bk - \bp_{1n} } F_n(\bp_1, \ldots, \bp_n) \tilde{\delta}_0(\bp_1)\ldots \tilde{\delta}_0(\bp_n), \\
v_i(\bk) &= - i f \mathcal H \frac{k_i}{k^2} \sum_n \int_{\bp_1\ldots \bp_n} \delta^D_{\bk - \bp_{1n} } G_n(\bp_1, \ldots, \bp_n) \tilde{\delta}_0(\bp_1)\ldots \tilde{\delta}_0(\bp_n). \notag
\label{eqn:ept_kernels}
\end{align}
However, despite the apparent differences LPT and EPT are formally equivalent (see e.g.\ the discussion in ref.~\cite{McQWhi16}). In particular, by solving for the observed matter overdensity
\begin{equation}
    1 + \delta(\bx) = \dtq \delta_D(\bx - \bq - \bPsi), \quad (2\pi)^3 \delta_D(\bk) + \delta(\bk) = \dtq e^{- i \bk \cdot (\bq + \bPsi)},
    \label{eqn:lpt_dens}
\end{equation}
order-by-order in the linear initial conditions, one recovers the expressions of EPT, and similarly for velocity statistics by weighting the integral above by appropriate functions of the velocity $\dot{\bPsi}(\bq)$. Nonetheless, the exponentiated displacements in Equation~\ref{eqn:lpt_dens} can be used to motivate resummations of particular contributions to the nonlinear density due to long-wavelength (IR) displacements \cite{SenZal15,VlaWhiAvi15}, which can lead to dramatic differences with the predictions of (pure) EPT, as we will see later.  A proper treatment of these IR displacements is important for cosmological inference.

\subsubsection{Modeling biased tracers}
\label{sssec:bias}

The fact that cosmological surveys generally do not observe the underlying matter distribution but rather tracers of the nonlinear density field such as halos and galaxies presents an additional complication in mapping theory to observations. In PT one approaches this problem by perturbatively expanding the large-scale component of the galaxy and halo field that responds to the short-wavelength (UV) galaxy and halo formation physics via the so-called bias coefficients (see e.g.\ ref.\ \cite{Des16} for a review, and recent ref. \cite{Fujita+:2020} for a direct construction based on the equivalence principle). Once again the treatment of bias in LPT and EPT, though ultimately equivalent, are subtly different; we will now describe them in turn.

In the Lagrangian approach the positions of discrete tracers like galaxies and halos are assumed to be drawn according to a distribution depending on local initial conditions such that their overdensities in their initial (Lagrangian) coordinates are given by
\begin{alignat}{2}
F[\delta_0(\bq), s_{0,ij}(\bq), ..., \nabla \delta_0(\bq)]  &=  1 &&+ \delta_g(\bq,\tau_0)\nonumber \\
&= 1 &&+ b_1 \delta_0(\bq) + \frac{1}{2}b_2 \big(\delta_0^2(\bq) - \avg{\delta_0^2} \big) + b_s \big(s_0^2(\bq) - \avg{s_0^2}\big) \nonumber \\
& &&+ b_3\ O_3(\bq) +  \cdots + b_\nabla \nabla^2 \delta_0(\bq) + \epsilon(\bq),
\end{alignat}
where $s_0$ is the initial shear field\footnote{The inclusion of the initial shear and Laplacian information, in addition to the initial density, improves the ability to model assembly bias to the extent that this is encoded in the peak statistics (e.g.\ ref.~\cite{Dalal08}).} and we have included a representative third-order operator $O_3$ to account for the various degenerate contributions to the power spectrum at one-loop order \cite{McDRoy09}. Definitions for these quantities are given in Appendix \ref{app:others}.  Given this bias functional, these initial overdensities can then be mapped to the evolved overdensities of biased tracers via number conservation much like the nonlinear matter density:
\begin{align}
    1 + \delta_g(\bx,\tau) &= \dtq F(\bq)\ \delta_D(\bx - \bq - \bPsi(\bq,\tau)) \nonumber \\
    (2\pi)^3 \delta_D(\bk) + \delta_g(\bk) &= \dtq e^{i\bk \cdot (\bq+ \bPsi(\bq))} F(\bq). 
\end{align}
In this way, within LPT we have the apparent separation of clustering due to initial biasing in $F(\bq)$ and clustering due to nonlinear dynamics enforced by the equality $\bx = \bq + \bPsi$.

In the Eulerian approach, on the other hand, the galaxy overdensity is expressed in terms of a bias expansion based on present-day operators such as the nonlinear density $\delta(\bx)$. 
Here we adopt the biasing scheme of ref.~\cite{McDRoy09}, where up to third order a biased tracer field is expanded in terms of the nonlinear Eulerian fields as
\begin{equation}
\delta_h = c_1 \delta + \frac{c_{2}}{2} \delta^2+c_{s} s^2 + \frac{c_{3}}{6}\delta^3+c_{1s}\delta s^2 + c_{st} st + c_{s3} s^3+c_\psi \psi,
\label{eqn:ept_bias}
\end{equation}
where $s^2= s_{ij}s_{ij}$, $s^3= s_{ij}s_{jl}s_{li}$ and $st= s_{ij}t_{ij}$, and the shear operators are defined as
\begin{equation}
\psi = \eta-\frac{2}{7}s^2 +\frac{4}{21}\delta^2,
~~ s_{ij} = \left(\frac{\partial_i\partial_j}{\partial^2}-\frac{1}{3}\delta_{ij}\right)\delta,
~~ t_{ij} = \left(\frac{\partial_i\partial_j}{\partial^2}-\frac{1}{3}\delta_{ij}\right)\eta,
~~ \eta = \theta -\delta.
\end{equation}
In the above bias expansion we also implicitly assume subtraction of mean field values like $\left\langle\delta^2 \right\rangle$.

Despite formal differences, the bias schemes in LPT and EPT can in fact be mapped to one another via the appropriate linear transformations of the bias parameters (see e.g.\ refs.~\cite{Cha12,Saito14}). Indeed, these two approaches are a subset of a more general scenario in which the response of tracer formation to the large-scale structure is local in space but not in time, requiring us to take into account the evolution of the density field in the neighborhood around a tracer's trajectory; fortunately, these time-dependent responses have been shown to be perturbatively factorizable and equivalent to either LPT or EPT \cite{Sen14,Des16,Abidi18}. For our purposes, at one loop we have that the rotation\footnote{In performing this rotation we have implicitly assumed that the contributions from $c_3$ and $b_3$ degenerate with linear bias have been removed.} between the Lagrangian and Eulerian bases can be accomplished by (see e.g.\ ref.~\cite{Des16})
\begin{align}
    c_1 &= 1 + b_1 \nonumber \\
    c_2 &= b_2 + \frac{8}{21} b_1, \quad c_s = b_s - \frac{2}{7} b_1 \nonumber \\
    c_3 &= b_3 + a b_1 
    \label{eqn:bias_map}
\end{align}
where we have used $b$ and $c$ to distinguish between the Lagrangian and Eulerian bias parameters, respectively, and $a$ is a constant depending on which third-order bias parameter one chooses.  For instance, choosing the third order operator to be $st = s_{ij} t_{ij}$ we obtain $c_{st} = b_{st} + \frac{1}{3} b_1$. Beyond being necessary to complete the correspondence between LPT and EPT, these bias mappings can also be of practical use; for example there is some evidence that higher order Lagrangian bias is small for halos in N-body simulations and the higher-order Eulerian bias parameters are generated primarily by evolution \cite{Abidi18,Schmittfull19,Modi19}.  The Eulerian $c_n$ thus tend towards those predicted by ``local'' Lagrangian bias, allowing us to set useful restrictions on the Eulerian biases in EPT analyses.

\subsubsection{Derivative Corrections and Stochastic Contributions}

In addition to the bias operators discussed in the previous subsection, one also needs to consider terms from the derivative expansion and contributions arising purely from the coupling of short modes (stochastic contributions). 
In this paper, we follow the standard approach in the literature (see, e.g., ref.~\cite{Des16} for a review) and add the leading order derivative contributions in the galaxy field of the form $(\partial/k_\star)\delta$ (in the appropriate coordinates for LPT and EPT). In the power spectrum, these terms generically result in contributions of the form $(k^2/k_\star^2) P_{\rm lin}$ (or $(k^2/k_\star^2) P_{\rm Zel}$ in case of LPT). In most of the velocity moment power spectra, these terms are degenerate with the counterterm contributions at one-loop order. We explicitly account for these in each of the moments discussed below and finally combine them in the redshift space power spectrum. 

Stochastic contributions,  in the RSD power spectrum as well as velocity moments,  can come in two forms. First, we should add the pure noise field $\epsilon$ to our density expansion, which captures the galaxy field component uncorrelated with the long density fields and is characterized by scale-independent autocorrelations (shot noise). The second type of stochastic contributions appear as small-scale counterterms of the contact velocity correlators of the form $\left\langle v^n(x) \right\rangle$ that feature prominently in the higher velocity moments. These terms are traditionally labeled as ``Finger of God'' terms \cite{Jackson72}.  They reflect the non-linear structure of the redshift space mapping, encapsulating the feedback of small-scale (non-perturbative) velocity modes on the correlators on large scales.

It is important to note that `perturbative' operators carry the bulk of the cosmological dependence, while stochastic terms mostly parameterize the part of the signal that is decorrelated with the linear density fluctuations and consequently with the initial conditions. Thus, once stochastic parameters \edit{dominate}, it can be taken as an indication that little cosmological signal is left to be extracted from these scales. \edit{However, it is important to distinguish between pure stochastic terms, such as shot noise, and FoG-like contributions due to stochastic velocities; the latter behave like counterterms with shapes that depend nontrivially on large-scale modes. Similarly,  higher derivative terms can show a significant correlation with long-wavelength fluctuations and thus, in principle, can also carry cosmological information.} However, heavy reliance on these terms can, in practice, lead to many approximate degeneracies and thus can quickly reduce the amount of information available from the scale dependence of the correlations of interest. In the rest of this section, we shall see how velocity moments exhibit this behavior, with higher moments displaying stronger reliance on stochastic and derivative contributions.

\subsection{Velocity Correlators in LPT and EPT}

Having reviewed the essential ingredients of LPT and EPT, our goal in this subsection is to provide expressions for the pairwise velocity moments at one loop in both formalisms. In LPT, these can be naturally computed as derivatives of the generating functional in Equation \ref{eqn:gen_func}, which can be written as
\begin{equation}
    M(\bJ,\bk) = \frac{k^3}{2\pi^2} \dtq e^{i\bk \cdot \bq}\ \langle F(\bq_1) F(\bq_2)\ e^{i \bk \cdot \Delta + i \bJ \cdot \dDelta} \rangle_{\bq = \bq_1 - \bq_2},
\end{equation}
where $\Delta = \bPsi_1 - \bPsi_2$ and $\dDelta$ is its time derivative, and which has the additional benefit that derivatives with respect to $\bJ$ are automatically Galilean invariant. In EPT, on the other hand, the pairwise velocity moments are most straightforwardly computed by decomposing them into density-velocity correlators
\begin{equation}
P_{LL'} (k, \mu)
\equiv
\left< \big(1 + \delta\big)* u^{L}_{\hat n}
\Big| \big(1 + \delta\big)* u^{L'}_{\hat n}\right>',
\end{equation}
where, for brevity, we introduce the primed expectation values to denote expectation values with Dirac delta function dropped and a bar notation to indicate the arguments, i.e.\ $
\left< A | B \right> \equiv
\langle A(\bk) B(\bk') \rangle =  (2\pi)^3 \delta_D (\bk + \bk')\  \langle A(\bk) B(\bk') \rangle'$. Working at one loop in perturbation theory yields non-zero zeroth through fourth velocity moments, which we will now describe in detail.

\subsubsection{Zeroth Moment: Power Spectrum}

\begin{figure}
    \centering
    \includegraphics[width=\textwidth]{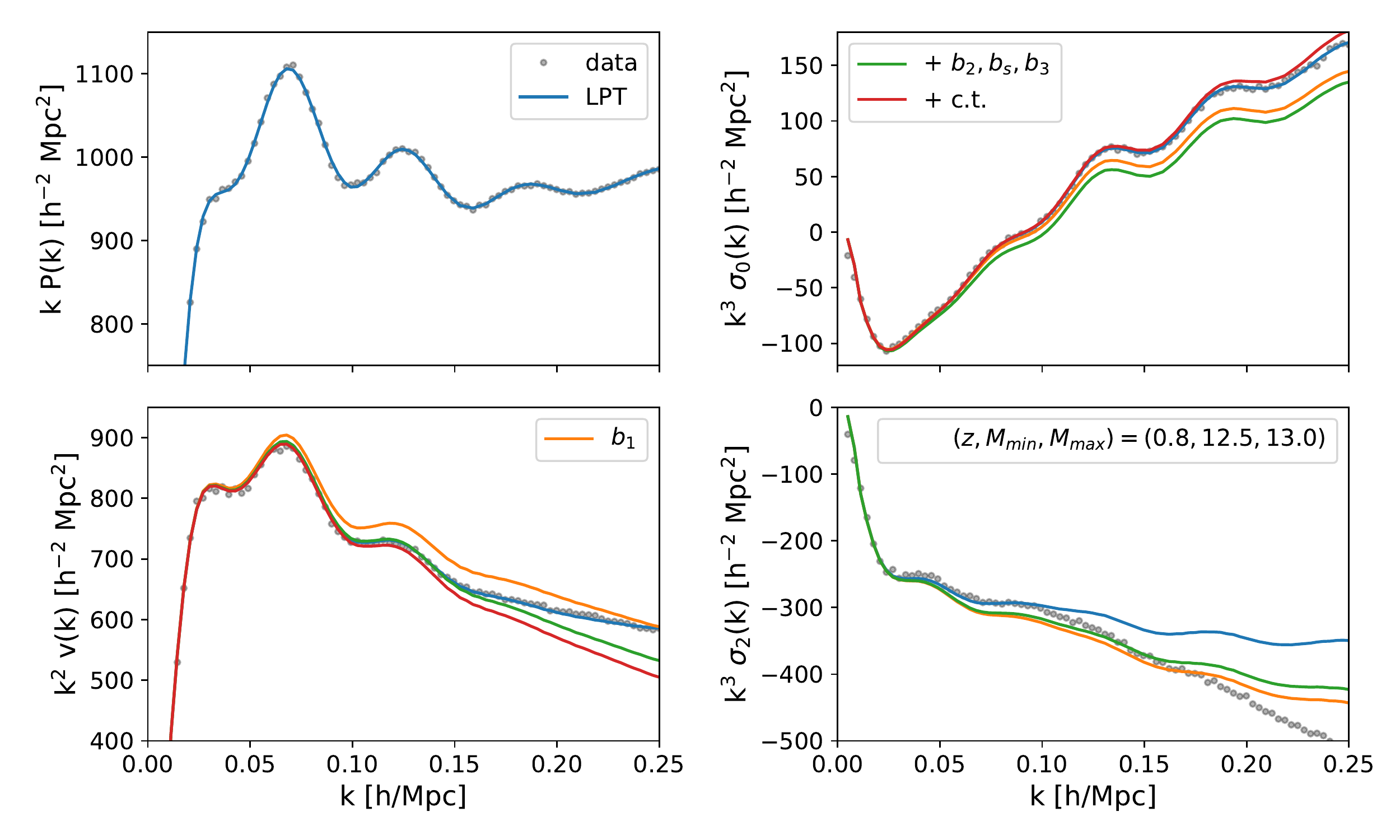}
    \caption{Fits to zeroth ($P(k)$, top left), first ($v(k)$, bottom left), and second ($\sigma$, right column) halo pairwise velocity moment spectra measured from simulations (gray points) in one-loop Lagrangian perturbation theory (blue) for the fiducial mass bin and redshift. The second moment is split into its monopole and quadrupole for ease of presentation. The contributions from sequentially adding linear bias (orange), nonlinear bias (green) and counterterms (red) are also shown as separate curves. The full model (blue) differs from the red curves by stochastic contributions (though they are identical for $\sigma_2$, for which we do not include any stochastic corrections in the lower right panel). We do not include the separate contributions to the power spectrum as the stochastic contribution contributes significantly at all scales. Our model fits these velocity statistics at the percent level out to $k = 0.25 \kMpc$, except for $\sigma_2$ which is only fit to around $k = 0.1 \kMpc$ (see text for discussion).
    }
    \label{fig:vel_lpt}
\end{figure}

In LPT, the zeroth moment pairwise velocity spectrum, i.e.\ the real-space power spectrum $P(k)$, is given by
\begin{align}
    P(k) = &\dtq e^{i\bk \cdot \bq}\ e^{-\frac{1}{2}k_ik_j A^{\rm lin}_{ij}} \Big\{ 1  - \half k_i k_j A^{\rm loop}_{ij} + \frac{i}{6}k_i k_j k_k W_{ijk} \nonumber \\
    & +2 i b_1 k_i U_i - b_1 k_i k_j A^{10}_{ij} + b_1^2 \xi_{\rm lin} + i b_1^2 k_i U^{11}_i - b_1^2 k_i k_j U^{\rm lin}_i U^{\rm lin}_j\nonumber \\
    &+ \frac{1}{2}b_2^2 \xi_{\rm lin}^2  + 2i b_1 b_2 \xi_{\rm lin} k_i U^{\rm lin}_i  - b_2 k_i k_j U^{\rm lin}_i U^{\rm lin}_j + i b_2 k_i U^{20}_i  \nonumber \\
    &+ b_s (-k_i k_j \Upsilon_{ij} + 2i k_i V^{10}_i) + 2 i k_i b_1 b_s V^{12}_i + b_2 b_s \chi + b_s^2 \zeta \nonumber \\
    &+ 2 i b_3 k_i U_{b_3,i} + 2 b_1 b_3\theta + \alpha_P k^2 + ... \Big\} + R_h^3.
\label{eqn:pkr}
\end{align}
The ``1'' in the first line gives the (linear) Zeldovich prediction \cite{Zel70} for matter power spectrum $P_{\rm Zel}$. The first line gives the one-loop matter power spectrum in LPT, while the second to fifth lines give contributions successively including the linear, quadratic, shear and third-order biases. The final line also includes a counterterm, $\alpha_P k^2$ and stochastic term $R_h^3$. The Lagrangian correlators due to third-order bias $U_{b_3}$ and $\theta$ are defined in Appendix~\ref{app:third_order_bias}; the other various Lagrangian-field correlators (e.g.\ $U_i, A_{ij}, W_{ijk}$ etc.) are defined\footnote{Note that there is an erroneous factor of two in the expression for $V^{10}$ in Eq.~D.17 of ref.~\cite{VlaCasWhi16} . The correct prefactor should be $-\hq_i/7$ not $-2\hq_i/7$.} in \cite{Mat08a,Mat08b,CLPT,VlaWhiAvi15, VlaCasWhi16}. Some quantities, such as $U_i = U^{\rm lin}_i + U^{\rm loop}_i$, contain contributions at both linear and one-loop levels, which we will use the ``lin'' and ``loop'' sub- or superscripts to denote when separated. 

Lagrangian perturbation theory in principle includes a much larger set of effective contributions \cite{PorSenZal14,VlaWhiAvi15} ---  including derivative bias $b_\nabla$ \cite{VlaCasWhi16} --- however, all of these contributions to the real-space power spectrum are proportional to $k^2 P_{\rm Zel}(k)$ at one-loop order (counting $\alpha_P$ as itself first order), so we will summarize their effect by one counterterm only. Finally, the autocorrelation of the stochastic modes gives a ``shot-noise'' contribution $R_h^{3} \sim \bar{n}^{-1}$, where $\bar{n}$ is the number density of tracers \cite{McDRoy09,Perko16,Des16}. 

\begin{figure}
    \centering
    \includegraphics[width=\textwidth]{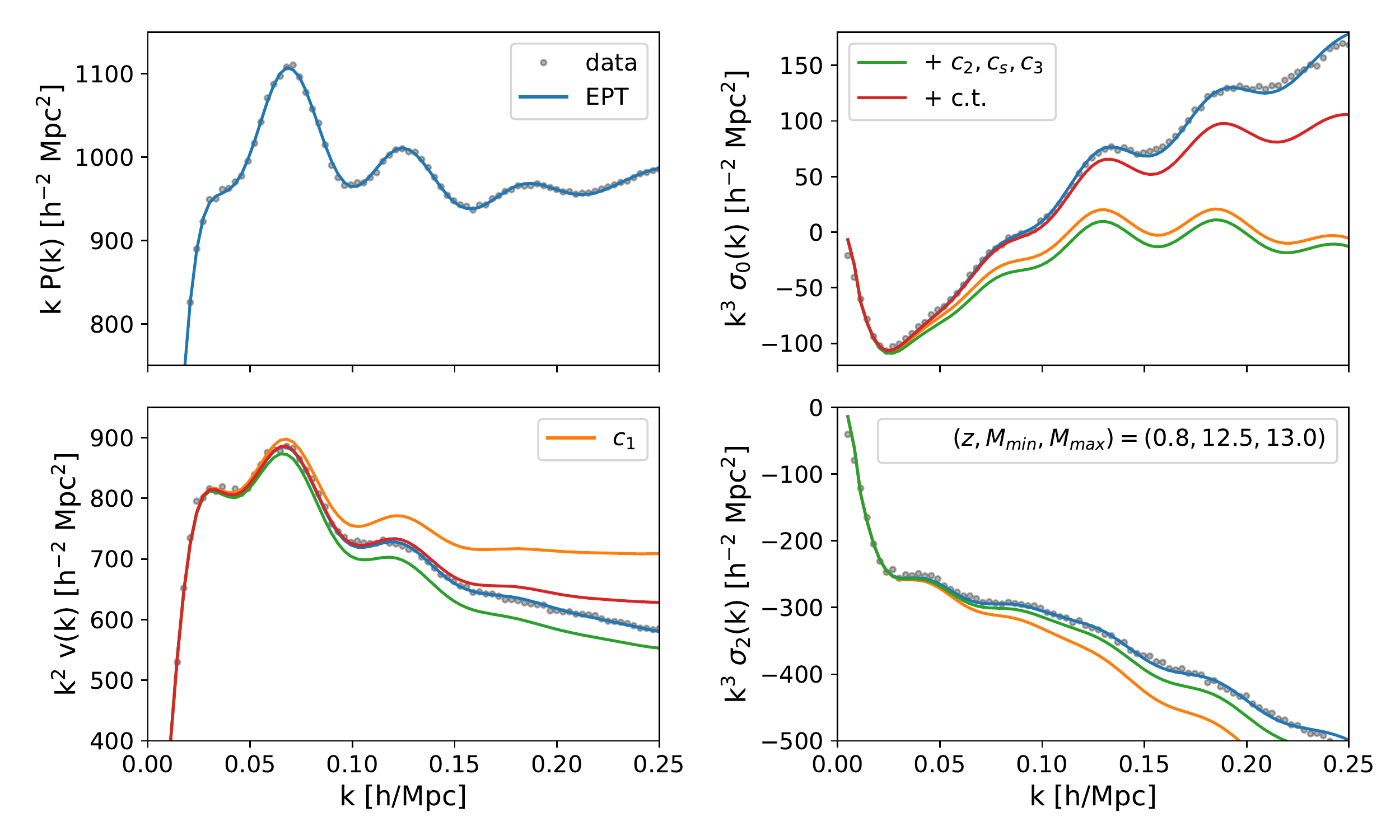}
    \caption{Same as Figure~\ref{fig:vel_lpt} but for EPT. Note that, in the lower right panel, there is almost no (numerical) difference between the green and blue lines, the former of which differs from the full prediction of EPT by a counterterm; we have not included any stochastic contributions in $\sigma_2$.
    }
    \label{fig:vel_ept}
\end{figure}

In EPT, on the other, hand we have
\begin{align}
P(k) = c_1^2 P_{\rm lin}(k) &+ \int_\bp \bigg[  2 c_1^2 \big[ F_2 (\bp, \bk-\bp) \big]^2 + 2 c_1 c_2 F_2(\bp,\bk - \bp)  + 4 c_1 c_s ~ F_2(\bp,\bk - \bp) S_2(\bp,\bk - \bp) \notag\\
&\hspace{2.3cm} + \frac{c_2^2}{2} + 2 c_2 c_s  S_2(\bp,\bk - \bp) + 2 c_s^2 \big[ S_2(\bp,\bk - \bp) \big]^2 \bigg] P_{\rm lin}(p) P_{\rm lin}(|\bk - \bp|)  \notag\\
&\hspace{0.0cm} + 6 c_1 P_{\rm lin}(k)  \int_\bp  \Big( c_1 F_3 (\bp, -\bp, \bk) + c_3 S_{\psi} (\bp, -\bp, \bk) \Big) P_{\rm lin}(p) 
+ c^{(0)}_0 \frac{k^2}{k^2_*} P_{\rm lin}(k)\notag \\
&\hspace{0.0cm} + {\rm ``const_0"} \quad .
\label{eqn:pkr_EPT}
\end{align}
Many of the third order bias operators listed in Section~\ref{sssec:bias} do not contribute explicitly to the one-loop power spectrum, and only one non-vanishing independent contribution remains.  The details of the EPT derivations for this and the velocity statistics below are given in Appendix \ref{app:ept}. In addition, in EPT an explicit IR-resummation is required to tame the effects of long-wavelength modes, which is described in  Appendix \ref{app:IR_resumm} for all velocity moments and implicitly performed in all our EPT results.

In addition to the ``deterministic'' bias parameters there is one counter term (with coefficient $c_0^{(0)}$) that is required to regularize the one-loop, $P_{13}$-like terms and is degenerate with the derivative bias contribution. In general, for counterterm we will use the $c^{(\ell)}_n$ thus notation taking into account that different angular dependence can have different counterterm contributions.
In addition to these terms there is a constant shot noise contribution obtained by correlating the purely stochastic component of the halo field with itself (labeled $\mathrm{const}_0$ in the above).

Fits to the power spectrum extracted from N-body data, along with fits for other velocity statistics using a single, consistent set of bias parameters, are shown in Figures~\ref{fig:vel_lpt} and \ref{fig:vel_ept}. As shown in the top-left panels of the two figures, both LPT and EPT provide good fits to the data past $k \sim 0.25 \kMpc$, beyond which the shot noise accounts for an increasingly large share of the total power, reaching more than $35\%$ of the total power by $k = 0.2 \kMpc$. Setting the third-order Lagrangian parameter $b_3 = 0$, as discussed in Section~\ref{sssec:bias}, does not qualitatively change our results.

\subsubsection{First Moment: Pairwise Velocity Spectrum}

The pairwise velocity spectrum, the Fourier transform of $v_i(\textbf{r}) \equiv \Xi_i(\textbf{r})$, is given in LPT by\footnote{Note that our expression for term proportional to $b_s$ differs from that in ref.~\cite{VlaCasWhi16} by a factor of two.}
\begin{align}
    \textbf{v}_i(\textbf{k}) = &\dtq e^{i\bk \cdot \bq}\ e^{-\frac{1}{2}k_ik_j A^{\rm lin}_{ij}} \Big\{ i k_j \dot{A}_{ji} - \frac{1}{2} k_j k_k \dot{W}_{jki} \nonumber \\
    &+ 2 b_1 \dot{U}_i + 2 b_1^2 i k_j U^{\rm lin}_j \dot{U}^{\rm lin}_i + \big(2ib_1 k_k U^{\rm lin}_k + b_1^2 \xi_{\rm lin} \big)\ i k_j \dot{A}^{\rm lin}_{ji} + 2i b_1 k_j \dot{A}^{10}_{ji} + b_1^2 \dot{U}^{11} \nonumber \\
    &+ 2 \big( i b_2 k_j U^{\rm lin}_j + b_1 b_2 \xi_{\rm lin} \big) \dot{U}^{\rm lin}_i  + b_2 \dot{U}^{20} + 2b_s ( \dot{V}^{10}_i  + ik_j \dot{\Upsilon}_{ji}) + 2 b_1 b_s \dot{V}^{12}_i  \nonumber \\
    &+ 2 b_3 \dot{U}_{b_3,i} + \alpha_v k_i + ... \Big\} + R_h^4 \tilde{\sigma}_v k_i.
\label{eqn:vk}
\end{align}
Here again the first two lines give the matter and density bias contributions, while the third line contains contributions due to shear bias and an effective correction $\sim \alpha_v k_i P_{\rm Zel}$. The latter regulates, for example, UV sensitivities in $\dot{A}_{ij} = \dot{A}_{ij}^{\rm LPT} + \bar{\alpha}_v \delta_{ij} + \cdots$ and is contracted with the wavevector $k_i$ in the velocity spectrum. By symmetry, $\bv_i(\bk)$ must be imaginary and point in the $\bk$ direction, so we can decompose it as $\bv_i(\bk) = i v(k) \hk_i$. Explicit expressions for $v(k)$, written as a sum of Hankel transforms, are provided in Appendix~\ref{app:Hankel}. 

As in the case of the power spectrum, while there are in principle several more counterterms and derivative bias contributions in addition to the one indicated (e.g.\  $\sim \langle \dDelta_i \nabla^2 \delta \rangle$ or $\langle \nabla_i \delta_1 \delta_2 \rangle$), all such contributions Fourier transform to $\sim k_i P_{\rm lin}(k)$ at lowest order and as such we account for them using only one effective correction, $\alpha_v$. The final term, $R_h^4 \tilde{\sigma}_v k_i$, is the leading order stochastic contribution due to the correlation between the stochastic density and velocity, $\avg{\epsilon(\bq_1) \epsilon_i(\bq_2} \sim R_h^3 \tilde{\sigma}_v \nabla_i \delta_D(\bq)$ \cite{Perko16,Des16}, which can be approximated as a Dirac-$\delta$ derivative on large scales. 

Similarly to the density auto power spectrum, in EPT we have contributions from all the bias operators introduced previously. We have 
\begin{align}
 \textbf{v}_i(\textbf{k}) 
 =& - 2 i c_1 \frac{k_i}{k^2}P_{\rm lin}(k) \\
& - 2 i \int_\bp ~\bigg[ \frac{k_i}{k^2} \Big( 2 c_1 F_2 (\bp, \bk-\bp) +  c_2 +  2 c_s S_2 (\bp, \bk-\bp) \Big) G_2 (\bp, \bk-\bp) \notag\\
&\hspace{2.5cm} + \frac{p_i}{p^2} \Big( 2 c_1^2 F_2 (\bp, \bk-\bp) +  c_1 c_2 + 2 c_1 c_s S_2 (\bp, \bk-\bp) \Big) \bigg] P_{\rm lin}(p) P_{\rm lin}(|\bk - \bp|) \notag\\
& - 2 i P_{\rm lin}(k) \int_\bq ~ \bigg[ 3 \frac{k_i}{k^2} \Big( c_1  F_3 (\bp, -\bp, \bk) + c_1  G_3 (\bp, -\bp, \bk) +  c_3 S_{\psi} (\bp, -\bp, \bk) \Big) \notag\\
&\hspace{5.5cm} + 2 c_1^2 \Big( \frac{p_i}{p^2} F_2 (\bp, - \bk) + \frac{(\bk - \bp)_i}{(\bk - \bp)^2} G_2 (\bp, - \bk) \Big) \bigg] P_{\rm lin}(p) \Bigg] \notag\\
&- i c^{(0)}_1 \frac{\hat k_i}{k^2_*} P_{\rm lin}(k) + {\rm ``const_1"}  k_i
\cdots \notag
\end{align}
where $c_1^{(0)}$ is the coefficient of the counterterm, and the $\mathrm{const}_1$
is the leading stochastic velocity contribution.

A comparison to $v(k)$ from N-body data is shown in the bottom-left panels of Figure~\ref{fig:vel_lpt} and \ref{fig:vel_ept}. Both formalisms give a good fit to the data past $k = 0.2\kMpc$, though as noted in Section~\ref{sec:nbody}, comparing the theory to the N-body data at large scales suggests that the simulations slightly under-predict velocities (by one or two percent). The stochastic contribution accounts for a significant fraction of the power in both fits at high wavenumber ($k > 0.1 \kMpc$) that cannot be accounted for by the other bias parameters or counterterms.  Not fiting for it leads to oscillatory residuals due to a mismatch between the BAO and overall broadband amplitude.

\subsubsection{Second Moment: Pairwise Velocity Dispersion Spectrum}

The pairwise velocity dispersion spectrum, $\Xi_{ij} \equiv \sigma_{12,ij}^2$, is given in LPT by
\begin{align}
    \sigma^2_{12,ij}(\bk) = &\dtq e^{i\bk \cdot \bq} e^{-\frac{1}{2}k_ik_j A^{\rm lin}_{ij}} \Big\{ \ddot{A}_{ij} + i k_n \ddot{W}_{nij} + \left(2 i b_1 k_n U^{\rm lin}_n + b_1^2 \xi_{\rm lin}\right) \ddot{A}^{\rm lin}_{ij} \nonumber \\
    &- k_n k_m \dot{A}^{\rm lin}_{ni} \dot{A}^{\rm lin}_{mj} + 2 (b_1^2 + b_2) \dot{U}^{\rm lin}_i \dot{U}^{\rm lin}_j + 2i k_n b_1 \big(\dot{A}^{\rm lin}_{ni} \dot{U}^{\rm lin}_j + \dot{A}^{\rm lin}_{nj} \dot{U}^{\rm lin}_i \big) \nonumber \\
    & + 2b_1 \ddot{A}^{10}_{ij} + 2 b_s \ddot{\Upsilon}_{ij} + \alpha_\sigma \delta_{ij} + \beta_\sigma \xi^2_{0,L} \left(\hq_i \hq_j - \third \delta_{ij}\right) + \cdots \Big\} + R_h^3 s_v^2 \delta_{ij}.
\label{eqn:sigma12}
\end{align}
The velocity dispersion spectrum can be decomposed into a number of possible bases such as the parallel-perpendicular basis, $\sigma^2_{ij} = \sigma_{||}(k) \hat{k}_i \hat{k_j} + \frac{1}{2}\sigma_{\bot}(k)(\delta_{ij} - \hat{k}_i \hat{k_j})$, or the Legendre basis, $\sigma_{ij} = \sigma_{0}(k) \delta_{ij} + \frac{3}{2}\sigma_{2}(k)(\hat{k}_i \hat{k_j} - \third \delta_{ij})$. These scalar components, expressed as Hankel transforms, are detailed in Appendix~\ref{app:Hankel}.

Unlike the zeroth and first moments, the second moment ($\sigma^2_{ij}$) requires two counterterms: $\alpha_\sigma$ and $\beta_\sigma$. The latter contribution is proportional to the $j_2$ Hankel transform of the linear power spectrum, $\xi^2_{0,{\rm lin}}$ (Appendix~\ref{app:fftpt}), and cancels UV sensitivities in the non-isotropic component of $A^{\rm 1-loop}_{ij}$. These contributions can alternatively be parametrized as counterterms $\sim \alpha_0 P_{\rm lin}(k)$ and $\alpha_2 P_{\rm lin}(k)$ to the velocity-dispersion monopole ($\sigma_0$) and quadrupole ($\sigma_2$), respectively. Finally, we include an isotropic stochastic contribution $R_h^3 s_v^2 \delta_{ij}$.  Such a term can, for example, arise from the disconnected part of the second moment
\begin{equation}
    \sigma^2_{12}(\bk) \ni \int d^3\textbf{r}\ e^{i \bk \cdot \textbf{r}}\ \sigma_v^2 \delta_{ij} \langle (1 + \delta_1)(1 + \delta_2) \rangle = \sigma_v^2 P_{\rm NL}(k)\ \delta_{ij} \ni \sigma_v^2 R_h^3 \delta_{ij}
\end{equation}
where $\sigma_v^2$ is a contact term coming from evaluating the average velocity squared at a point and $P_{\rm NL}$ is the full nonlinear real-space power spectrum including a constant stochastic contribution $R_h^3$ (selectively resumming only these terms yields the exponential damping formula for FoG).  Our treatment of this stochastic contribution differs from much of the literature \cite{Perko16,Des16}; this is of no consequence when fitting the redshift-space power spectrum, since its contribution there is degenerate with that of the stochastic component to $v(k)$, but makes a signficant difference when studying pairwise velocities on their own.

It is useful to note the relations between the parameters for $\sigma_{12}$ in Fourier and configuration space, the latter as presented in ref.~\cite{VlaCasWhi16}. While the bias contributions are identical, up to Fourier transforms, there are important differences in the counterterms and bias parameters. Firstly, the corresponding expression for the pairwise velocity dispersion in configuration space contains two isotropic counterterms in the curly brackets $\{ \cdots \}$ in Equation 3.10 of ref.~\cite{VlaCasWhi16}, corresponding to our Equation~\ref{eqn:sigma12}.  These are $A_\sigma \delta_{ij}+B_\sigma \xi_{\rm lin} \delta_{ij}$, which both result at lowest order in contributions to $\sigma^2_{12}(\textbf{r})$ proportional to the linear correlation function $\xi_{\rm lin}$, and thus in Fourier space to a counterterm $\propto P_{\rm lin}(k)$.  For this reason, in Fourier space we have chosen to summarize them using one counterterm $\alpha_\sigma$. However, we note that the constant counterterm proportional to $\delta_{ij}$ stems in part from the contribution of small-scale velocities to the $q \rightarrow \infty$ limit of $\sigma_{12}$, which shows up as a point-contraction of the stochastic velocities
\begin{equation}
    \avg{ (1 + \delta_1)(1 + \delta_2) \Delta \bu_i \Delta \bu_j } \ni \sigma_\epsilon^2 \delta_{ij}\ (1 + \xi(\textbf{r})).
    \label{eqn:sigma12_ct}
\end{equation}
Roughly speaking, this $\sigma_\epsilon^2$ is the asymptotic value for the stochastic component of the halo velocity $\sigma_{\epsilon, ij} = \avg{ \Delta \epsilon_i \Delta \epsilon_j}$ at scales $q > R_h$ above the halo scale. This contribution to the configuration-space velocity dispersion is closely related to the Fourier-space stochastic contribution $R_h^3 s_v^2$ to $\sigma^2_{12}(\bk)$, which is just the large scale ($k \lesssim R_h^{-1}$) limit of the Fourier-transform of $\sigma^2_\epsilon$.  
There are therefore two free parameters in $\sigma^2_{12}$ characterizing isotropic effective and stochastic contributions in both real and Fourier space; if in addition the fit is performed in both spaces, it is important to note that the counterterms in configuration space sum to that in Fourier space, i.e.\ $\alpha_\sigma = A_\sigma + B_\sigma$, while $s_v^2$ remains independent, leaving us with three parameters total. This may be especially relevant in predicting statistics for upcoming kSZ surveys.

Moving on to the EPT formulation of the velocity dispersion correlators, we find only up to second
order bias parameters contributing to the velocity dispersion (c.f.\ the density auto power spectrum and pairwise velocity spectrum).  This is consistent with our LPT analysis.  In EPT we have 
\begin{align}
\sigma^2_{12,ij}(\bk) =
& -  2\frac{k_i k_j}{k^4} P_{\rm lin} (k)\\
& - 2 \int_{\bp} \Bigg[ 
\Big(2 c_1 F_2 (\bp, \bk-\bp) + c_2 + 2 c_s  S_2 (\bp, \bk-\bp) \Big) \frac{p_i (\bk - \bp)_j}{p^2 (\bk - \bp)^2} +  2\frac{k_i k_j}{k^4} G_2 (\bp, \bk-\bp)^2 \notag\\
&\hspace{1.7cm} + 4c_1 \frac{k_i p_j}{k^2 p^2} G_2 (\bp, \bk-\bp) + c_1^2 \frac{p_i}{p^2} \bigg(\frac{p_j}{p^2}+\frac{(\bk-\bp)_j} {(\bk - \bp)^2} \Bigg) \Bigg] P_{\rm lin}(p) P_{\rm lin}(|\bk - \bp|) \notag\\
&- 4 P_{\rm lin}(k) \int_{\bp} ~ \Big[ 3\frac{k_i k_j}{k^4} G_3 (\bp,-\bp, \bk) \notag\\
&\hspace{2.9cm} + 2 c_1 \left( \left( \frac{k_i}{k^2} + \frac{p_i}{p^2}\right) \frac{(\bk-\bp)_j} {(\bk - \bp)^2}G_2 (-\bp, \bk)+\frac{k_i p_j}{k^2 p^2} F_2 (-\bp, \bk) \right) \bigg] P_{\rm lin}(p) \notag\\
& + 2 c_1^2 P_{\rm lin}(k) \delta^K_{ij} \sigma_{\rm lin}^2 - 2\left(c^{(0)}_2 \delta^K_{ij} + c^{(2)}_2 \frac{k_i k_j}{k^2} \right)\frac{1}{k^2_*} P_{\rm lin}(k) + ``{\rm const_2}", \notag
\end{align}
where $c_2^{(0)}$ and $c_2^{(2)}$ are two counterterm coefficients corresponding to different angular dependency,
$\sigma_{\rm lin}$ is the linear velocity dispersion, and we have one isotropic stochastic contribution, $``{\rm const_2}"$.

Fits of LPT and EPT to $\sigma_{0,2}$ are shown in the right column of Figures~\ref{fig:vel_lpt} and \ref{fig:vel_ept}. While both theories give an excellent fit to $\sigma_0$ to similar scales as the real-space power spectrum, the fit to $\sigma_2$ is only good up to $k \sim 0.1\kMpc$ in LPT.
As we will discuss in more depth in Section~\ref{sec:comparison}, this is partly due to particularities of the resummation scheme in LPT, which keeps all linear displacements exponentiated.  In principle, this could be somewhat mitigated by adopting an alternative IR-resummation  scheme or considering higher order corrections in the current scheme.  However, such a strategy would require some changes in the formalism above, and the overall effect on the redshift space power spectrum due to these differences in $\sigma_2$ is negligible.  Thus we shall not pursue this strategy.  We also note that the fit to $\sigma_2$ on large scales suggests that the velocities in the N-body simulations are somewhat underpredicted compared to theory\footnote{The fit to $\sigma_0$ is less succeptible to this systematic due to a floating stochastic contribution to its amplitude.}, consistent with our expectations of their systematic error.

\subsubsection{Higher Moments}

Finally, let us give expressions for the third and fourth moments despite them not figuring prominently in our redshift-space model. In one-loop LPT these are given by
\begin{align}
    \gamma_{ijk} &= \dtq  e^{i\bk \cdot \bq -\frac{1}{2}k_ik_j A_{ij}} \Big\{ \dddot{W}_{ijk} + i k_l \dot{A}^{\rm lin}_{l\{ i} \ddot{A}^{\rm lin}_{jk\}} + 2 b_1 \dot{U}^{\rm lin}_{\{i} \ddot{A}^{\rm lin}_{jk\}} + \alpha_\gamma  \frac{k_{\{i} \delta_{jk\}}}{k^2} + \beta_\gamma \frac{k_i k_j k_k}{k^4} \Big\} \nonumber \\
    \kappa_{ijkl} &= \dtq  e^{i\bk \cdot \bq -\frac{1}{2}k_ik_j A_{ij}} \Big\{ \ddot{A}^{\rm lin}_{\{ij} \ddot{A}^{\rm lin}_{kl\}} + \alpha_\kappa \frac{k_{\{i} k_j \delta_{kl\}}}{k^4} \Big\} + R_h^3 s_\kappa^4 \delta_{\{ij} \delta_{kl\}}.
\end{align}
We see that at this perturbative order only the $b_1$ bias parameter contributes to the the third velocity moment, while the fourth moment has purely velocity contributions and does not depend on deterministic bias parameters. The expressions above also require the necessary counterterms and stochastic contributions, together with the pure FoG contributions. 

In EPT, at one-loop, we equivalently have contributions to both third and fourth velocity moments. For the third moment we have
\begin{align}
\tilde{\Xi}^{(3)}_{ijl}
&= 
 12 i \int_{\bp} ~ \left(
 \frac{k_{\{i} p_j (\bk - \bp)_{l\}}}{k^2 p^2 (\bk - \bp)^2}
G_2 (\bp, \bk-\bp) 
+ c_1 \frac{p_{\{i} p_j (\bk - \bp)_{l\}}}{p^4 (\bk - \bp)^2}
\right)
P_{\rm lin}(p) P_{\rm lin}(\bk - \bp) \\
&~~~+24 i P_{\rm lin}(k) \int_{\bp} ~
\frac{k_{\{i} p_j (\bk - \bp)_{l\}}}{k^2 p^2 (\bk - \bp)^2}
G_2 (\bp, -\bk) 
P_{\rm lin}(p) \notag\\
&~~~ - 12 i c_1  \frac{\delta_{\{ij}k_{l\}}}{k^2}P_{\rm lin} \sigma_{\rm lin}^2
+ 6 i \left(c^{(0)}_3 \delta_{\{ij} + c^{(2)}_3 \hat k_{\{i} \hat k_{j} \right) 
\frac{k_{l\}}}{k^2}
\frac{1}{k_\star^2}
P_{\rm lin} + \ldots,\notag
\end{align}
while the fourth velocity moment is given by
\begin{align}
\tilde{\Xi}^{(4)}_{ijlm}
&= 12\int_{\bp} ~
\frac{p_{\{i} p_j (\bk - \bp)_l (\bk - \bp)_{m\}}}{p^4 (\bk - \bp)^4}
P_{\rm lin}(p) P_{\rm lin}(|\bk - \bp|) \\
&~~~ - 24 \left( \sigma_{\rm lin}^2 - c^{(2)}_4
\right) \frac{\delta_{\{ij} k_lk_{m\}}}{k^4} \frac{1}{k^2_\star} P_{\rm lin}(k)
 + ``{\rm const_4}" \delta_{\{ij} \delta_{lm\}} + \ldots. \notag
\end{align}
We note that the structure of these velocity moments in LPT and EPT is quite similar, with equivalent counterterm and stochastic contribution structure. 
Further details of the one-loop EPT contributions to higher moments are discussed in Appendix \ref{app:ept}.

\subsection{Comparing LPT and EPT}
\label{sec:comparison}

In the previous section, we described the predictions for the pairwise velocity moments within two formalisms, LPT and EPT, at one-loop in perturbation theory. A comparison of Figs.~\ref{fig:vel_lpt} and \ref{fig:vel_ept} shows that LPT and EPT both perform comparably well for the power spectrum, once IR resummation is taken into account.  The pairwise velocity and velocity dispersion monopole likewise show a similar level of agreement for both LPT and EPT. Note however, that in the latter spectrum essentially all of the power at $k > 0.1 \kMpc$ comes from the counterterm and stochastic contributions in EPT, unlike in LPT where the contributions due to large-scale modes and deterministic bias qualitatively match the spectral shape. In both cases the power due to stochastic contributions (shot noise) becomes increasingly significant towards the highest $k$s plotted, with the models correctly accounting for the mild non-linearity at intermediate $k$. 
However, significant differences appear in the predictions of LPT and EPT for the second moment, $\sigma^2_{12}$, particularly in the broadband shape of the quadrupole, $\sigma_2$. Our goal in this section is to compare and contrast the LPT and EPT models described in the previous sections with these differences in mind.

As we have already noted, the two formalisms are equivalent, term-by-term, when Taylor-series expanded in powers of the linear power spectrum and differ only in the treatment of IR displacements, which are canonically included order-by-order in (non-resummed) EPT but manifestly resummed via the exponential $\exp(-k_i k_j A^{\rm lin}_{ij}/2)$ in LPT. Within LPT, we can therefore recover analagous EPT results by expanding this exponential--- indeed, by splitting the linear displacements into long and short modes separated by an infrared cutoff $k_{\rm IR}$ we can recover a spectrum of theories between LPT and EPT. Specifically, writing $A^{\rm lin}_{ij} = A^{<}_{ij} + A^{>}_{ij}$, where the less-than indicates displacement two-point functions calculated by smoothing out long modes via a Gaussian filter $\exp(-(k/k_{\rm IR})^2/2)$ and the greater-than denotes all the remaining power, we have  generically for velocity moments
\begin{equation}
   \tilde{\Xi}^{(n)}(\bk) = \dtq e^{i\bk \cdot \bq - \frac{1}{2}k_i k_j A^{<}_{ij}(\bq)}\ \Big(1 - \frac{1}{2} k_i k_j A^{>}_{ij} + \frac{1}{8} k_i k_j k_k k_l A^{>}_{ij} A^{>}_{kl} + \mathcal{O}(P_{\rm lin}^3) \Big) \Big\{ \cdots \Big\}.
   \label{eqn:lpt_to_spt}
\end{equation}
where the $\{ \cdots \}$ indicate the terms in curly brackets in Eqs.~\ref{eqn:vk} and \ref{eqn:sigma12}.  Taking $k_{\rm IR} \rightarrow 0$ and keeping the product of the round and curly brackets to second order yields one-loop EPT.
This implies that the differences between the LPT and EPT predictions for the velocity moments, and $\sigma^2_{12}$ in particular, in both BAO wiggles and broadband shape must be due to the selective resummation of $A_{ij}$, i.e.\ to differences at $\ge 2$-loop order.

Let us briefly mention a technical detail in the above mapping between EPT and LPT. In addition to expanding the linear displacement two-point function $A_{ij}$, in order to make the low $k_{\rm IR}$ limit of LPT agree with EPT, one needs to use the bias-parameter mapping in Equation~\ref{eqn:bias_map}. A useful feature of this mapping is that, while LPT contains the same number of bias parameters as EPT, the contributions of these biases to various statistics are organized rather differently. For example, since $c_1^2 = 1 + 2 b_1 + b_1^2$, the `1' term in LPT is equal to the $c_1^2$ term and the $b_1$ term is twice the $c_1^2$ term at leading order. We can take advantage of these differences to, for example, compute the third-order bias contribution in EPT using those from the biases in LPT up to second order alone. Specifically, we can write for the third-order bias contribution to the power spectrum
\begin{equation}
    a P_{c_1 c_3} = 2 P_{b_1^2} - P_{b_1} - \frac{8}{21} P_{b_1 b_2} + \frac{2}{7} P_{b_1 b_s} + \mathcal{O}(P_{\rm lin}^3)
\end{equation}
and similarly for the third-order bias contribution to $v(k)$:
\begin{equation}
    a v_{c_3} = v_{b_1} - v_{1} - v_{b_1^2} - \frac{8}{21}(v_{b_2} - v_{b_1 b_2}) + \frac{2}{7}(v_{b_s} - v_{b_1 b_s}) + \mathcal{O}(P_{\rm lin}^3).
\end{equation}
We have checked these identities numerically.

\begin{figure}
    \centering
    \includegraphics[width=\textwidth]{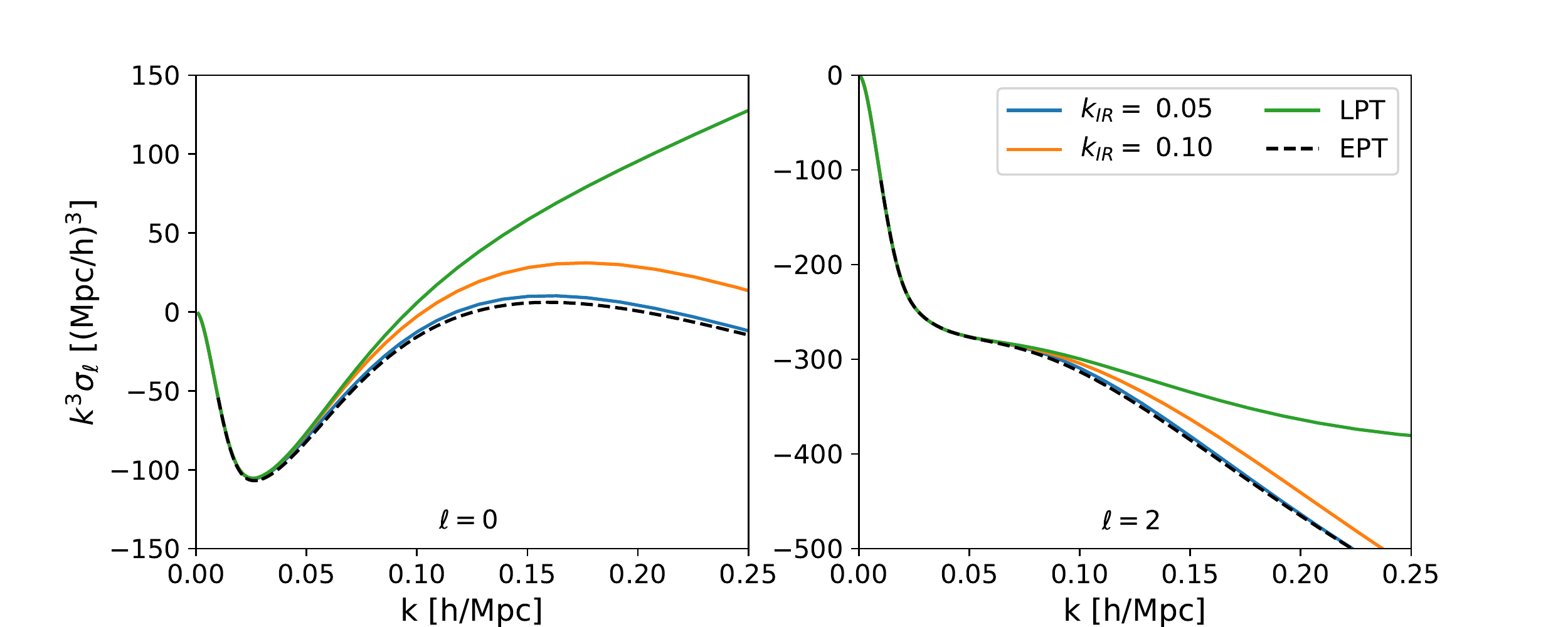}
    \caption{The monopole ($\ell=0$) and quadrupole ($\ell=2$) of $\sigma^2_{12}(k)$ predicted by 1-loop PT (Eq.~\ref{eqn:lpt_to_spt}) for several cutoffs, $k_{\rm IR}$, using a ``no-wiggle'' version of our fiducial power spectrum.  The amplitude of $\sigma_\ell$ at high $k$ is strongly affected by the choice of IR resummation in Eq.~\ref{eqn:lpt_to_spt}, indicating that 2-loop contributions may be important for density-weighted velocity dispersion.  }
\label{fig:sigma_kir}
\end{figure}

To look at the effects of IR resummation, let us begin with the broadband. Figure~\ref{fig:sigma_kir} shows the monopole and quadrupole of the second moment $\sigma^2_{12}$ for a range of cutoffs, $k_{\rm IR}$, computed using a no-wiggle version of our fiducial power spectrum, which we use in this section only to isolate broadband effects.  As expected, the EPT prediction is recovered in in the limit of vanishing $k_{\rm IR}$, while LPT represents the $k_{\rm IR} \rightarrow \infty$ limit. It is notable that the two limits predict dramatically different broadband shapes at even intermediate wavenumbers. For example, EPT predicts the monopole to have close-to-vanishing power at $k \sim 0.2 \kMpc$, where LPT predicts $k^3 \sigma_0$ to have significant power increasing with $k$; conversely, EPT predicts a more significant (more negative) quadrupole compared to LPT. These differences are particularly noteworthy because LPT shows excellent agreement with the $\sigma_0$ measured from simulations while under-predicting $\sigma_2$ at small scales (Fig.~\ref{fig:vel_lpt}), and conversely for EPT (Fig.~\ref{fig:vel_ept}), where essentially all of the power at $k\simeq 0.1\kMpc$ and beyond in $\sigma_0$ is accounted for by the stochastic and counterterms.

\begin{figure}
    \centering
    \includegraphics[width=\textwidth]{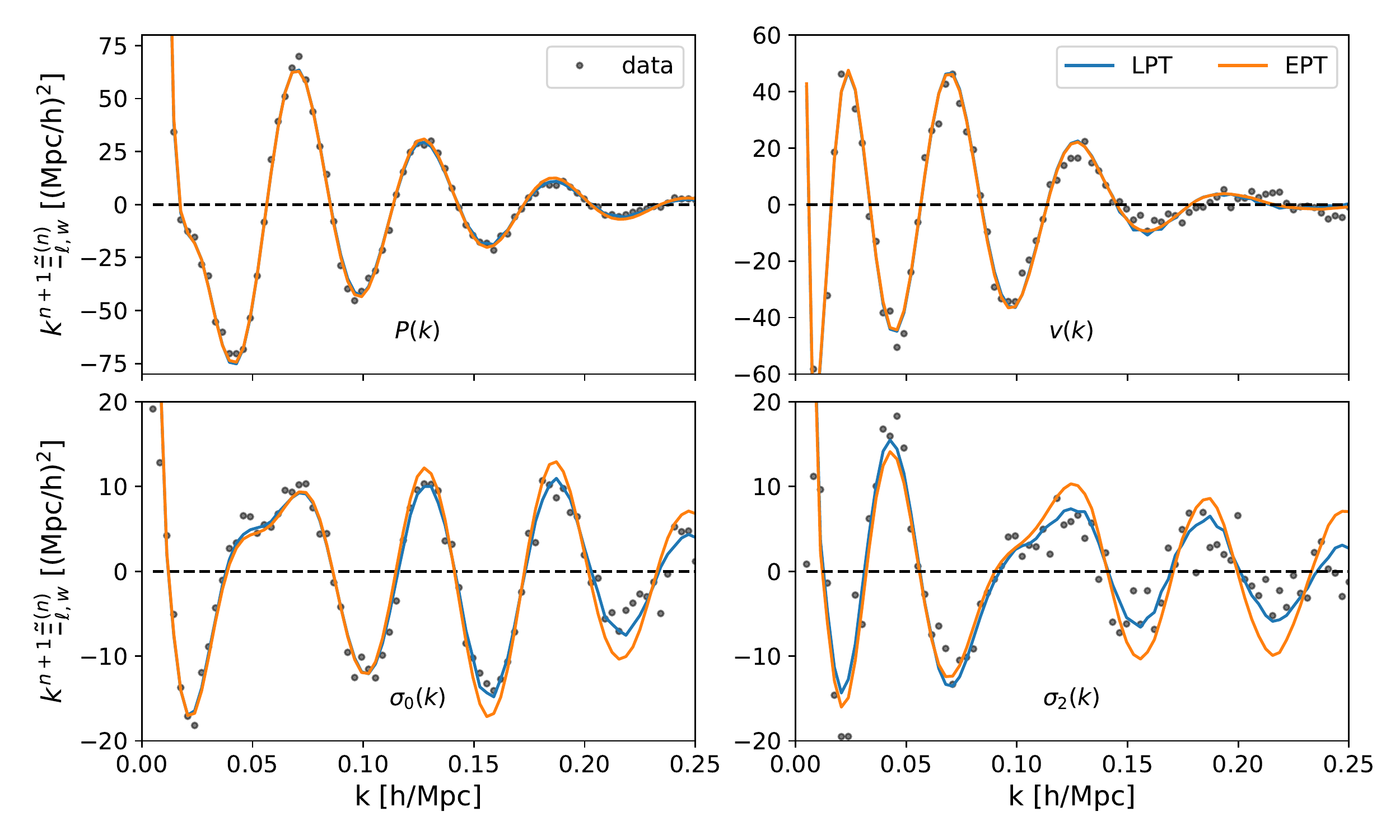}
    \caption{Oscillatory component of the real-space power spectrum (top left), pairwise velocity spectrum (top right) and the monopole and quadrupole (bottom left and right) of the velocity dispersion spectrum $\sigma^2_{\rm 12}$ in LPT and EPT compared to N-body data (dots). The smooth component subtracted from the data is computed using a Savitsky-Golay filter, and the theory signals are supplemented with a quartic polynomial in $k$ to improve agreement with the broadband-subtracted data. While the power spectrum and pairwise velocity show excellent agreement between LPT and EPT even when the fitted independently, the oscillatory signals in the velocity dispersion spectra differ significantly, with EPT underdamped compared to LPT. Notably, unlike in the lower velocity moments the dominant oscillations in $\sigma^2_{12}$ are due to one-loop effects, whose damping seem to be more naturally captured by the IR-resummation in LPT when compared to data (black dots).
    }
    \label{fig:sigma_w}
\end{figure}

In addition to the above, EPT and LPT also make different predictions for the BAO feature. In Figure~\ref{fig:sigma_w} we have plotted \edit{$P(k)$, $v(k)$ and} the monopole and quadruopole of $\sigma^2_{12}$ with smooth broadbands---estimated using a Savitsky-Golay filter\footnote{We use a quintic filter linear in $k$ with width of $0.25 \kMpc$, but note that our results are relatively robust to this choice as we are only concerned with the oscillatory components, modding out any residual broadband with a smooth polynomial fit.}---subtracted off. 
The blue and orange lines show the predictions of LPT and EPT modulo a quartic polynomial in $k$ which we fit to the data. Evidently, the IR resummation inherent in one-loop LPT provides and excellent description for the oscillatory component in the second moment, while the resummation scheme we have employed for EPT underpredicts the requisite nonlinear damping. On the other hand, \edit{the upper two panels} show that the two formalisms produce far better agreement for both the zeroth and first moments. This is likely in part due to the dominance of the one-loop $b_1$ contributions noted in the previous paragraph, which account for most of the oscillatory signal shown in both panels; indeed, we note that the (significantly smaller) damped linear BAO wiggles are more-or-less exactly out of phase with the nonlinear wiggles shown \cite{Baldauf2015, Blas2016, Peloso2017, Ivanov2018}.

The size of the one-loop terms and the divergence between one-loop LPT and EPT at even intermediate $k$ for $\sigma^2_{12}$ can heuristically be used to gauge the magnitude of higher-order ($\ge 2$-loop) corrections, and suggests that density-weighted pairwise velocity statistics may be significantly more nonlinear than the density-only real-space power spectrum. For example, direct inspection of bias contributions to $\sigma_2$ indicates that while the leading-order contribution is due to matter velocities only, the largest numerical contribution comes from $b_1$ at one loop.  Indeed, at $k = 0.1 \kMpc$ the one-loop $\sigma_2$ predicted by our EPT model has $50 \%$ extra power compared to linear theory and $100 \%$ by $k = 0.15 \kMpc$. In this case the level of agreement between the 1-loop EPT and N-body results suggests that the two-loop contributions happen to be small for $\Lambda$CDM power spectra of the amplitude we consider, so that the additional contributions included in the IR resummation by LPT are worsening the agreement with the N-body results.  We have been unable to find a symmetry that would explain why the 2-loop contribution to $\sigma_2$ should be small, so it could be that this is a numerical coincidence where 1-loop EPT is `accidentally' performing better than expected for this particular power spectrum shape and normalization. Indeed, for $\sigma_0$ the one-loop terms in EPT --- which are dominated by the stochastic and counterterms --- account for a $100 \%$ difference compared to linear theory by $k = 0.1 \kMpc$, suggesting that velocities at even these intermediate scales are subject to large nonlinearities.
As suggested by Fig.~\ref{fig:components_mu}, and we discuss further below, a detailed modeling of $\sigma_2$ is not necessary in order to obtain an accurate measure of the redshift-space power spectrum, $P(k,\mu)$, so we have not attempted to further improve the performance of either LPT or EPT for this statistic.

Before leaving the velocity statistics and turning to the redshift-space power spectrum, it is worth noting that our results have direct implications for the use of velocities (either from peculiar velocity surveys or kSZ measurements) as cosmological probes.  In particular, the relative size of the perturbative contributions (green lines in Figs.~\ref{fig:vel_lpt} and \ref{fig:vel_ept}) and the stochastic or counter terms (blue lines) can be taken as a proxy for where cosmological information dominates over small-scale information (e.g.\ about astrophysics).  For $\sigma_{ij}^2$, in particular, it appears that the cosmological information is confined to reasonably small $k$, which argues that high resolution observations of this statistic will not be necessary if the goal is inference about cosmological parameters.

\section{All Together Now: the Redshift-Space Power Spectrum in PT}
\label{sec:rsd_pk}

\begin{figure}
    \centering
    \includegraphics[width=\textwidth]{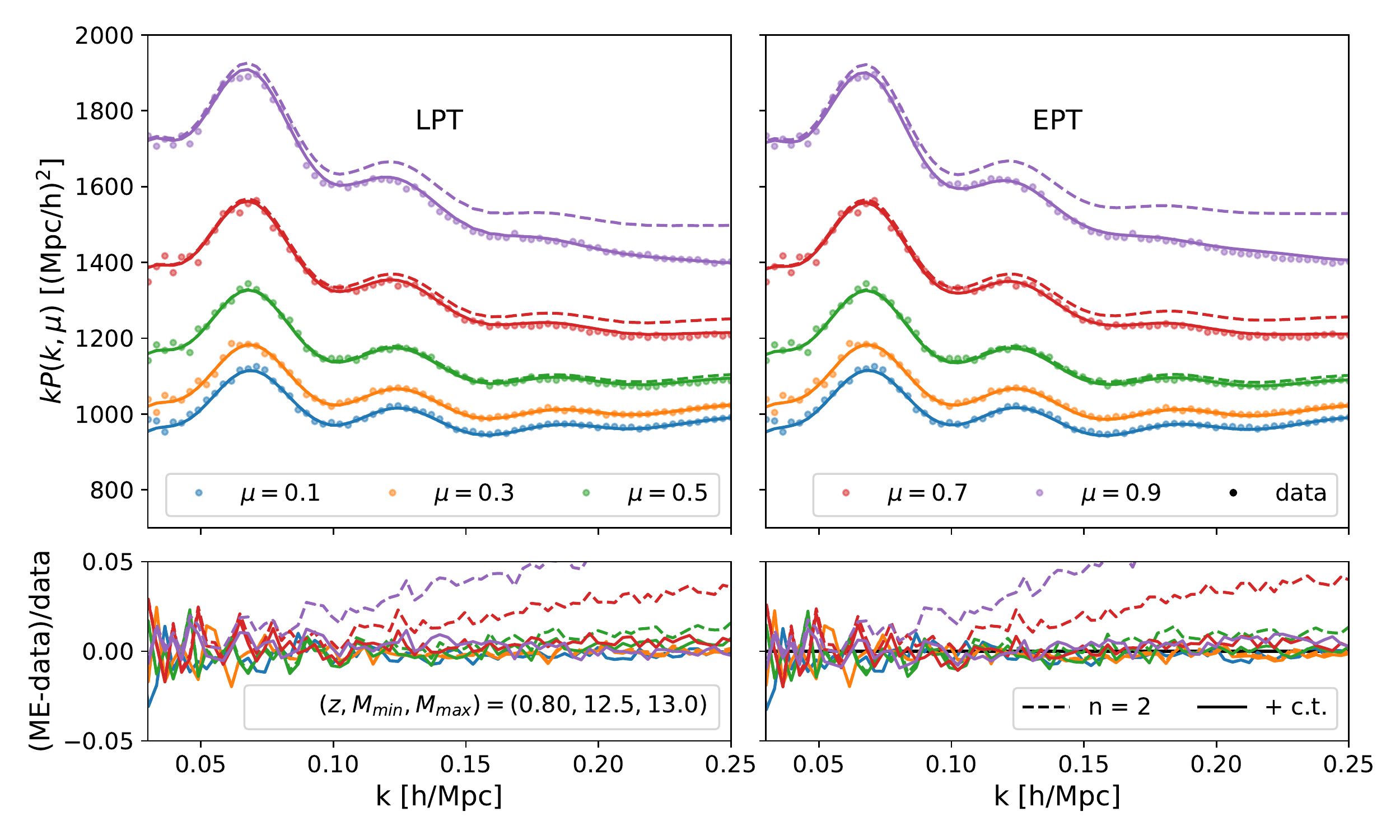}
    \caption{A comparison of the halo power spectrum wedges ($0.0<\mu<0.2$, $\cdots$, $0.8<\mu<1.0$) measured in the N-body simulations (points) to the predictions from PT models where the first two velocity moments are calculated using LPT (left) and EPT (right) and the third moment is approximated using a counterterm ansatz (lines; Eq.~\ref{eqn:me_master}).  The upper panel shows the measurements, while the lower panel shows the fractional differences.  We have chosen to show the $12.5<{\rm lg}M<13.0$ mass bin at $z=0.8$ though the other masses and redshifts behave similarly.  The dashed lines show the PT contributions excluding the $n = 3$ counter term, while the solid lines show the results of the full model.  Note the addition of these terms significantly improves the model for high $\mu$ while the improvement is much more modest for low $\mu$.
    }
    \label{fig:wedges_lpt}
\end{figure}

\begin{figure}
    \centering
    \includegraphics[width=\textwidth]{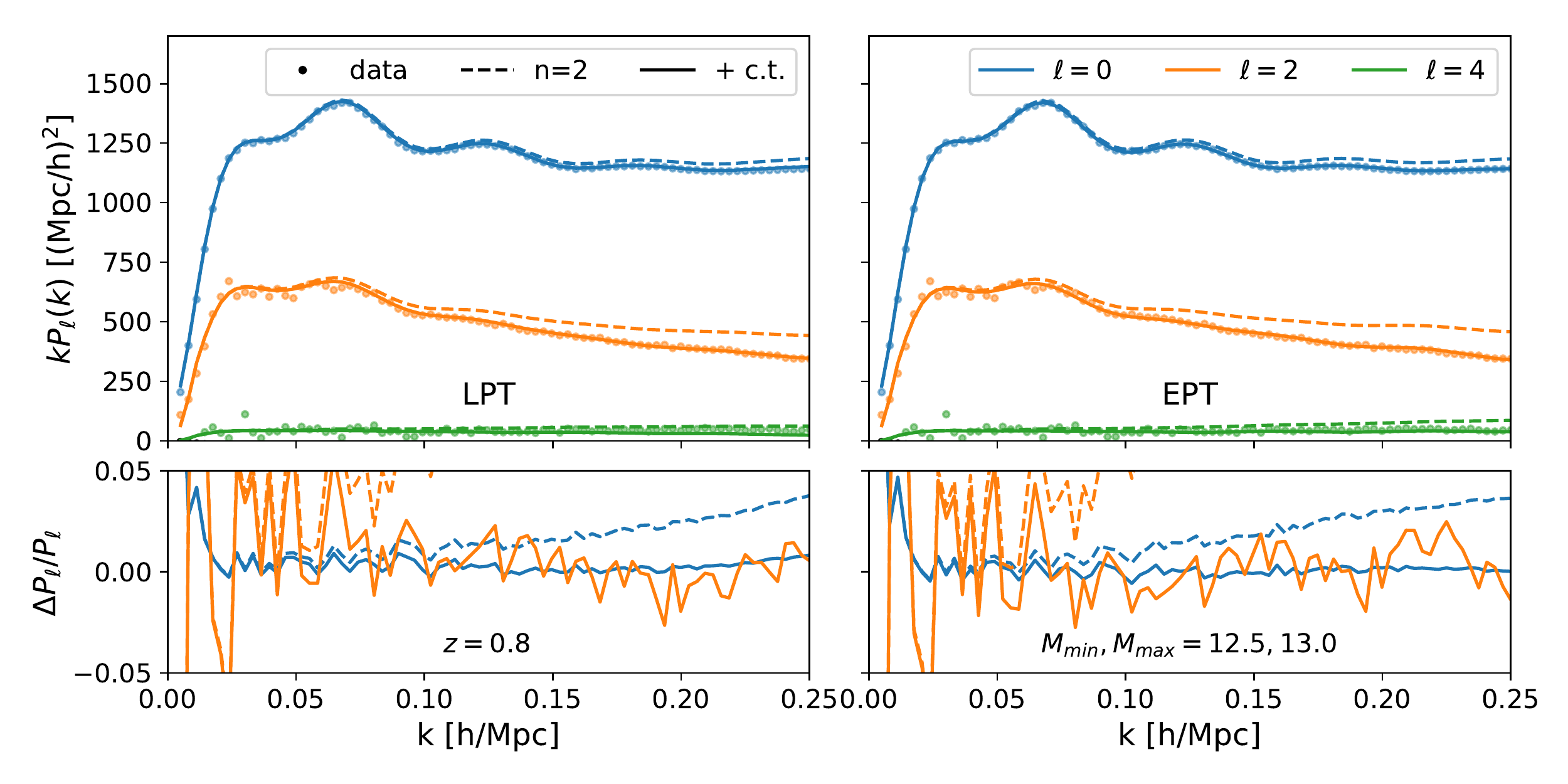}
    \caption{A comparison of the halo power spectrum multipoles measured in the N-body simulations (points) to the predictions from our LPT (left) and EPT (right) models (lines; Eq.~\ref{eqn:me_master}).  The upper panel shows the measurements, while the lower panel shows the fractional difference.  The dashed lines show the PT contributions excluding the $n=3$ counterterm, while the solid lines show the results of the full model.  Note the addition of these terms significantly improves the model for $\ell>0$, even more dramatically than in Fig.~\protect\ref{fig:wedges_lpt}.
    In interpreting these differences it is important to bear in mind that the N-body data contain systematics that can bias results at the few-percent level---indeed it clearly under-predicts the quadrupole by around $2\%$ around $k = 0.05 \kMpc$ compared to both LPT and EPT--- and that for any observation the error on the quadrupole and hexadecapole are dominated by the monopole contribution and are therefore fractionally much larger than for the monopole--- hexadecapole errors are not plotted in the bottom panel for this reason.
    }
    \label{fig:pls_lpt}
\end{figure}

Sections~\ref{sec:vel_exp} and \ref{sec:spectra} examined the convergence of velocity expansions for the redshift-space power spectrum and how the required velocities can be computed using perturbation theory; in this section we combine these ingredients to produce a model of the redshift-space power spectrum based on 1-loop perturbation theory.

\subsection{Comparison for halos}

Figures~\ref{fig:wedges_lpt} and \ref{fig:pls_lpt} show the PT predictions for the redshift-space power spectrum wedges and multipoles using the bias parameters, counterterms and stochastic contributions determined from the fits in Figs.~\ref{fig:vel_lpt} and \ref{fig:vel_ept}, together with the moment expansion approach. Figure \ref{fig:wedges_lpt} demonstrates that these parameters give an excellent fit, agreeing with the data at the percent level even for the highest $\mu$ wedges. It is worth noting that the redshift-space distortions captured by the quasilinear velocities is highly nontrivial, and a naive multiplication of the real-space power spectrum by the factor $(b+f\mu^2)^2$ yields $P(k,\mu)$ that is $5\%$ away from the data even at $k = 0.1 \kMpc$ and $\mu = 0.5$.

Figure \ref{fig:pls_lpt} tells a similar story to Fig.~\ref{fig:wedges_lpt}, though with some caveats.  The monopole, $P_0$, remains well-fit by both the LPT and EPT models.  The same is not true of the quadrupole, which is both noisier and possibly biased.  However, recall there is some evidence that the simulations with derated timesteps may not be converged. Indeed, the data quadrupole for $k < 0.1 \kMpc$ suggests that the simulations under-predict the value of velocities by around two percent compared to perturbation theory.
For such $k$ the best fitting LPT and EPT models are in excellent agreement, being dominated by linear theory, but differ visibly with the N-body quadrupole (the contribution of the monopole to each wedge reduces the visibility of this effect substantially in Fig.~\ref{fig:wedges_lpt}).
As mentioned earlier, we cannot rule out a systematic error in the N-body simulations of several per cent and so we take this difference as a rough estimate of the size of the systematic error in $P_2$.

The only remaining free parameter in our model once the power spectrum and first two velocity moments are fit is the coefficient of the counterterm $\propto k^2 \mu^4 P(k)$, which we argued at the end of Section~\ref{sec:vel_exp} was a good stand-in for the higher-order velocity statistics not explicitly included in our model. Indeed the input value, which we fit by eye, is comparable in magnitude to the contribution from the dipole of the third moment divided by the linear power spectrum. In the spirit of perturbation theory, our philosophy in adjusting this parameter was to increase agreement at low $k$ and $\mu$ rather than minimize errors across the board, even at high $\mu$ where the convergence of the velocity expansions is poor. The model without this counterterm is shown in the dashed lines. Absent this counterterm our model still describes the power spectrum wedges with $\mu \leq 0.5$ at the percent level out to $k = 0.25 \kMpc$, with errors rapidly growing towards higher $\mu$ such that $\mu = 0.7$ is 5\% off at a similar wavenumber; however, the strong angular dependence of the errors means that the quadrupole is more than 10\% away from the data at $k = 0.25\kMpc$. This validates our approach of modeling the redshift-space power spectrum using perturbative models of the first two velocity moments together with the counterterm ansatz for the third moment.

It is important to note, however, that many of the velocity parameters are degenerate for analyses of the redshift-space power spectrum only. In the moment expansion, all the one-loop counterterms in the velocity statistics ultimately take the form $k^2 \mu^{2n} P_{\rm Zel}(k)$ [or $k^2 \mu^{2n} P_{\rm lin}(k)$] at leading order when combined to form the power spectrum. For example, both the counterterm for $\sigma_{2}$ and the third moment take the form $k^2 \mu^4 P(k)$. Similarly, the stochastic contributions will tend to contribute as $(k\mu)^{2n}$. Within the moment expansion we can thus write
\begin{align}
    P^{\rm ME}_s(\bk) = &\, \Big( P(k) + i (k \mu) v_{12,\hn} (\bk) - \frac{(k \mu)^2}{2} \sigma^2_{12,\hn \hn}(\bk) + ...\ \Big)^{\rm PT} \nonumber \\
    &\, + \Big(\alpha_0 + \alpha_2 \mu^2 + \alpha_4 \mu^4 + \cdots\ \Big)\ k^2 P_{\rm lin,Zel}(k) + R_h^3\ \Big(1 + \sigma_v^2 (k\mu)^2 + \cdots\ \Big),
    \label{eqn:me_master}
\end{align}
where $(...)^{\rm PT}$ refers to contributions due only to large scale gravitational dynamics and nonlinear bias parameters computed in either EPT or LPT (with the $k^2P_{\rm lin,Zel}$ being the linear or Zeldovich power spectra in each case, respectively).  This leads to a redshift-space power spectrum with 9 free parameters (4 bias, 3 counterterms, 2 stochastic) with a similar structure of effective corrections as found in the EPT analyses of refs.~\cite{DAmico19,Ivanov19}\footnote{ \edit{Indeed, Equation~\ref{eqn:me_master} is equivalent, up to details of IR resummation and choices of marginal EFT parameters, to the models in those works, with similar ranges of applicability. Specifically, compared to ref.~\cite{DAmico19} we do not including the next-order real-space stochastic correction $\propto k^2$ but include a counterterm $k^2 \mu^6 \PL$ to account for UV dependence in the fourth moment, while compared to ref.~\cite{Ivanov19} we include a superset of 1-loop effective corrections but omit the 2-loop FoG correction in their Equation 3.10, which we do not require for good fits at the velocity level.} }. If the corrections due to third-order bias ($b_3, c_3$) can be set by assuming the Lagrangian bias $b_3 = 0$, as noted in Section~\ref{sssec:bias}, then this is reduced to 8 free parameters. On the other hand, if we wish to include the full one-loop expressions for the third and fourth moments, which possess their own effective and stochastic corrections, two additional non-degenerate parameters are needed, bringing the total up to 11. The aforementioned degeneracy is less manifest in the Fourier streaming model due to the nonlinear composition of the cumulants (and similarly in the configuration-space Gaussian streaming model); however, due to the high degree of quantitative agreement between the ME and FSM expansions at the data level, the various counterterms and stochastic contributions will nontheless be highly degenerate, and as such should not all be fit. Indeed, it should be sufficient to expand these effective contributions as in Equation~\ref{eqn:me_master}, though doing so will break the structure of the streaming model, strictly speaking. Finally, while our model for $P(k,\mu)$ includes five free parameters for counterterms and stochastic effects a condensed set of terms can be used if fitting to more restricted summary statistics. For example, since the counterterms are of the form $k^2 \mu^{2n} P_{\rm lin,Zel}(k)$ they contribute to each multipole proportional to $k^2 P_{\rm lin,Zel}$.  When fitting only the monopole and quadrupole (as in refs.~\cite{Ivanov19,DAmico19,Colas19}) one should fit only for two summary contributions $P_{\ell, \rm c.t.} = \alpha_\ell k^2 P_{\rm Zel},$ though doing so necessarily obscures some of the structure in $P(k,\mu)$ which is poorly fit using only two counterterms.  On the other hand, since we include only two purely stochastic terms, nondegenerate in their contribution to the monopole and quadrupole, they can be separately included even when fitting only for those two statistics.

In Section \ref{sec:spectra} we noted that the predictions of LPT and EPT for $\sigma_{ij}$ differed, and that they appeared to depend upon higher order contributions.  The fact that both the LPT and EPT models do well at describing $P(k,\mu)$ in Fig.~\ref{fig:wedges_lpt} is thus surprising at first sight.  As shown in Section \ref{sec:vel_exp} (Fig.~\ref{fig:components_mu}), however, the errors in $\sigma_2$ are highly suppressed in $P(k,\mu)$ except near $\mu\approx 1$ and so this theoretical uncertainty is subdominant when predicting redshift-space clustering. Furthermore, for realistic galaxy samples we expect the role of stochastic velocities, i.e.\ fingers of god, to be even more significant than the halo sample studied in the figures above; these velocities further increase the role of the monopole $\sigma_0$ relative to $\sigma_2$. This also justifies our choice of modeling for $\sigma_2$, where we do not spend further effort in improving the LPT and EPT modeling, as was argued in Sec.~\ref{sec:spectra}.

\subsection{Comparison for mock galaxies}

\begin{figure}
    \centering
    \includegraphics[width=\textwidth]{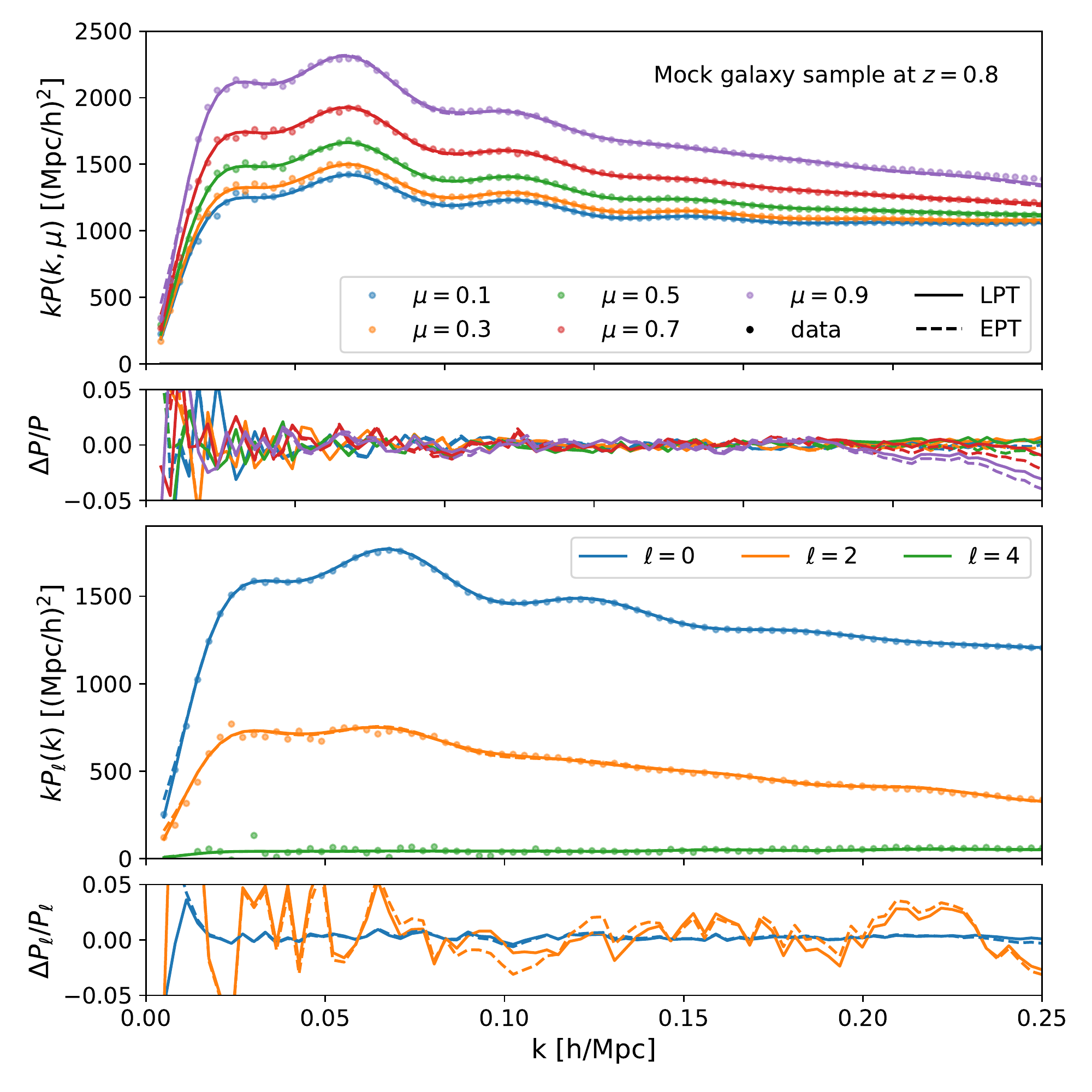}
    \caption{A comparison of the (top) power spectrum wedges ($0.0<\mu<0.2$, $\cdots$, $0.8<\mu<1.0$) and (bottom) multipoles measured for our mock galaxy sample at $z\simeq 0.8$ (points) to the predictions from our PT models (lines; Eq.~\ref{eqn:me_master}).  The upper panel shows the measurements while the lower panel shows the fractional differences.
    }
    \label{fig:gal_sum}
\end{figure}

As a further test of our power spectrum model, in Figure~\ref{fig:gal_sum} we fit our RSD model in Equation~\ref{eqn:me_master} on the mock sample of galaxies embedded into the N-body data using a halo occupation distribution as described at the end of Section~\ref{sec:nbody}. Galaxy samples present a more realistic and stringent test for our model as they are affected by the virial motions of satellite galaxies and indeed, fits to the satellite velocity statistics require significantly larger counterterms (see the discussion around Eq.~\ref{eqn:sigma12_ct}) and stochastic contributions, particularly for the monopole $\sigma_0$ of the second moment and a slightly reduced range-of-fit ($k\sim\sigma_v^{-1}$) compared to the halo case. Nonetheless, at the power spectrum level our model fits the power spectrum wedges $P(k,\mu)$ at the percent level at least up to $k = 0.25 \kMpc$ for all but the highest $\mu$-bin ($\mu = 0.9$), where unlike in the halo case the suppression of power by random velocities towards high $k$ is evident and which begins to deviate from our model prediction at $k = 0.15\kMpc$, reaching 3 percent off by $k = 0.2 \kMpc$. Similarly, our model yields a significantly sub-percent-level fit to the galaxy power spectrum monopole on perturbative scales, while the quadrupole begins to deviate around where the highest-$\mu$ wedge does at around  $k \sim 0.15 \kMpc$. We also checked that, assuming Gaussian covariances and letting the growth $f$ vary, our model (Equation~\ref{eqn:me_master}) can fit the redshift-space power spectrum directly to nearly identical, sub-percent precision over a wide range of scales and recover the growth rate to $1\%$, consistent with the systematic error of the simulations themselves.

\subsection{Fingers of God and stochastic terms}

Despite the above, the size and structure of stochastic velocities and finger-of-god effects, particularly for the specific galaxy samples that will be observed by upcoming spectroscopic surveys, remains one of the biggest limitations of (perturbative) models of redshift-space distortions. It is thus worth discussing the pros and cons of various approaches to tame these effects, in particular the effective-theory parametrization of finger-of-god effects in EFT models such as ours (and those used in refs.~\cite{Ivanov19,Colas19,DAmico19}) compared to more conventional FoG models such as \cite{Peacock92,Park94,Peacock94,TNS10,RSDmock}. As discussed in ref.~\cite{Bella18}, the main difference between these approaches is that traditional\footnote{An intermediate case is represented by ref.~\cite{Hand17}, who assume a functional form with many free parameters.} models assume strict forms for FoG effects (e.g.\ exponential or Lorentzian damping) depending on a small set of parameters, while EFT parametrizations such as ours are restricted only by symmetry arguments and thus in principle span the entire allowed space of FoG models, such that the former could be preferred if they well-describe observed FoGs. Assuming specific FoG models necessarily implies setting strong restrictions on the structure of halo or galaxy velocities at small scales. For example, in the language of the moment expansion approach, assuming Gaussian damping $\propto \exp(-k^2\mu^2\sigma_{\rm FoG}^2)$ \edit{, as in the TNS model \cite{TNS10}, } is equivalent to requiring that the effects of higher-order moments of virial velocities be described by the same paramter, $\sigma_{\rm FoG}$, as the lower moments. A similar, but more EFT-minded, approach could be to input these restrictions as priors (e.g.\ based on fits to simulations) while enabling fitting the full set of allowed parameters to a given order.  The priors would reduce the statistical impact of the additional free parameters and a comparison of the posterior to the prior would allow us to tell if the observed data were in tension with the assumptions.  This is especially important since the velocities of galaxies with complex selections can have significantly more structure than usually assumed in mock catalogs.

Finally we note that the impact of redshift errors, which also affect the line-of-sight clustering signal, can be partly compensated by having a very flexible finger-of-god model such as we have introduced above.  If there is reason to suspect that redshifts are not being accurately estimated in a survey, this could argue for broader priors on these terms than might otherwise arise just from dynamical studies of galaxy orbits in observations or simulations.

\subsection{IR resummation}

Let us comment on the role of large-scale (IR) displacements in the velocity-expansion approach to redshift-space distortions. While these large-scale modes have essentially linear dynamics, their presence results in the nonlinear damping of spatially-localized features such as baryon acoustic oscillations (BAO) that can be complicated to capture perturbatively in Eulerian treaments \cite{SenZal15,Ivanov19}. On the other hand, a convenient feature of Lagrangian perturbation theory is that it naturally includes a resummation of these bulk displacements, making it a natural candidate with which to understand the nonlinear damping of the BAO feature in both real and redshift space \cite{Mat08a,CLPT,Vlah16}. By extension, our LPT calculations of the real-space velocity moments naturally resums these modes.

However, the combination of these velocity moments into the redshift-space power spectrum breaks the resummation of IR \textit{velocities} while keeping the isotropic displacements resummed. Within the framework of LPT, bulk velocities can be naturally resummed by promoting Lagrangian displacements to redshift space using matrix multiplication $\Psi^{(n)}_i \rightarrow R_{ij}^{(n)} \Psi^{(n)}_j = (\delta_{ij} + f \hn_i \hn_j) \Psi^{(n)}_j$ \cite{Mat08a}. This transformation takes the exponentiated linear displacements $A_{ij} \rightarrow R_{in} R_{im} A_{nm}$, naturally endowing the resummed exponential with the angular structure of redshift-space distortions \cite{Mat08a,CLPT}. For example, the isotropic part of $A_{ij} = X(q) \delta_{ij} + Y \hq_i \hq_j$, given by $\Sigma^2(q) = (X + Y/3)$, becomes multiplied by $ (1 + f(2+f)\mu^2) $ under this transformation. Expanding order-by-order in the velocities as we have done in this paper, and thus in the growth rate $f$, necessarily breaks this structure. The procedure to capture all the IR modes, including the velocity contributions, in purely LPT framework has been outlined in \cite{VlaWhi19}. We intend to return to that in future work.  

In the EPT framework, an approximate but pragmatic way of handling these IR modes has been developed, relying on the wiggle/no-wiggle split. The essential feature of this IR resummation procedure is the decomposition and isolation of the wiggle part (caused by the baryon acoustic oscillations) of linear power spectra, and the damping of oscillatory components due to the wiggles by an appropriate factor dependent on the IR displacements to be resummedsee  \cite{Baldauf2015,Vlah16,Blas2016,Ding2017,Ivanov2018} (details in Appendix~\ref{app:IR_resumm}). This procedure, however, relies on several approximation steps, from the details of the wiggle/no-wiggle splitting to ensuring that subleading corrections can be neglected at the order of interest. As highlighted earlier, and in contrast to EPT, LPT performs the resummation of long displacement modes directly and does not rely on any of these approximation steps. LPT thus constitutes a natural environment to understand the various approximation levels undertaken in the wiggle/no-wiggle splitting procedure, and thus provides the bridge from the direct and exact treatment of IR modes to the comprehensive and intuitive picture provided by the simplicity of the wiggle/no-wiggle splitting result. 

These characteristics and differences of the LPT and EPT in the treatment of the IR modes are highlighted even further once the possible additional, beyond BAO, oscillatory features of the power spectrum are considered. Such oscillatory features can be produced by, e.g., primordial physics, and are also affected by the long displacements in a similar manner to the BAO, exhibiting damping and smoothing that can again be captured by performing IR resummation \cite{Vlah16,Beutler19,Vasudevan19}. The evident advantage of LPT in this scenario is that this resummation is performed automatically without the need for further engagement or analysis, finding saddle points etc. 

Despite the incomplete resummation of IR velocities as described above, however, as shown in Figure~\ref{fig:wedges_lpt}, in Fourier space the velocity expansions are nonetheless able to capture the anisotropic BAO wiggles to high accuracy. We can attempt to estimate the effects of the missing bulk contributions to the higher velocity moments as follows. Within the context of LPT we can write, for example for the lowest-order $b_1^2$ contribution to the redshift-space power spectrum \cite{Vlah16}
\begin{equation}
    P_{b_1^2}(\bk) = \dtq e^{i\bk \cdot \bq - \frac{1}{2} k_i k_j R_{in} R_{jm} A_{nm}(\bq)} \xi_{\rm lin}(q) = e^{-\frac{1}{2}k^2 \Sigma^2(r_s, \mu)} P_{w}(k) + P_{\rm smooth}.
    \label{eqn:pwnl}
\end{equation}
Equation~\ref{eqn:pwnl} can be understood as follows: since the linear correlation function $\xi_{\rm lin}$ has a prominent BAO ``bump'' at $r_s$, it picks out the exponentiated damping factor at $q = r_s$ such that the bump is smoothed by $\Sigma^2(r_s,\mu) = (1 + f(2+f)\mu^2)\ \Sigma^2(r_s)$ in Fourier space, while the correlation without the bump gets affected smoothly since it has no preferred scale. The separation into a smooth component and the BAO feature is commonly used in the literature and known as the wiggle/no-wiggle split \cite{ESW07,Xu12,Vlah16}, but LPT makes an exact prediction for the damping through the resummation of linear modes at the BAO scale. In particular, we can now understand how the BAO feature is affected if we neglect the effect of bulk velocities at $n^{\rm th}$ order in the moment expansion. Noting that the $n^{\rm th}$ velocity moment contributes to the power spectrum proportional to $f^n$, we can expand the exponential in Equation~\ref{eqn:pwnl} as
\begin{equation}
    P_{w, NL}(k) = e^{-\frac{1}{2} k^2 \Sigma_0^2} \Big[ 1 - (k\mu)^2 \Sigma_0^2 f + \frac{1}{2} (-(k\mu)^2 \Sigma_0^2 + (k\mu)^4 \Sigma_0^4 ) f^2 + \cdots\ ) \Big] P_w(k)
\end{equation}
where the coefficients of $f$ and $f^2$ correspond to contributions from the first two velocity moments in Equation~\ref{eqn:me_master}. Using the moment expansion to $n = 2$ is equivalent to Taylor-expanding in $f$ and keeping only two terms. However, a corollary of the above is that the damping beyond these terms necessarily scales strongly with $\mu$ (and $k$), making it negligible for all but the highest $\mu$ wedges --- and indeed any residual anistropic wiggles in our fits to the redshift-space power spectrum from simulations must be well within the errors of these measurements, which are themselves tighter than state-of-the-art spectroscopic surveys like DESI. 

Nonetheless, while being almost undetectable in Fourier space these errors will accumulate in configuration space to produce deviations from measurements noticeable by eye, particularly in the quadrupole, so our current strategy will need to be modified for configuration space analyses.  The two most obvious options to this end are (1) to compute $P_s(\bk)$ for the broadband using $P_{nw}$ only and add in the exponential damping factor for $P_w$ by hand, as has been done in recent analyses in the EFT framework \cite{Ivanov19} or (2) to use the Gaussian streaming model (GSM; \cite{VlaCasWhi16}) for configuration space analyses employing the same bias parameters and counterterms for velocity statistics in configuration space. The latter is an attractive option because the velocity expansions in Fourier space and in the GSM can be computed within the same dynamical framework employing consistent bias parameters and counterterms\footnote{And keeping in mind the relation between stochastic terms in Fourier and configuration space.}, though the GSM captures the IR displacements almost perfectly (see Appendix~\ref{app:gsm}) while the Fourier space methods can more easily capture the broadband effects of the IR displacements. A more complete but laborious approach would be to compute the power spectrum with both linear displacements and velocities resummed as in Convolution Lagrangian Perturbation Theory \cite{CLPT}; we intend to return to this in the near future.

\section{Conclusions}
\label{sec:conclusions}

The upcoming generation of spectroscopic surveys and CMB experiments promise to deliver unprecedented information about galaxy velocities on cosmological scales, either indirectly through the anisotropic clustering of observed galaxies due to redshift-space distortions or directly through the kinetic Sunyaev-Zeldovich effect or peculiar velocity surveys. Velocity and density statistics provide us with complementary information about structure formation, which can further be combined with probes such as weak lensing and allow us to test the predictions of $\Lambda$CDM and general relativity on the largest scales.

Our goal in this paper is to consistently model both real-space velocity spectra and the redshift-space power spectrum of biased tracers (e.g.\ galaxies) within one-loop perturbation theory.  The redshift-space power spectrum, $P(k,\mu)$, can be understood as an expansion in the line-of-sight wavenumber, $k_\parallel=k\mu$, multiplying $n^{\rm th}$-order pairwise velocity spectra.  After describing the four $(4 \Gpc)^3$ N-body simulations we compare to in Section \ref{sec:nbody}, we begin in Section \ref{sec:vel_exp} by using the N-body halo velocity statistics to test the convergence of two Fourier-space velocity expansions of the redshift-space power spectrum, the moment expansion approach and the Fourier streaming model. The expansions show good quantitative agreement with the $P(k,\mu)$ measured from the same set of simulations when the first three pairwise velocity moments are included, reaching percent-or-below levels of agreement on scales of interest to cosmology except when $\mu \approx 1$ (i.e.\ close to the line-of-sight) where the agreement is slightly worse at small scales, though still at the percent level or below for $k < 0.2 \kMpc$ for our fiducial halo sample.  Including higher moments ($n = 4$) fails to significantly improve the expansion, indicating slow convergence at scales where the nonlinear velocities of halos become dominant. We find that the redshift-space power spectrum can be modeled at the percent level on perturbative scales by using a counterterm ansatz for moments beyond $n = 2$ valid in precisely this scenario.

In Section~\ref{sec:spectra} we model the real-space power spectrum and the first two velocity moments within one-loop Lagrangian and Eulerian perturbation theory, comparing them to N-body simulations and highlighting their salient features and differences in the final subsection. Our model employs effective field theory (EFT) corrections to the nonlinear dynamics as well as a bias scheme including shear and cubic contributions as well as derivative bias degenerate with the counterterms. We find that, when the appropriate counterterms and stochastic corrections are included, one-loop LPT and EPT can model the zeroth and first velocity moments (the real-space power spectrum and pairwise velocity) to comparable scales for both broadband shape and oscillatory features.  For the second moment (velocity dispersion) LPT shows a more limited range-of-fit while EPT relies on one-loop terms of the same order as linear theory and slightly under-predicts the damping of oscillatory features in the one-loop terms, suggesting that the velocity dispersion spectrum is subject to significant non-linearity even at intermediate scales.  In general we find the higher moments to be ``more non-linear'' and to have larger contributions from stochastic and counter terms as we move up the hierarchy.

Finally, in Section~\ref{sec:rsd_pk} we combine the velocity expansions and velocity modelling to obtain a model of the redshift-space power spectrum in one-loop perturbation theory. Using the bias parameters and effective corrections derived from the data statistics in addition to the aforementioned counterterm ansatz for contributions from velocity moments beyond $n = 2$ yields a percent-level fit to the halo power spectrum for all wedges out to $k = 0.25\kMpc$ at $z = 0.8$, with similar performance for the multipoles. As a further test, we analyzed a sample of mock galaxies using the same procedure, and found qualitatively similar results despite significantly more pronounced stochastic terms (expected due to virial motions of satellites) and a slightly decreased range of fit at higher $\mu$ and in the quadruople as a result. In addition, we conducted a fit directly to the power spectrum wedges for this sample, assuming Gaussian covariances and letting both $f$ and the bias parameters to float, and found that our model recovers the growth rate to within the systematic error of the simulations themselves with no loss of accuracy.

Our {\tt python} code {\tt velocileptors} to compute the aforementioned one-loop velocity statistics and redshift-space power spectrum in both EPT and LPT is publically available. For completeness, the code includes all terms up to the fourth pairwise velocity moment in both formalisms as well as modules to combine them using the moment expansion in both formalisms, full IR resummation as in Equation~\ref{eqn:full_ir} in EPT, and the Fourier and Gaussian streaming models in LPT. The LPT code takes slightly more than a second to compute the all relevant statistics to sub-percent precision on perturbative scales, while the EPT code takes slightly less. We make abundant use of the FFTLog algorithm throughout and compute one-loop EPT terms via manifestly Galilean invariant Hankel transforms inspired by the Lagrangian bias expansion (Appendix~\ref{app:Hankel}).

The structure of the moment expansion implies that the theoretical error should be a strong function of $\mu$, which can be taken as an argument in favor of modeling power spectrum wedges, $P(k,\mu)$, over multipoles, $P_\ell(k)$.  The importance of both counterterms and stochastic terms in the velocity statistics suggests that the cosmological information in $P(k,\mu)$ at high $k$ and $\mu$ is less than one might naively think, since it is precisely in this regime that these non-cosmological contributions become an appreciable fraction of the total power.  It is also at higher $k$ and $\mu$ that non-trivial behavior of FoG models and observational redshift errors would be expected to impact the measurements the most.

We close by noting some possible near-term applications. Firstly, our model naturally includes precision modelling of cosmological velocities at quasilinear scales and will be directly applicable to upcoming kSZ and peculiar velocity surveys \cite{SimonsObs,CMBS4,Howlett17,Graziani20}.  While we have focused our predictions on velocity statistics in real-space, the conversion to redshift space can be straightforwardly obtained by the appropriate $f$ derivatives of the redshift-space power spectrum \cite{Sug16}, which are themselves predicted by the model as linear combinations of density-weighted pairwise velocity statistics.  In this regard the increasing importance of counterterms and stochastic terms as we move higher in the moment hierarchy suggests that much of the cosmological information in velocity surveys will be contained on large scales.

In terms of redshift-space distortions, \edit{our model includes a superset of effective corrections at 1-loop level and} is similar in many respects to those recently used to analyze BOSS data in ref.~\cite{Ivanov19,DAmico19,Colas19,Chudaykin++:2020,DAmico20} or the ``blind challenge'' of ref.~\cite{Nishimichi20}.  An obvious next step from the present analysis would be to analyze those data with the formalism described in this work. Our model should likewise be competitive in analyses of future high-redshift galaxy surveys like HETDEX \cite{Hetdex}, DESI \cite{DESI}, Euclid \cite{EUCLID18} and even futuristic LBG \cite{Ferraro19} or 21-cm \cite{Slosar19} surveys, though as discussed in Section~\ref{sec:vel_exp} the applicable range of scales will likely be limited more by the scale of stochastic velocities ($k_{\rm FoG} \sim \sigma_v^{-1}$), or FoGs, than by the nonlinear wavenumber $k_{NL}$ at higher redshifts.  This was demonstrated already in EFT analyses of BOSS, where $k_{\rm FoG} \sim 0.2 \kMpc$, though specific FoG properties will depend on the galaxies sampled by each survey, and will be particularly interesting in the context of high-redshift 21-cm surveys where stochastic velocities are relatively small \cite{VN18} but observations are naturally limited by foregrounds to higher $\mu$. Finally, the aforementioned probes can be combined with upcoming lensing surveys. By letting the gravitational slip \cite{Jain10,Joyce15,Ame18} float as a free parameter like the linear growth rate $f$, this will let us test the predictions of General Relativity on cosmological scales.  By providing a model which can simultaneously fit all of the relevant statistics we enable a principled statistical analysis that can avoid taking ratios of noisy data points.

\acknowledgments
We would like to thank Emanuele Castorina and Marko Simonovi\'c for useful discussions during the preparation of this manuscript. We thank Yin Li for making the {\tt mcfit} package public and similarly thank Chirag Modi for sharing his {\tt CLEFT} code, as well as helping us check the {\tt velocileptors} for numerics and factors of two. S.C.~ thanks the CERN theory group for its hospitality while part of this work was being completed.
S.C.~ is supported by the National Science Foundation Graduate Research Fellowship (Grant No.~DGE 1106400) and by the UC Berkeley Theoretical Astrophysics Center Astronomy and Astrophysics Graduate Fellowship.
M.W.~is supported by the U.S. Department of Energy and the NSF.
This research has made use of NASA's Astrophysics Data System and the arXiv preprint server.
This research used resources of the National Energy Research Scientific Computing Center (NERSC), a U.S. Department of Energy Office of Science User Facility operated under Contract No. DE-AC02-05CH11231.

\appendix

\section{Velocity moments and RSD power spectrum in  Eulerian PT}
\label{app:others}

\subsection{Third-Order Bias Expansion in EPT and LPT}
\label{app:third_order_bias}

In this paper we extend the expressions for the real-space power and pairwise-velocity spectra found in \cite{VlaCasWhi16} to include contributions from third-order bias operators. In principle, going to third order in bias requires an additional four bias parameters (see e.g. \cite{Des16}), however as shown in \cite{McDRoy09} for EPT at one-loop order many of these contributions are either zero or amount to re-definitions of the linear bias parameter $b_1$. The remaining contributions are all degenerate and can be combined into a single (EPT or LPT) third-order bias parameter $c_3$ or $b_3$. In this subsection we will review the bias expansion in EPT and provide details for including the effects of third-order bias in LPT predictions of the velocity moments.

In order to evaluate these velocity moment correlators in Eulerian PT we adopt the biasing scheme of ref.~\cite{McDRoy09} in Equation~\ref{eqn:ept_bias} up to third order, which we repeat here for convenience:
\begin{equation}
\delta_h = c_1 \delta + \frac{c_{2}}{2} \delta^2+c_{s} s^2 + \frac{c_{3}}{6}\delta^3+c_{1s}\delta s^2 + c_{st} st + c_{s3} s^3+c_\psi \psi,
\end{equation}
where $s^2= s_{ij}s_{ij}$, $s^3= s_{ij}s_{jl}s_{li}$ and $st= s_{ij}t_{ij}$, and the shear operators are defined as
\begin{equation}
\psi = \eta-\frac{2}{7}s^2 +\frac{4}{21}\delta^2,
~~ s_{ij} = \left(\frac{\partial_i\partial_j}{\partial^2}-\frac{1}{3}\delta_{ij}\right)\delta,
~~ t_{ij} = \left(\frac{\partial_i\partial_j}{\partial^2}-\frac{1}{3}\delta_{ij}\right)\eta,
~~ \eta = \theta -\delta.
\end{equation}
As usual we assume subtraction of mean field values like $\left\langle\delta^2 \right\rangle$.
In Fourier space, the second and third order shear operators are given by the kernels in momentum space
\begin{align}
S^{(2)}_2(\bk_1,\bk_2) &= \frac{(\bk_1\cdot \bk_2)^2}{k_1^2 k_2^2} - \frac{1}{3}, \\
S^{(3)}_2(\bk_1,\bk_2,\bk_3) &= 2 
S_2\left( \bk_1, \bk_2 + \bk_3 \right)
F_2\left( \bk_2, \bk_3 \right) , \notag\\
S^{(3)}_3(\bk_1,\bk_2,\bk_3) &= \frac{(\bk_1\cdot \bk_2)(\bk_2\cdot \bk_3)(\bk_3\cdot \bk_1)}{k_1^2 k_2^2 k_3^2} 
- \frac{(\bk_1\cdot \bk_2)^2}{ 3 k_1^2 k_2^2} - \frac{(\bk_1\cdot \bk_3)^2}{ 3 k_1^2 k_3^2}  - \frac{(\bk_3\cdot \bk_2)^2}{ 3 k_3^2 k_2^2} 
+ \frac{2}{9}, \notag\\
S^{(3)}_{st}(\bk_1,\bk_2,\bk_3) &= \frac{2}{7} S_2\left( \bk_1, \bk_2 + \bk_3 \right) \left[ S_2\left( \bk_2, \bk_3\right) - \frac{2}{3} \right], \notag\\
S^{(3)}_{\psi}(\bk_1,\bk_2,\bk_3) &= G_3\left( \bk_1, \bk_2, \bk_3 \right) - F_3\left( \bk_1, \bk_2, \bk_3 \right) 
- \frac{4}{7} \left( S_2\left( \bk_1, \bk_2 + \bk_3 \right) - \frac{2}{3} \right) F_2\left( \bk_2, \bk_3\right). \notag
\end{align}
Given that in this paper we are interested only in two-point statistics, many of the third order bias operators listed above do contribute to the one-loop power spectrum in degenerate manner. After shot-noise renormalization only one non-vanishing independent contribution remains. The relevant correlators in one-loop EPT for the real-space power spectrum are:
\begin{align}
\left< \delta_{\rm lin} | \left[\delta s^2 \right]^{(3)} \right>' &= \left< \delta_{\rm lin} | [s^3]^{(3)} \right>'  = 0, \\
\left< \delta_{\rm lin} | [\delta^2]^{(3)} \right>' &= \frac{68}{21} \avg{\delta_{\rm lin}^2} \PL(k), \notag
\\
\left< \delta_{\rm lin} | \psi^{(3)} \right>' &= 
3 P_{\rm lin}(k)  \int_\bp S_{\psi} (\bp, -\bp, \bk) P_{\rm lin}(p), \notag\\
&=\frac{16}{21}\left(
\left< \delta_{\rm lin} | [st]^{(3)} \right>'
+\frac{16}{63}
\left< \delta^2_{\rm lin} \right>'
\PL
\right) \notag \\
&=-\frac{16}{105}\left(
\left< \delta_{\rm lin} | (s^2)^{(3)} \right>'
-\frac{136}{63}
\left< \delta^2_{\rm lin} \right>'
\PL
\right), \notag
\end{align}
The corresponding contributions to the pairwise velocity, due to the correlator $\avg{\delta_h | v_i}$ can be obtained by simply multiplying these terms by $\frac{i k_i}{k^2}$.

The above degeneracies also exist in one-loop LPT. In particular, for two third-order EPT operators $O_3(\bx)$ and $O'_3(\bx)$ such that $\avg{\delta_{\rm lin}|O_3} = \avg{\delta_{\rm lin}|O'_3 + A \avg{\delta_{\rm lin}^2} \delta_{\rm lin}}$ at one-loop, we also have the degeneracies
\begin{align}
    &\avg{ O_3(\bq_1) \Delta_i } = \avg{ O_3'(\bq_1) \Delta_i } + A \avg{\delta_{\rm lin}^2} U^{\rm lin}_i(\bq), \\
    &\avg{ O_3(\bq_1) \delta_{\rm lin}(\bq_2)} = \avg{ O'_3(\bq_1) \delta_{\rm lin}(\bq_2)} +  A \avg{\delta_{\rm lin}^2} \xi_{\rm lin}(q)\notag
\end{align}
in one-loop LPT. In the main body of the paper we will thus summarize these contributions with the third-order parameter $O_3(\bq) = s_{ij}(\bq) t_{ij}(\bq) + \frac{16}{63} \avg{\delta_{\rm lin}^2}$. This introduces the additional correlators
\begin{align}
    &U_{b_3,i}(\bq) = U_{b_3}(q)\ \hq_i = \avg{ O_3(\bq_1) \Delta_i } = - \int \frac{dk\ k}{2\pi^2} R_{b_3}(k) j_1(kq)  \\
    &\theta(\bq) = \avg{ O_3(\bq_1) \delta_{\rm lin}(\bq_2) } = \int \frac{dk\ k^2}{2\pi^2} R_{b_3}(k) j_0(kq). \notag
\end{align}
An explicit formula for $R_{b_3}$ expressed as a Hankel transform is given in Appendix~\ref{app:Hankel}. Finally, the expression for the pairwise velocity spectrum requires the time derivative of $U_{b_3}$, which is given by $\dot{U}_{b_3} = \avg{O_3 \dDelta} = f U_{b_3}.$

\subsection{Eulerian moment expansion}
\label{app:ept}

In this section we give a 
short overview of the Eulerian 
moment expansion framework for RSD 
based on the distribution function model 
\cite{SelMcD11, Vla12, Vla13}, using one-loop, Eulerian effective PT to compute the components. 
These results, after including IR-resummation, are equvivalent to those recently used in refs.~\cite{DAmico19, Ivanov19}. 

The velocity moments are combined to give the RSD power spectrum as in Eq.~\eqref{eqn:moment_expansion}.
Up to one-loop we need to consider the contributions of several velocity moments
\begin{align}
\tilde{\Xi}^{(0)}_{\hat n}(\bk) &= P_{00}(k), \\
\tilde{\Xi}^{(1)}_{\hat n}(\bk) &= P_{01}(k,\mu) - P^*_{01}(k,\mu) \notag\\
&= 2 i {\rm Im}[P_{01}(k,\mu)], \notag\\
\tilde{\Xi}^{(2)}_{\hat n}(\bk) &= P_{02}(k,\mu) - 2 P_{11}(k,\mu) + P^*_{02}(k,\mu) \notag\\
&= 2 {\rm Re}[P_{02}(k,\mu) - P_{11}(k,\mu)], \notag\\
\tilde{\Xi}^{(3)}_{\hat n}(\bk) &= P_{03}(k,\mu) - 3 P_{12}(k,\mu) + 3 P^*_{12}(k,\mu) - P^*_{03}(k,\mu) \notag\\
&= 2 i {\rm Im}[P_{03}(k,\mu) - 3 P_{12}(k,\mu)], \notag\\
\tilde{\Xi}^{(4)}_{\hat n}(\bk) &= P_{04}(k,\mu) - 4 P_{13}(k,\mu) + 6 P_{22}(k, \mu) - 4 P^*_{13}(k,\mu) + P^*_{04}(k,\mu) \notag\\
&= 2 {\rm Re} [P_{04}(k,\mu) - 4 P_{13}(k,\mu) + 3 P_{22}(k, \mu) ]. \notag
\end{align}
where the component spectra $P_{LL'}$ are the cross-correlations 
of different velocity moments 
\begin{equation}
P_{LL'} (k, \mu)
\equiv
\left< \big(1 + \delta\big)* u^{L}_{\hat n}
\Big| \big(1 + \delta\big)* u^{L'}_{\hat n}\right>'
\equiv
\left< \big(1 + \delta(\bk)\big)* u^{L}_{\hat n}(\bk) ~
\big(1 + \delta(\bk')\big)* u^{L'}_{\hat n}(\bk')\right>',
\end{equation}
where, for brevity, we introduce the primed expectation values to denote expectation values with Dirac delta function dropped, i.e.\ $
\left< A | B \right> \equiv
\langle A(\bk) B(\bk') \rangle =  (2\pi)^3 \delta_D (\bk + \bk')\  \langle A(\bk) B(\bk') \rangle'$. 

Note that $P_{LL'}  = P_{L'L}^*$, so, 
without loss of generality, we can assume $L \leq L'$.
See ref.~\cite{VlaWhi19} for a more detailed connection between the moment expansion and streaming models. 

\textbf{Contributions to $\Xi^{(0)}$} arise from only $P_{00}$. This is the usual real space halo-halo power spectrum.
We have
\begin{align}
\label{eq:Xi_0_spt}
\left< \delta_h | \delta_h \right>' =
c_1^2 P_{\rm lin}(k) &+ 2 c_1^2 \int_\bp \bigg( \big[ F_2 (\bp, \bk-\bp) \big]^2 P_{\rm lin}(|\bk - \bp|) + 3 F_3 (\bp, -\bp, \bk) P_{\rm lin}(k) \bigg) P_{\rm lin}(p) \\
&\hspace{0.0cm} + 2 c_1 c_2  \int_\bp F_2(\bp,\bk - \bp) P_{\rm lin}(p) P_{\rm lin}(|\bk - \bp|)  \notag\\
&\hspace{0.0cm} + 4 c_1 c_s  \int_\bp ~ F_2(\bp,\bk - \bp) S_2(\bp,\bk - \bp) P_{\rm lin}(p) P_{\rm lin}(|\bk - \bp|) \notag\\
&\hspace{0.0cm} + \frac{c_2^2}{2}  \int_\bp P_{\rm lin}(p) P_{\rm lin}(|\bk - \bp|)  \notag\\
&\hspace{0.0cm} + 2 c_2 c_s  \int_\bp ~ S_2(\bp,\bk - \bp) P_{\rm lin}(p) P_{\rm lin}(|\bk - \bp|) \notag\\
&\hspace{0.0cm} + 2 c_s^2  \int_\bp \big[ S_2(\bp,\bk - \bp) \big]^2 P_{\rm lin}(p) P_{\rm lin}(|\bk - \bp|) \notag\\
&\hspace{0.0cm} + 6 c_1 c_3 P_{\rm lin}(k)  \int_\bp S_{\psi} (\bp, -\bp, \bk) P_{\rm lin}(p)
\tag{in EPT}
\end{align}
where the third-order bias operators can be combined into a single term with the coefficient $c_3$.
Counter terms that are required to regularise 
the one loop $\Xi^{(0)}$ terms 
are of form $(k^2/k^2_\star) P_{\rm lin}(k)$
($k_\star$ is a characteristic proto-halo size scale)
and thus degenerate with the 
derivative bias contribution. 
Besides these there is 
a constant shot noise contributions
obtained by correlating 
the stochastic component of the halo density 
field $\epsilon_h(\bk)$, and we neglect the higher derivative stochastic terms. 
Thus the total $\Xi^{(0)}$ expression reads
\begin{equation}
\label{eq:Xi_0_eft}
\tilde{\Xi}^{(0)}_{\rm 1-loop}(k)
= {\rm \eqref{eq:Xi_0_spt}}
+ c^{(0)}_0 \frac{k^2}{k^2_*} P_{\rm lin}(k)
+ {\rm ``const_0"}
+ \ldots,
\end{equation}
where ${\rm ``const_0"} = \left< \epsilon_h | \epsilon_h \right>'$, and $c^{(0)}_0$
is the leading derivative counterterm.
In general, for counterterms we will use the notation
$c^{(\ell)}_m$ taking into account that different angular dependences
can have different counterterm contributions.

\textbf{Contributions to $\Xi^{(1)}$} arise from only the $P_{01}$ term. This gives us 
\begin{align}
\left< \delta_h | (1+\delta_h) * v_\parallel \right>' 
&\approx  \left< \delta_h | v_\parallel \right>'  +  \left< \delta_h | \delta_h * v_\parallel \right>', 
\end{align}
where in one-loop EPT we have
\begin{align}
\label{eq:dv_spt}
\left< \delta_h | v_\parallel \right>' 
&\approx - i \frac{\mu}{k} \bigg( c_1 P_{\rm lin}(k) + 2 c_1 \int_\bp ~ F_2 (\bp, \bk-\bp)  G_2 (\bp, \bk-\bp) 
P_{\rm lin}(p) P_{\rm lin}(|\bk - \bp|) \\
&\hspace{2.9cm} + 3 c_1 P_{\rm lin}(k) \int_\bp ~ \bigg[ F_3 (\bp, -\bp, \bk) + G_3 (\bp, -\bp, \bk) \bigg] P_{\rm lin}(p) \notag\\
&\hspace{2.9cm} + c_2 \int_\bp ~ G_2 (\bp, \bk-\bp) P_{\rm lin}(p) P_{\rm lin}(|\bk - \bp|) \notag\\
&\hspace{2.9cm} + 2 c_s \int_\bp ~ S_2 (\bp, \bk-\bp) G_2 (\bp, \bk-\bp) P_{\rm lin}(p) P_{\rm lin}(|\bk - \bp|) \notag\\
&\hspace{2.9cm} + 3 c_{3} P_{\rm lin}(k)  \int_\bp S_{\psi} (\bp, -\bp, \bk) P_{\rm lin}(p) \bigg) \tag{in EPT},
\end{align}
and
\begin{align}
\label{eq:ddv_spt}
\left< \delta_h | \delta_h * v_\parallel \right>' 
&\approx - 2 i 
\Bigg( 
c_1^2 \int_\bp ~ \frac{p_\parallel}{p^2} F_2 (\bp, \bk-\bp) P_{\rm lin}(p) P_{\rm lin}(|\bk - \bp|) \\
&\hspace{2.5cm} + c_1^2 P_{\rm lin}(k) \int_\bp ~ \bigg[ \frac{p_\parallel}{p^2} F_2 (\bp, - \bk) + \frac{(\bk - \bp)_\parallel}{(\bk - \bp)^2} G_2 (\bp, - \bk) \bigg] P_{\rm lin}(p) \notag\\
&\hspace{2.5cm} + c_1 c_2 \frac{1}{2} \int_\bp ~ \frac{p_\parallel}{p^2} P_{\rm lin}(p) P_{\rm lin}(|\bk - \bp|) \notag\\
&\hspace{2.5cm} + c_1 c_s \int_\bp ~ \frac{p_\parallel}{p^2} S_2 (\bp, \bk-\bp) P_{\rm lin}(p) P_{\rm lin}(|\bk - \bp|), \notag
\Bigg) \tag{in EPT}.
\end{align}
Note that, due to the angular symmetry, $c_2$ and $c_s$ do not contribute to the tadpole diagrams, i.e.\ to the $P_{13}$-like terms. 

Counter terms that are needed to regularise the one-loop $\Xi^{(1)}$ terms scale as $\mu (k^2/k^2_\star) P_{\rm lin}(k)$ and are again degenerate with the  derivative bias contribution. 
Collecting all the contributions  to $\Xi^{(1)}$ we have
\begin{equation}
\label{eq:Xi_1_eft}
\tilde{\Xi}^{(1)}_{\rm 1-loop}(k)
= 2\left[{\rm \eqref{eq:dv_spt}}
+ {\rm \eqref{eq:ddv_spt}}
\right]
- i c^{(0)}_1 \frac{\mu k}{k^2_\star} P_{\rm lin}(k)
+ \ldots.
\end{equation}

\textbf{Contributions to $\Xi^{(2)}$} arise from two correlators $P_{02}$ and $P_{11}$.
$P_{02}$ starts to contribute at one-loop order, while $P_{11}$ also has a linear contribution. 
We can write for the former
\begin{align}
\label{eq:P02_spt}
\left< \delta_h | (1+\delta_h) * v^2_\parallel \right>' &\approx  \left< \delta_h | v^2_\parallel \right>'  + \left< \delta_h | \delta_h \right>'
 \left< v^2_\parallel \right>, \\
&=  
- 2c_1 \int_{\bp} \frac{p_\parallel (\bk - \bp)_\parallel}{p^2 (\bk - \bp)^2} F_2 (\bp, \bk-\bp) P_{\rm lin}(p) P_{\rm lin}(|\bk - \bp|)\notag\\
&~~~ - 4c_1 P_{\rm lin}(k) \int_{\bp} \frac{p_\parallel (\bk - \bp)_\parallel}{p^2 (\bk - \bp)^2} G_2 (\bp, -\bk) P_{\rm lin}(p)\notag\\
&~~~ - c_2 \int_{\bp} \frac{p_\parallel (\bk - \bp)_\parallel}{p^2 (\bk - \bp)^2} P_{\rm lin}(p) P_{\rm lin}(|\bk - \bp|)\notag\\
&~~~ - 2 c_s \int_{\bp} \frac{p_\parallel (\bk - \bp)_\parallel}{p^2 (\bk - \bp)^2} S_2 (\bp, \bk-\bp) P_{\rm lin}(p) P_{\rm lin}(|\bk - \bp|)\notag\\
&~~~ + c_1^2 P_{\rm lin}(k) \sigma_{\rm lin}^2, \tag{in EPT}
\end{align}
and for the latter
\begin{align}
\label{eq:P11_spt}
\left<(1+\delta_h) * v_\parallel | (1+\delta_h) * v_\parallel \right>'  
&\approx  \left< v_\parallel | v_\parallel \right>' + 2 \left< v_\parallel | \delta_h v_\parallel \right>' +  \left< \delta_h v_\parallel | \delta_h v_\parallel \right>'  \\
&= \frac{\mu^2}{k^2}
\bigg( P_{\rm lin} + 2 \int_{\bp} \big[G_2 (\bp, \bk-\bp)\big]^2 
P_{\rm lin}(p) P_{\rm lin}(|\bk - \bp|) \notag\\
&\hspace{1.7cm}
+6 P_{\rm lin}(k) \int_{\bp} ~ G_3 (\bp,-\bp, \bk) P_{\rm lin}(p)
\bigg) \notag\\
&~~~
+4c_1\frac{\mu}{k}
\bigg(\int_{\bp} ~\frac{p_\parallel}{p^2}G_2 (\bp, \bk-\bp) P_{\rm lin}(p) P_{\rm lin}(|\bk - \bp|) \notag\\
&\hspace{2.0cm}+ P_{\rm lin}(k) \int_{\bp} ~ \bigg[\frac{(\bk-\bp)_\parallel} {(\bk - \bp)^2}G_2 (-\bp, \bk)+\frac{p_\parallel}{p^2} F_2 (-\bp, \bk)\bigg] 
P_{\rm lin}(p)\bigg)\notag\\
&~~~
+c_1^2\int_{\bp} ~
\frac{p_\parallel}{p^2}
\Bigg(
\frac{p_\parallel}{p^2}
+
\frac{(\bk-\bp)_\parallel}
{(\bk - \bp)^2}
\Bigg)
P_{\rm lin}(p) P_{\rm lin}(|\bk - \bp|). \tag{in EPT}
\end{align}
The second contribution in $P_{02}$  (i.e.\  $\propto c_1^2 P_{\rm lin} \sigma_{\rm lin}^2$) and the last term in $P_{11}$ ensure IR cancelation in the soft limit. 

A new feature of the $\Xi^{(2)}$ correlator is 
that it has terms with isotropic angular dependence, $\mu^0$, as well as $\mu^2$ dependence.
Both of these have one-loop terms that need 
to be regularised and thus we have to 
introduce counterterms of form 
$(k^2/k^2_\star)P_{\rm lin}(k)$ for each of these
two angular dependencies. 
The contribution to the isotropic angular dependence comes primarily due to the small-scale 
velocity dispersion contribution in the $P_{02}$
term, i.e.\ $P_{00} \sigma^2 \ni 
c_1^2 P_{\rm lin}(k) \sigma_{\rm non-lin}^2$. 
Here $\sigma_{\rm non-lin}$ encapsulates the
non-perturbative, small-scale, contribution 
to the halo velocity dispersion.

In addition to these derivative terms the
$\Xi^{(2)}$ correlator contains both isotropic and anisotropic stochastic contribtuions. For example, the UV dependence of the $c_2$ contribution to $P_{02}$ needs to be renormalized by a constant, isotropic contribution proportional to $\delta_{ij}$. Moreover, $\Xi^{(2)}$ will generically inherit a 
stochastic contribution via $P_{00}$ 
in \eqref{eq:P02_spt}. 
We can write
\begin{equation}
 \left< \delta_h | \delta_h * v^2_\parallel \right>'
 \ni 
 \left< \epsilon_h | \epsilon_h \right>'
 \left< v^2_\parallel \right>
=
 ``{\rm const_0}"
 \left(\sigma_{\rm lin}^2 + \sigma_{\rm non-lin}^2 \right),
\end{equation}
where, in the last line, we split the halo velocity dispersion into the linear component 
and the residual non-linear component 
coming from small-scales.
However a similar contribution can 
be obtained from the last term in 
$P_{11}$, where we again have
\begin{equation}
 \left< \delta_h * v_\parallel
 | \delta_h * v_\parallel \right>'
 \ni 
{\rm FT}\Big[
\left< \epsilon_h | \epsilon_h \right>
\left< v_\parallel | v_\parallel \right>
\Big]'
= 
 \left< \epsilon_h | \epsilon_h \right>'
 \left< v^2_\parallel \right>
=
 ``{\rm const_0}"
 \left(\sigma_{\rm lin}^2 + \sigma_{\rm non-lin}^2 \right),
\end{equation}
and the two shot noise contributions exactly cancel in the sum. In the more general case, the power spectrum of stochastic field $\epsilon_h$ can have some nontrivial scale dependence, 
i.e.\ $\left< \epsilon_h | \epsilon_h \right>' = P_\epsilon(\bk)$.
In that case, the above discussed cancellation is no longer exact and we have
\begin{equation}
 \left< \delta_h | \delta_h * v^2_\parallel \right>'
-
\left< \delta_h * v_\parallel
 | \delta_h * v_\parallel \right>'
 \ni 
 \int_{\bp} \Big(
 P_\epsilon(\bk)- P_\epsilon(\bk - \bp)
 \Big) P_{vv}(\bp)
\approx {\rm ``const_2"}
+ \cdots,
\end{equation}
where $P_{vv}(\bp) = 
\left< v_\parallel|v_\parallel \right>'$
is the halo velocity spectrum.
It is also instructive to investigate a polynomial scale dependence of the stochastic power spectrum, $P_\epsilon(\bk)= a_0 + a_2 k^2 + a_4 k^4 + \ldots$.
In that case it follows that the noise contribution takes the simple form 
\begin{equation}
 \int_{\bp} \Big(
 P_\epsilon(\bk)- P_\epsilon(\bk - \bp)
 \Big) P_{vv}(\bp)
\approx {\rm ``const^{(0)}_{2,0}}"
+ {\rm ``const^{(0)}_{2,2}"} k^2 
+ {\rm ``const^{(2)}_{2,2}"} \mu^2 k^2 + \cdots,
\end{equation}
from which it follows that only the isotropic part obtains a shot noise like contribution.  Collecting all the contributions we get
\begin{equation}
\label{eq:Xi_2_eft}
\tilde{\Xi}^{(2)}_{\rm 1-loop}(k)
= 2 \big[
{\rm \eqref{eq:P02_spt}}
-  {\rm \eqref{eq:P11_spt}}
\big]
- 2\left(c^{(0)}_2+c^{(2)}_2 \mu^2 \right)
\frac{1}{k^2_\star} P_{\rm lin}(k)
+ ``{\rm const_2}"+ \ldots.
\end{equation}

\textbf{Contributions to $\Xi^{(3)}$} arise from two correlators, $P_{03}$ and $P_{12}$, both of which contribute at one-loop
\begin{align}
\label{eq:P_03_spt}
\left< \delta_h | (1+\delta_h) * v^3_\parallel \right>' &\approx 3 P_{01} \sigma^2 \\
&= - 3 i \frac{\mu}{k}c_1 P_{\rm lin} \sigma_{\rm lin}^2, \tag{in EPT}
\end{align}
\begin{align}
\label{eq:P_12_spt}
\left<(1+\delta_h) * v_\parallel | (1+\delta_h) * v^2_\parallel \right>'  &\approx  \left< v_\parallel | v^2_\parallel \right>' 
+ \left< v_\parallel | 
\delta_h v^2_\parallel \right>'
+ \left< \delta_h v_\parallel | v^2_\parallel \right>'  \\
&= - \frac{i}{k} \bigg( 
2 \mu \int_{\bp} ~
\frac{p_\parallel (\bk - \bp)_\parallel}{p^2 (\bk - \bp)^2}
G_2 (\bp, \bk-\bp) 
P_{\rm lin}(p) P_{\rm lin}(|\bk - \bp|) \notag\\
&~~~+4 \mu P_{\rm lin}(k) \int_{\bp} ~
\frac{p_\parallel (\bk - \bp)_\parallel}{p^2 (\bk - \bp)^2}
G_2 (\bp, -\bk) 
P_{\rm lin}(p) \notag\\
&~~~+2 k c_1 \int_{\bp} ~
\frac{p^2_\parallel (\bk - \bp)_\parallel}{p^4 (\bk - \bp)^2}
P_{\rm lin}(p) P_{\rm lin}(\bk - \bp) \notag\\
&~~~ - c_1\mu P_{\rm lin} \sigma_{\rm lin}^2 \bigg), \tag{in EPT}
\end{align}
The combination of $P_{03}$ and the last two terms of $P_{12}$ ensure IR cancelation in the soft limit. 
Collecting all the one-loop contributions
we get
\begin{equation}
\label{eq:Xi_3_eft}
\tilde{\Xi}^{(3)}_{\rm 1-loop}(k)
= 2 \big[
{\rm \eqref{eq:P_03_spt}}
-  3{\rm \eqref{eq:P_12_spt}}
\big]
+ 6 i \left(c^{(0)}_3 + c^{(2)}_3 \mu^2 \right) 
\frac{\mu}{k}
\frac{1}{k_\star^2}
P_{\rm lin} + \ldots.
\end{equation}

\textbf{Contributions to $\Xi^{(4)}$} can be approximated by a contribution giving zero lag (which we can consider as a non-perturbative contribution) multiplied by the lower order moments.  Heuristically we can write:
\begin{align}
\label{eq:P_04_spt}
\left< \delta_h | (1+\delta_h) * v^4_\parallel \right>' &\approx 3 P_{02} \sigma^2
\approx 3 P_{00} \sigma^4 \\
& = 3 c_1^2 P_{\rm lin} \sigma_{\rm lin}^4, \tag{in EPT}\\
\left<(1+\delta_h) * v_\parallel | (1+\delta_h) * v^3_\parallel \right>' &\approx 3 P_{11} \sigma^2 \label{eq:P_13_spt}\\
&= 3 \frac{\mu^2}{k^2}P_{\rm lin} \sigma_{\rm lin}^2,
\tag{in EPT}\\
\left<(1+\delta_h) * v^2_\parallel | (1+\delta_h) * v^2_\parallel \right>'  &\approx  \left< v^2_\parallel | v^2_\parallel \right>' + P_{00} \sigma^4  
\label{eq:P_22_spt}\\
&\approx 
2\int_{\bp} ~
\bigg[
\frac{p_\parallel (\bk - \bp)_\parallel}{p^2 (\bk - \bp)^2}
\bigg]^2 
P_{\rm lin}(p) P_{\rm lin}(|\bk - \bp|)
\notag\\
&\hspace{0.5cm}+ c_1^2 P_{\rm lin} \sigma_{\rm lin}^4 \tag{in EPT}
\end{align}
The only proper one-loop contributions in the fourth moment come from $P_{13}\propto P_{\rm lin} \sigma_{\rm lin}^2$ and $P_{22} \sim \left< v^2_\parallel | v^2_\parallel \right>'$,
terms that also exhibit a degree of IR cancelation in the soft limit in their $\mu^2$ angular dependence.  
Similar cancelation also appears at the two-loop level for the $\mu^0$ angular dependence where all of the 
$c_1^2 P_{\rm lin} \sigma_{\rm lin}^4$ terms cancel in the IR limit. 

Similarly to the case of $\Xi^{(2)}$, we can show that the scale dependence of the stochasticity field generates a shot-noise contribution in the $\Xi^{(4)}$ term, even though to show these explicitly a two-loop calculation is formally required. 
However, treatment of the shot noise terms on equal perturbative footing as the deterministic fields might not generally be justified and thus even an indication of the presence of such stochastic terms could serve as a justification for adding a shot noise contribution. 
These stochastic terms would be suppressed by $(\mu k)^4$ factors in the total power spectrum. 

Collecting all the one-loop contributions we get
\begin{equation}
\label{eq:Xi_4_eft}
\tilde{\Xi}^{(4)}_{\rm 1-loop}(k)
= 2 \big[
{\rm \eqref{eq:P_04_spt}}
-  
4{\rm \eqref{eq:P_13_spt}}
+  
3{\rm \eqref{eq:P_22_spt}}
\big]
+ 24 c^{(2)}_4 \frac{\mu^2}{k^2} \frac{1}{k^2_\star} P_{\rm lin}(k)
+ ``{\rm const_4}" + \ldots.
\end{equation}

\subsection{Eulerian redshift-space power spectrum}
\label{app:rsd_spt}

Using the moment expansion of the redshift-space power spectrum given in Eq.~\eqref{eqn:moment_expansion}
we obtain the one-loop result
\begin{align}
\label{eq:Euler_RSD_PS}
   P^s_{\rm 1-loop}(\bk) &= \tilde{\Xi}^{(0)}(\bk) + i k \mu \tilde{\Xi}^{(1)}(\bk) - \frac{1}{2} k^2 \mu^2 \tilde{\Xi}^{(2)} (\bk)- \frac{i}{6} k^3 \mu^3 \tilde{\Xi}^{(3)}(\bk) + \frac{1}{24} k^4 \mu^4 \tilde{\Xi}^{(4)} (\bk) \\
&= \bigg[  {\rm \eqref{eq:Xi_0_spt}} + 2 i f k \mu \left[ {\rm \eqref{eq:dv_spt}} + {\rm \eqref{eq:ddv_spt}} \right] - f^2 k^2 \mu^2 \big[ {\rm \eqref{eq:P02_spt}} -  {\rm \eqref{eq:P11_spt}} \big] \notag\\
  &~~ - \frac{i}{3} f^3 k^3 \mu^3 \big[ {\rm \eqref{eq:P_03_spt}} -  3{\rm \eqref{eq:P_12_spt}} \big] + \frac{1}{12} f^4 k^4 \mu^4 \big[ {\rm \eqref{eq:P_04_spt}} - 4{\rm \eqref{eq:P_13_spt}} + 3{\rm \eqref{eq:P_22_spt}} \big] \bigg]_{\rm EPT} \notag\\
  &~~+  \Big( c_0 + f c_1 \mu^2 + f^2 c_2 \mu^4  + f^3 c_3 \mu^6 \Big) \frac{k^2}{k^2_*} P_{\rm lin}(k) 
   + \left( 1 + s_1 f^2 \mu^2 \frac{k^2}{k_\star^2} + s_2 f^4 \mu^4 \frac{k^4}{k_\star^4} \right) {\rm ``const"} .  \notag
   \label{eqn:full_ir}
\end{align}
The first two lines above refer to the EPT expressions of the given one-loop power spectra, while in the last line counter terms and stochastic contributions are listed.
The counter terms are redefined so that $c_0 = c^{(0)}_0$, $c_1=c^{(0)}_1  + f c^{(0)}_2 $, $c_2  = c^{(2)}_2 + f c^{(0)}_3 $ and $c_3 = c^{(2)}_3 + f c^{(2)}_4$. 
This ensures that all the UV sensitive $P_{13}$ terms are under control.
In the last line we have also defined the stochastic parameters  $ {\rm ``const"} = {\rm ``const_0"}$, 
$s_1 = -\frac{1}{2} k_\star^2 \frac{``{\rm const_2}"}{{\rm ``const_0"}} $ and $s_1 = \frac{1}{24} k_\star^{4} \frac{``{\rm const_4}"}{{\rm ``const_0"}} $.
In the result above we have neglected higher derivative contributions to the stochasticity. 
This result, up to the couple of different choices for the counter terms and stochastic contributions, agree with recent references \cite{DAmico19, Ivanov19}.

\subsection{IR resummation of Velocity Moments and RSD power spectrum}
\label{app:IR_resumm}

Eulerian perturbation theory expands density and velocity fields, and correlators thereof, in powers of long wavelength modes that are assumed to be small. However, this assumption does not hold for long displacement modes that can have order one contributions and thus should be resummed, i.e.\ treated non-perturbatively. Given that in the equal time correlators most of the effects of such long wavelength displacements cancel out standard Eulerian PT is still an operational framework. However, the presence of the BAO feature on fairly large scales makes it more prone to these displacements and thus it is of interest to handle these non-perturbative contributions.  The procedure for handling these long modes goes under the name of IR-resummation, and is most naturally done in Lagrangian perturbation theory \cite{ESW07,Mat08a,SenZal15,VlaWhiAvi15,Vlah16,Ding2017}.  However, in Eulerian perturbation theory results can also be resummed in order to obtain the equivalent behavior \cite{Baldauf2015, Blas2016, Peloso2017, Ivanov2018}.

In the Eulerian framework the most pragmatic rendering of these IR-resummation procedures relies on splitting the linear power spectrum into smooth and oscillatory parts, $P_{\rm lin}(k) = P^{\rm nw}_{\rm lin} + P^{\rm w}_{\rm lin}$.  The choice of splitting is in many ways arbitrary.  The displacement resummation is taken to act on $P^{\rm w}_{\rm lin}$ alone, and is usually applied to produce the real-space power spectrum.  However, the procedure can be generalised to any velocity moment power spectrum giving
\begin{align}
   \tilde{\Xi}^{(n),IR}_{\rm 1 - loop}(\bk) 
   &= \tilde{\Xi}^{(n), {\rm nw} }_{\rm lin}(\bk) + e^{-\frac{1}{2} \Sigma^2 k^2} \left( 1 + \tfrac{1}{2} \Sigma^2 k^2 \right) \tilde{\Xi}^{(n),{\rm w}}_{\rm lin}(\bk)  \\
   &\hspace{2.2cm} + \tilde{\Xi}^{(n)}_{loop}(\bk) \left[ P_{\rm lin}(\bk) \to P^{\rm nw}_{\rm lin}(\bk) + e^{-\frac{1}{2} \Sigma^2 k^2} P^{\rm w}_{\rm lin}(\bk)  \right] \notag\\
   &\approx \tilde{\Xi}^{(n), {\rm nw} }_{\rm 1 - loop}(\bk) + e^{-\frac{1}{2} \Sigma^2 k^2} \left( 1 +\tfrac{1}{2} \Sigma^2 k^2 \right) \tilde{\Xi}^{(n),{\rm w}}_{\rm lin}(\bk) 
                     +  e^{-\frac{1}{2} \Sigma^2 k^2} \left( \tilde{\Xi}^{(n)}_{loop}(\bk) - \tilde{\Xi}^{(n),{\rm nw}}_{loop}(\bk) \right) \notag
\end{align}
where the label ``$loop$" stands for the next-to-linear-order correction in PT while the label ``1 - loop" stands for the total one-loop result, i.e.\  a sum of linear and next-to-linear orders. In the above $\Sigma$ is the estimated dispersion of the long wavelength displacement contributions:
\begin{equation}
\label{eq:IR_dispersion}
    \Sigma^2 = \int_0^\Lambda \frac{dk}{3 \pi^2} \big[ 1 - j_0 \left( k r_{\rm bao} \right) + 2 j_2 \left( k r_{\rm bao} \right) \big] P_{\rm lin} (k), 
\end{equation}
and $\Lambda$ is the scale of the IR mode split. In practice, $\Lambda$ can be chosen to be arbitrarily large given that the integral is naturally saturated by the power law drop of $P_{\rm lin}$ at high $k$.

For redshift space power spectra, in addition to the long wavelength displacements one can also resum
long wavelength velocity modes. This introduces slight change to the expression above, making the total redshift space dispersion $\Sigma_{\rm s}$ dependent on the angle to the line of sight, i.e.
\begin{equation}
    \Sigma_s^2 (\mu) = \left[ 1 + f(f+2)\mu^2 \right] \Sigma^2,
\end{equation}
where the $\Sigma^2$ is given by Eq.~\eqref{eq:IR_dispersion}.  The power spectrum becomes
\begin{align}
  P^{s,IR}_{\rm 1 - loop}(\bk) 
   &= P^{s, {\rm nw} }_{\rm lin}(\bk) + e^{-\frac{1}{2} \Sigma_s^2(\mu) k^2} \left( 1 +\tfrac{1}{2} \Sigma_s^2(\mu) k^2 \right) P^{s, {\rm w} }_{\rm lin}(\bk)  \\
   &\hspace{2.cm} + P^{s}_{loop}(\bk) \left[ P_{\rm lin}(\bk) \to P^{\rm nw}_{\rm lin}(\bk) + e^{-\frac{1}{2} \Sigma_s^2(\mu) k^2} P^{\rm w}_{\rm lin}(\bk)  \right] \notag\\
   &\approx  P^{s, {\rm nw} }_{\rm 1 - loop}(\bk) + e^{-\frac{1}{2} \Sigma_s^2(\mu) k^2} \left( 1 +\tfrac{1}{2} \Sigma_s^2(\mu) k^2 \right) P^{s, {\rm w} }_{\rm lin}(\bk) 
                     +  e^{-\frac{1}{2} \Sigma_s^2(\mu) k^2} \left( P^s_{loop}(\bk) - P^{s,{\rm nw}}_{loop}(\bk) \right), \notag
\end{align}
where the wiggle and no-wiggle $P^{s}_{\rm 1 - loop}$ (and similarly the $P^{s}_{loop}$ by dropping the linear Kaiser part) are given by Equation \eqref{eq:Euler_RSD_PS}.

\section{Gaussian Streaming Model}
\label{app:gsm}

The Gaussian Streaming Model (GSM), like the ME and FSM described in the main body of the text, is yet another way to expand the exponential in Equation~\ref{eqn:rsd_power_spectrum}. However, it differs from the two aforementioned models in that it is a cumulant expansion in configuration space \cite{VlaCasWhi16,VlaWhi19}. Our goal in this section is to explain why this structure makes it particularly easy to handle the effects of bulk (IR) velocities within the GSM.

We begin by reviewing the derivation of the GSM as presented in ref.~\cite{VlaCasWhi16}. The exponential in Equation~\ref{eqn:rsd_power_spectrum} can be expanded using configuration space statistics as
\begin{equation}
    \avg{(1+\delta_g(\bx_1)) (1+\delta_g(\bx_2)) e^{i\bk\cdot \Delta\bu}} = \left(1 + \xi_g(r)\right) \exp\left\{ \sum_{n=0}^{\infty} \frac{i^n}{n!} k_{i_1} ... k_{i_n} C^{(n)}_{i_1 ... i_n}(\textbf{r}=\bx_1 - \bx_2) \right\} \nonumber
\end{equation}
where, for example, the first two configuration-space cumulants are given by
\begin{align}
    C^{(1)}_i(\textbf{r}) &= (1 + \xi_g)^{-1} \Xi^{(1)}_i(\textbf{r}) \nonumber \\
    C^{(2)}_{ij}(\textbf{r}) &= (1 + \xi_g)^{-1} \Xi^{(2)}_{ij}(\textbf{r}) - C^{(1)}_i C^{(1)}_j,
\end{align}
and can be straightforwardly interpreted as the mean and variance of the density-weighted pairwise velocity. Truncating the cumulant expansion at second order and Fourier transforming yields the intuitive form \cite{ReiWhi11}
\begin{equation}
    1 + \xi_s(\textbf{s}) = \int \frac{dy}{\sqrt{2\pi C^{(2)}_\parallel}}\ \left(1 + \xi_g\right) \exp\left\{ \frac{\left(s_\parallel - y - \mu C^{(1)}_\parallel\right)^2}{2 C^{(2)}_\parallel}  \right\}.
\end{equation}

As shown in Section~\ref{sec:vel_exp}, truncating the velocity expansion at second order yields rather imperfect fits to the redshift-space power spectrum, especially towards small scales and large $\mu$, and thus does not yield a good model for the power spectrum broadband.  However, the configuration space structure is particularly suited to the close-to-Gaussian statistics of the large-scale bulk motions critical to describing the BAO feature. This is because, roughly speaking, $C^{(n>2)}$ do not contribute to BAO damping, which can be attributed to
\begin{itemize}
    \item The correlation function is much smaller than unity, such that $1 + \xi_g \approx 1$.
    \item For the nonlinear damping of the BAO we only need to consider the Gaussian statistics of the linear $\Delta$ and $\Delta \bu$, such that higher moments factorize via Wick's contraction.
\end{itemize}
For example, for the fourth cumulant we can write
\begin{equation}
    C^{(4)} \sim \avg{\Delta \bu \Delta \bu \Delta \bu\Delta \bu  } - 6 C^{(2)} C^{(2)} \approx 6  \avg{\Delta \bu \Delta \bu} \avg{\Delta \bu \Delta \bu}  - 6 C^{(2)} C^{(2)} \sim 0.
\end{equation}
Thus the truncation at second order, while not a good approximation for the power spectrum broadband in general, well-describes physics around the BAO scale. Note that this is not the case for the FSM, because the Fourier-space cumulants do not factor multiplicatively via Wick contractions.

\section{Wedges vs.~Multipoles}
\label{app:wedges}

While perturbation theory models of the power spectrum expand in $k_\parallel = k \mu$ and thus naturally predict the values of power spectrum wedges, analyses of actual spectroscopic surveys naturally produce power spectrum multipoles $P_\ell(k)$ \cite{Yamamoto06,Scoccimarro15,Bianchi15,Hand17}. Since $P(k,\mu)$ is a relatively smooth function of $\mu$, dominated in amplitude by the monopole and quadrupole, choosing to analyze the first few multipoles of the power spectrum vs.~wedges should amount to little more than a change of basis. However, as we have seen in Figures~\ref{fig:data_expansion} and \ref{fig:data_plms}, this choice of basis can make a dramatic and somewhat counterintuitive difference in the apparent goodness of fit or range of model validity, which we comment on briefly in this appendix.

Perhaps the most surprising aspect of multipoles vs.~wedges is that the errors on $\ell > 0$ do not have to be lesser in magnitude than wedge errors. Perhaps more importantly, even the quadruople can diverge from perturbative predictions while all but the highest $\mu$ wedges are predicted at the sub-percent level. This was already seen in Figures~\ref{fig:data_expansion} and \ref{fig:data_plms}; however, since in that case much of the monopole power at higher $k$ derives from shot noise it is worth considering a simpler example which emphasizes the point.

\begin{figure}
    \centering
    \includegraphics[width=\textwidth]{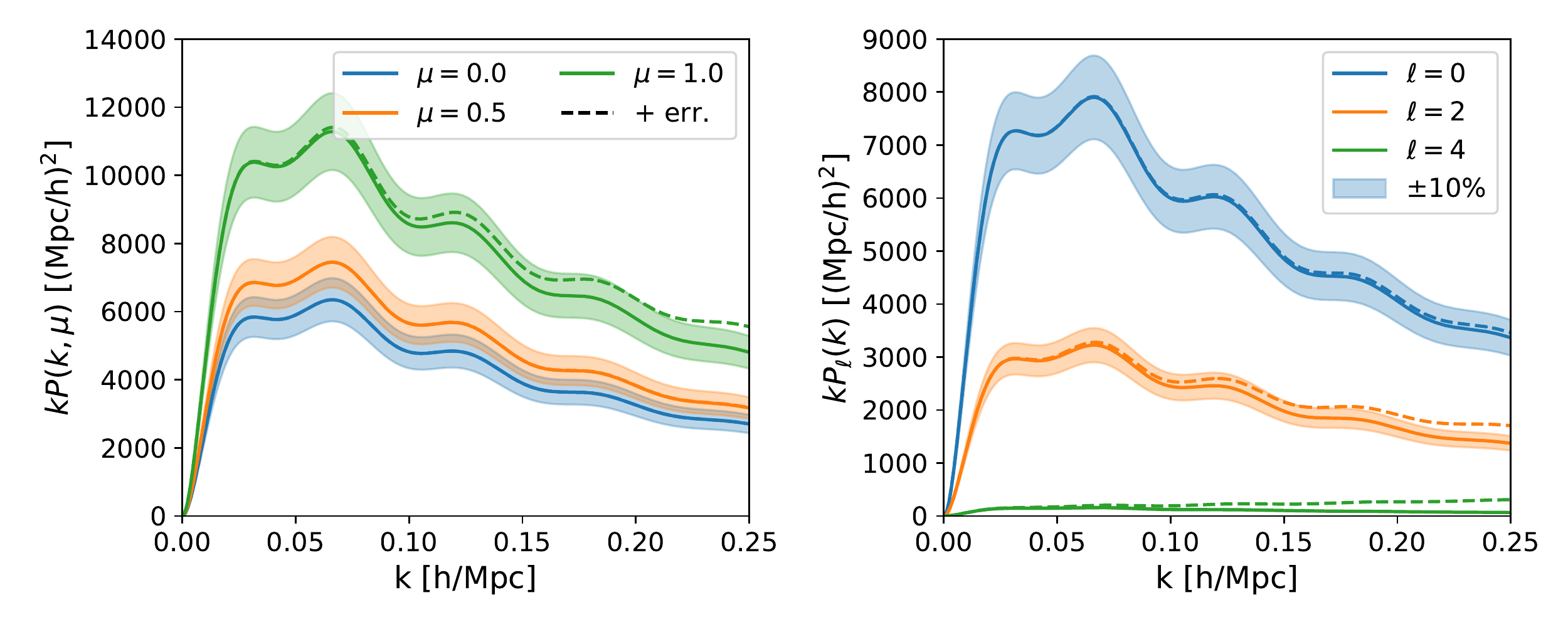}
    \caption{Toy model illustrating the different error properties of wedges (left) vs.~multipoles (right). In this example the ``theory'' is given by the Kaiser approximation with $b = 2$ and $f = 1$ while ``truth'' is given by Kaiser multiplied by $ 1 + k^2 \sigma^2 \mu^6$ normalized such that the power spectrum is $10\%$ away from theory at $k = 0.15 \kMpc$ and $\mu = 1$. While the $\mu = 0$ and $0.5$ wedges agree with theory at sub-percent level over the entire range shown, the quadrupole deviates from theory by more than $10\%$ already at $k = 0.2 \kMpc$, showing that fractional errors on the quadruople do not have to be less than or equal to those on the wedges.
    }
    \label{fig:plm_errors}
\end{figure}

Specifically, let us consider a shot-noise free example wherein our theory model is linear theory, $P(k,\mu) = (b + f\mu^2)^2 P_{\rm lin}(k)$, while ``truth'' is given by that model multiplied by $1+k^2\sigma^2\mu^6$, with $\sigma$ normalized to produce a $10\%$ error at $k = 0.2\kMpc$ and $\mu=1$. Such an error term is exactly what one might expect from the virial velocities in the fourth velocity moment which we do not model in this work. In this example, shown in Figure~\ref{fig:plm_errors}, we see that while the $\mu = 0$ and $0.5$ wedges are predicted by ``theory'' at sub-percent levels for all scales shown, the quadrupole already differs from theory by ten percent by $k = 0.15 \kMpc$. The mathematical reason for this is straightforward: unlike the monopole, the quadrupole is not a positive-definite average of power spectrum wedges. Indeed, the Legendre polynomial $\mathcal{L}_2(\mu) = \frac{1}{2}(3 \mu^2 - 1)$ will tend to pick up \textit{differences} in the error between $\mu = 0$, where errors are small and $\mathcal{L}_2$ is negative, and $\mu = 1$, where errors are maximal and $\mathcal{L}_2$ is positive. The situation is particularly acute for perturbative treatments of redshift space, which as we have shown expand order-by-order in $\mu$, making much better predictions perpendicular to the line-of-sight than parallel to it.

The error properties of multipoles vs.~wedges described above carry implications for data analysis. From an aesthetic standpoint, presenting data in terms of $P(k,\mu)$ has the slight advantage that fractional errors $\Delta P(k,\mu)/P$ roughly correspond to standard deviations in the Gaussian approximation while $\Delta P_\ell / P_\ell$ are hard to interpret as the errors for $P_{\ell > 0}$ are dominated by the monopole. Nonetheless, as the two statistics are connected by a basis transformation, the choice between them should in principle be irrelevant to data analysis as long as errors are properly taken into account, and theory errors\footnote{See ref.~\cite{Chudaykin19} for an example calculation of the theory error on multipoles. Their calculation uses a different ansatz for higher-order FoG effects than our $k^2 \mu^4 P(k)$, underscoring the difficulty of modeling FoG effects not incorporated into the base model.} for higher-order FoGs that scale strongly with $\mu$ will have the desired effect of down-weighting data from higher $\mu$. However, the magnitude and shape of the theory error can be hard to estimate for non-simulated samples, and a far more common choice in the literature is to adopt hard scale cuts $k_{\rm max}$ when fitting to theory (corresponding to infinite theoretical error beyond that scale). In this case, operating in wedges corresponds to defining an angular theshold $k_{\rm max}(\mu)$ where all but the highest $\mu$ wedge can be fit over most perturbative scales, while $\mu\approx 1$ has be cut off at much smaller $k$ due to virial motions, fingers-of-god and (for real surveys) redshift errors. On the other hand, operating within the multipole formalism means setting $k_{\rm max, \ell}$, which means much of the angular information carried by the lower-$\mu$ wedges will be lost by the scale-cut in the quadrupole due to contamination from the highest-$\mu$ bin (complementary discussions of the robustness of wedges can be found in refs.~\cite{Ross15,Ivanov19,Chudaykin19}). However, since redshift space power spectra are naturally measured as multipoles, an alternative approach beyond the strict wedges/multipoles dichotomy might be to weight multipoles in a scale-dependent fashion to minimize contamination by FoG effects at high $\mu$. Devising such an estimator is outside the scope of this work, though we note that related strategies have been suggested to deal with systematics in configuration space \cite{Rei14,Moh16} and plane-of-the-sky effects near $\mu = 0$ \cite{Hand17}.  The above suggests that rather than simply discarding one wedge or finding an orthogonal basis on the range $[-\mu_{\rm max},\mu_{\rm max}]$, the practical need to go through the multipole basis promotes an apodization or tapering of the wedges in $\mu$ to restrict the support in $\ell$.  Making predictions for apodized wedges presents no problems over the case of sharp-edged wedges.

\section{Fast Evaluation of LPT Kernels via FFTLog}
\label{app:fftpt}

One of the more time-consuming steps in computing LSS statistics beyond linear order in perturbation theory is the evaluation of one-loop integrals \cite{Mat08b,CLPT,VlaCasWhi16}. Recently, ref.~\cite{Schmittfull2016a} proposed a method to dramatically speed up these calculations by exploiting the underlying spherical symmetry of these integrals. To do so they note that these kernels can generally be decomposed into sums of integrals of the form
\begin{equation}
    \dtq q^{n_1} |\bk - \bq|^{n_2}\ P_{\ell}(\hq \cdot \widehat{k-q})\ \PL(q) \PL(|\bk-\bq|) = (-1)^\ell 4\pi \int dr\ r^2 j_0(kr) \xi^\ell_{n_1}(r) \xi^\ell_{n_2}(r)
    \label{eqn:fftpt_workhorse}
\end{equation}
where $P_\ell$ are Legendre polynomials and the generalized linear correlation functions are defined by
\begin{equation}
    \xi^\ell_n(r) = \int \frac{dk}{2\pi^2}\ k^{2+n} j_\ell(kr) \PL(k).
\end{equation}
The intuition behind Equation~\ref{eqn:fftpt_workhorse} is that, as the scalar-valued left hand side must be independent of the orientation of $\bk$, the angular integral in $\bq$ can be performed analytically, for example using the plane-wave expansion, to yield the spherical Bessel integrals in right-hand-side expression. Conveniently, these spherical Bessel integrals can be readily computed as Hankel transforms, which can in turn be efficiently computed using the FFTLog algorithm.

Ref.~\cite{Schmittfull2016a} applied the above-described method to LPT kernels relevant to the matter power spectrum, which can be written as
\begin{align}
    R_1(k) = k^2 \PL(k) \Bigg[ &\frac{8}{15} \int dr\ r\ j_0(kr) \xi^0_0 - \frac{16}{21}\int dr\ r\ j_2(kr) \xi^2_0 + \frac{8}{35}\int dr\ r\ j_4(kr) \xi^4_0 \Bigg] \nonumber \\
    R_2(k) = k^2 \PL(k) \Bigg[&-\frac{2}{15} \int dr\ r\ j_0(kr) \xi^0_0 - \frac{2}{21}\int dr\ r\ j_2(kr) \xi^2_0 + \frac{8}{35}\int dr\ r\ j_4(kr) \xi^4_0 \nonumber \\
    &+ \frac{2k}{5} \int dr\ r\ j_1(kr) \xi^1_{-1} - \frac{2k}{5} \int dr\ r\ j_3(kr) \xi^3_{-1} \Bigg]
\end{align}
and
\begin{align}
    Q_{1}(k)= 4\pi \int dr\ r^2 j_0(kr) \Bigg[ &\frac{8}{15}(\xi^0_0)^2 - \frac{16}{21}(\xi^2_0)^2 + \frac{8}{35}(\xi^4_0)^2 \Bigg] \nonumber \\
    Q_{2}(k)= 4\pi \int dr\ r^2 j_0(kr) \Bigg[ &\frac{4}{5}(\xi^0_0)^2 - \frac{4}{7}(\xi^2_0)^2 - \frac{8}{35}(\xi^4_0)^2 - \frac{4}{5} \xi^1_1 \xi^1_{-1} + \frac{4}{5}\xi^3_1 \xi^3_{-1} \Bigg] \nonumber \\
    Q_{3}(k)= 4\pi \int dr\ r^2 j_0(kr) \Bigg[ &\frac{38}{15}(\xi^0_0)^2 + \frac{68}{21}(\xi^2_0)^2 + \frac{8}{35}(\xi^4_0)^2 \nonumber \\
    &+ \frac{2}{3}\xi^0_2 \xi^0_{-2} - \frac{32}{5} \xi^1_1 \xi^1_{-1} + \frac{4}{3}\xi^2_2 \xi^2_{-2} -\frac{8}{5}\xi^3_1 \xi^3_{-1} \Bigg],
\end{align}
where the $r$ dependence of the generalized correlation functions is left implicit.
In this Appendix we complete this list by deriving Hankel-transform expressions for the remaining LPT kernels relevant to biased tracers up to one-loop order; these are:
\begin{align}
    Q_5(k) = 4\pi &\int dr\ r^2 j_0(kr) \Big[ \ \frac{2}{3}(\xi^0_0)^2 - \frac{2}{3}(\xi^2_0)^2 - \frac{2}{5}\xi^1_1 \xi^1_{-1} + \frac{2}{5}\xi^3_1 \xi^3_{-1} \Big] \nonumber \\
    Q_8(k) = 4\pi &\int dr\ r^2 j_0(kr) \Big[ \ \frac{2}{3}(\xi^0_0)^2 - \frac{2}{3}(\xi^2_0)^2 \Big] \nonumber \\
    Q_{s^2}(k)= 4\pi &\int dr\ r^2 j_0(kr) \Big[ -\frac{4}{15}(\xi^0_0)^2 + \frac{20}{21}(\xi^2_0)^2 - \frac{24}{35}(\xi^4_0)^2 \Big] \nonumber \\
    R_{b_3}(k) = \frac{8}{63} P(k)  &\int dr\  r\ \Big[   \frac{24 k^2}{5}  \xi^0_0 j_0(kr) - \frac{16k}{5} \xi^1_1 j_1(kr) -  \big( \frac{20k^2}{7} \xi^2_0 - 4 \xi^2_2 \big) j_2(kr) \nonumber \\
    &- \frac{24k}{5} \xi^3_1 j_3(kr) + \frac{72 k^2}{35} \xi^4_0 j_4(kr) \Big],
\end{align}
where the final kernel is defined such that
\begin{equation}
    \langle \delta_{\rm lin} | (st + \frac{16}{63} \sigma_\delta^2 \delta_{\rm lin}) \rangle = R_{b_3}(k).
\end{equation}
The relation of these kernels to physical quantities in LPT can be found in, for example, refs.~\cite{Mat08b,CLPT,VlaCasWhi16}.

\section{Hankel Transforms}
\label{app:Hankel}

In this section we give expressions for the Hankel transforms that give the $k$-space velocity moments in both LPT and EPT described in the main paper. We begin with LPT, from which we show that the expressions for EPT can be extracted as an especially simple limit. Similar approaches to evaluating the integrals in EPT are discussed in e.g.\ \cite{Schmittfull2016a,Schmittfull+:2016,Fang17,Simonovic18,Tomlinson20}. We differ from these mainly in that the FFTLog expressions are derived using the Lagrangian bias basis, which naturally organizes 1-loop contributions into combinations of linear generalized correlation functions $\xi_\ell^n$ that are automatically Galilean invariant. We set the linear growth rate $f = 1$ throughout this section, with the n$^{\rm th}$ velocity moment carrying an implicit factor $f^n$.

\subsection{LPT}
The integrals for velocity moments in LPT take the form
\begin{align}
    \dtq &e^{i\bk \cdot \bq - \frac{1}{2}k_i k_j A^{\rm lin}_{ij}}\ \mu^{m} f(q) \nonumber \\
    &= \sum_{n=0}^\infty 4\pi \int dq\ q^2\ e^{-\half k^2(X^{\rm lin} + Y^{\rm lin})} f^{m}_n(k^2 Y_{\rm lin}) \Big( \frac{kY_{\rm lin}}{q} \Big)^n f(q)\ j_n(kq),
    \label{eqn:lpt_hankel}
\end{align}
where $\mu = \hk \cdot \hq$ and we have used Equation~\ref{eqn:hankel_translate}; explicit expressions for $f^{m}_n$ are provided in Appendix~\ref{app:bessel}. The summands are Hankel transforms and can be efficiently evaluated using the FFTLog algorithm. In practice we find that this series converges quickly; for the matter contribution in the power spectrum (i.e. $f(q) = 1$) the series converges to sub-percent precision at $z = 0.8$ and $k = 0.25 \kMpc$ for typical cosmologies when $n_{\rm max} = 5$, with improving performance towards smaller wavenumbers and higher redshifts. Our expressions agree with those in ref.~\cite{VlaWhi19} up to shear and counterterms, as well as that for the power spectrum in ref.~\cite{VlaCasWhi16}, and we follow the conventions in refs.~\cite{CLPT,VlaCasWhi16} for the Lagrangian-space two-point functions ($U_i, A_{ij}, W_{ijk}$ etc.) though we correct for minor algebraic mistakes in a few cases. In this section only we will ignore the stochastic contributions and counterterms as they have trivial scale dependence.

\subsubsection{Real-Space Power Spectrum}

The real-space power spectrum expressed as an infinite sum of Hankel transforms was given in Appendix B of ref.~\cite{VlaCasWhi16}. As it is an important component of our model, we include it here for completeness:
\begin{align}
    P(k) &= \dtq  e^{i\bk \cdot \bq} e^{-\frac{1}{2}k_ik_j A^{\rm lin}_{ij}} \Big\{1 - \frac{1}{2} k^2 \Big(X^{\rm loop} + Y^{\rm loop} \mu^2\Big) - \frac{i}{6} k^3 \Big( \tilde{V} \mu + \tilde{T}\mu^3 \Big) \nonumber \\
    &+ b_1 \Big( 2 i k U \mu - k^2 ( X^{10} + Y^{10}\mu^2) \Big) + b_1^2 \Big( \xi_{\rm lin}^2 - k^2 U_{\rm lin}^2 \mu^2 + i k U^{11}\mu \Big) \nonumber \\
    &+ b_2 \Big(-k^2 U_{\rm lin}^2 \mu^2 + i k U^{20}\mu \Big) + 2 i b_1 b_2 k \xi_{\rm lin} U^{\rm lin} \mu + \frac{1}{2} b_2^2 \xi_{\rm lin}^2 \nonumber \\
    &+ b_s \Big( -k^2 (X_{s} + Y_{s}\mu^2) + 2 i k V^{10} \mu \Big)  + 2 i b_1 b_s k V^{12} \mu + b_2 b_s \chi + b_s^2 \zeta \nonumber \\
    &+ 2 i b_3 k U_{b_3} \mu + 2 b_1 b_3 \theta \Big\}
\end{align}
where for brevity we have defined we have defined $\tilde{V} = 3(2V_1 + V_3)$\footnote{Note that there is a typo in Equations B21, B22 of ref.~\cite{CLPT}, such that one should substitute $V_{1,3} \rightarrow V_{1,3} + S$, where $S(q)$ is defined in Equation B23, for the correct expressions.}, $\tilde{T} = 3 T$ and $\Upsilon_{ij} = X_s \delta_{ij} + Y_s \hq_i \hq_j$.

\subsubsection{Pairwise Velocity Spectrum}

This scalar decomposition of the pairwise velocity spectrum is given by $v_i(k) = i v(k) \hk_i$, with
\begin{align}
    v(k) = \dtq  &e^{i\bk \cdot \bq}\ e^{-\frac{1}{2}k_ik_j A^{\rm lin}_{ij}} \Big\{ k\Big(\dot{X} + \dot{Y}\mu^2\Big) + \frac{i k^2}{2}\Big(\dot{V} \mu + \dot{T} \mu^3\Big) \nonumber \\
    &+ 2 b_1\Big( i (k^2 U^{\rm lin}\dot{X}^{\rm lin} - \dot{U}) \mu + ik^2 U^{\rm lin} \dot{Y}^{\rm lin}\mu^3 + k(\dot{X}^{10} + \dot{Y}^{10} \mu^2) \Big) \nonumber \\
    & + b_1^2 \Big(k \xi_{\rm lin} \dot{X}^{\rm lin} +  k (\xi_{\rm lin} \dot{Y}^{\rm lin} +  2 U^{\rm lin} \dot{U}^{\rm lin}) \mu^2 - i \dot{U}^{11} \mu\Big) \nonumber \\
    &+ b_2 \Big(2 k U^{\rm lin} \dot{U}^{\rm lin} \mu^2 - i \dot{U}^{20}\mu\Big) - 2i b_1 b_2 \xi_{\rm lin} \dot{U}^{\rm lin} \mu  \nonumber \\
    &+ b_s \Big(-2i \dot{V}^{10} \mu + 2 k (\dot{X}_s + \dot{Y}_s \mu^2)\Big) - 2 i b_1 b_s \dot{V}^{12} \mu -2 i b_3 \dot{U}_{b_3} \mu \Big\} 
\end{align}
where we have followed the dot notation of Refs.~\cite{Wan14,VlaCasWhi16} such that each dotted quantity is proportional to $f$. We have used dots on the scalar components to denote the components of the vector quantities, e.g.\ $\dot{U}_i = \dot{U} \hq_i$. However, the three-indexed $\dot{W}_{ijk}$ has a somewhat more complicated structure than the one- or two-indexed quantities\footnote{Any one-indexed Lagrangian correlator must be proportional to $\hq_i$ and any two-indexed correlator must be a linear sum of $\delta_{ij}$ or $\hq_i \hq_j$ and therefore symmetric, whereas $W^{(112)}_{ijk}$ for example is not symmetric in all indices.} and we have chosen to summarize its contributions in terms of its contractions with $\hk$ alone, i.e. $\dot{V} = \frac{4f}{3}\tilde{V}$ and $\dot{V} = \frac{4f}{3}\tilde{T}$.

\subsubsection{Pairwise Velocity Dispersion Spectrum}

For the pairwise velocity dispersion we have chosen to compute the two contractions $\sigma^2_{\hk \hk}$ and $\sigma^2_{ii}$. These are related to the multipole moments via $\sigma^2_{ii} = 3 \sigma_0$ and $\sigma^2_{\hk\hk} = \sigma_0 + \sigma_2$.
\begin{align}
    \sigma^2_{\hk \hk} = \dtq  &e^{i\bk \cdot \bq}\ e^{-\frac{1}{2}k_ik_j A^{\rm lin}_{ij}}  \Big\{ \Big( (\ddot{X} - k^2 \dot{X}_{\rm lin}^2) + (\ddot{Y} - 2k^2 \dot{X}_{\rm lin} \dot{Y}_{\rm lin})\mu^2 - k^2 \dot{Y}_{\rm lin}^2 \mu^4 \Big) \nonumber \\
    &+ \frac{5f^2 ik}{3} \Big(\tilde{V} \mu + \tilde{T} \mu^3 \Big) \nonumber \\
    &+ 2b_1 \Big( ik( U^{\rm lin}\ddot{X}^{\rm lin} + 2 \dot{U}^{\rm lin}\dot{X}^{\rm lin}) \mu + ik(U^{\rm lin}\ddot{Y}^{\rm lin} + 2\dot{U}^{\rm lin}\dot{Y}^{\rm lin}) \mu^3 + \ddot{X}^{10} + \ddot{Y}^{10}\mu^2 \Big)  \nonumber \\
    &+ b_1^2 \Big( \xi_{\rm lin} \ddot{X}^{\rm lin} + (\xi_{\rm lin} \ddot{Y}^{\rm lin} + \dot{U}^{\rm lin}\dot{U}^{\rm lin}) \mu^2 \Big) + 2 b_2 \dot{U}^{\rm lin} \dot{U}^{\rm lin} \mu^2 + 2 b_s \Big( \ddot{X}_{s^2} + \ddot{Y}_{s^2}\mu^2 \Big) \Big\} \nonumber \\
    \newline \nonumber \\
    \sigma^2_{ii} = \dtq &e^{i\bk \cdot \bq}\ e^{-\frac{1}{2}k_ik_j A^{\rm lin}_{ij}}  \Big\{ \Big(3\ddot{X} + \ddot{Y} - k^2\dot{X}_{\rm lin}^2 - k^2(\dot{Y}_{\rm lin}^2+2\dot{X}_{\rm lin}\dot{Y}_{\rm lin})\mu^2 \Big) \nonumber \\
    &+ if^2 k (18 V_1 + 7 V_3 + 5 T)\mu \nonumber \\
    &+ 2 b_1 \Big( ik  (U^{\rm lin} (3\ddot{X}^{\rm lin} + \ddot{Y}^{\rm lin}) + (3\ddot{X}^{10}+\ddot{Y}^{10}) + 2 \dot{U}^{\rm lin}(\dot{X}^{\rm lin}+\dot{Y}^{\rm lin})) \mu \Big) \nonumber \\
    &+ b_1^2 \Big( \xi_{\rm lin} (3\ddot{X}^{\rm lin}+\ddot{Y}^{\rm lin}) + 2\dot{U}^{\rm lin}\dot{U}^{\rm lin} \Big) + 2 b_2 \dot{U}^{\rm lin}\dot{U}^{\rm lin} + 2 b_s (3\ddot{X}_{s^2} + \ddot{Y}_{s^2}) \Big\}.
\end{align}
Quantities with two dots are proportional to $f^2$. Once again, the time derivative $\ddot{W}_{ijk}$ is more complicated than the one- or two-indexed quantities, and in this case we have chosen to simply write them out as $f^2$ multiplied by the relevant un-dotted quantities\footnote{Note that Equation C.~11 in ref.~\cite{VlaCasWhi16} should instead be $\ddot{W}_{ijk} = f^2 (2 W^{(112)}_{ijk} + 2 W^{(121}_{ijk} + W^{211}_{ijk})$, i.e.\ the indices on the right-hand side should not be permuted with the order of solution.}.

\subsubsection{Higher Moments}
As usual we decompose $-i \gamma_{ijk} =  \third \gamma_1\ \hk_{\{i} \delta_{jk\}} + \gamma_3\ \hk_i \hk_j \hk_k$ where the scalar components can be derived from the contractions:
\begin{align}
    -i \hk_i \hk_j \hk_k \gamma_{ijk} = \gamma_1 + \gamma_3 = -i \dtq  &e^{i\bk \cdot \bq -\frac{1}{2}k_ik_j A^{\rm lin}_{ij}} \Big\{ \dddot{V} \mu + \dddot{T} \mu^3  \nonumber \\
    &+ 3ik \Big(\dot{X}^{\rm lin} \big(\ddot{X}^{\rm lin} + \ddot{Y}^{\rm lin}\mu^2 \big) + \dot{Y}^{\rm lin} \mu^2 \big(\ddot{X}^{\rm lin} + \ddot{Y}^{\rm lin}\mu^2 \big) \Big) \nonumber \\\
    &+  6 b_1 \dot{U}^{\rm lin} \mu \big(\ddot{X}^{\rm lin} + \ddot{Y}^{\rm lin}\mu^2\big) \Big\} \nonumber \\
    -i \hk_i \delta_{jk} \gamma_{ijk} = \frac{5}{3}\gamma_1 + \gamma_3 = -i \dtq  &e^{i\bk \cdot \bq -\frac{1}{2}k_ik_j A^{\rm lin}_{ij}} \Big\{ \Big(\frac{5}{3} \dddot{V} + \dddot{T} \Big)\mu \nonumber \\
    &+ ik \Big( \dot{X}^{\rm lin}\big(5\ddot{X}^{\rm lin} + (1+ 2\mu^2)\ddot{Y}^{\rm lin} \big)    + \dot{Y}^{\rm lin}\mu^2\big(5\ddot{X}^{\rm lin} + 3\ddot{Y}^{\rm lin} \big) \nonumber \\
    &+ 2 b_1 \dot{U}^{\rm lin} \big(5\ddot{X}^{\rm lin} + 3\ddot{Y}^{\rm lin}\big)\mu \Big) \Big\}. \nonumber
\end{align}
For the sake of brevity we have defined the triple-dotted quantities $\dddot{V} = 2 f^3 \tilde{V}$ and $\dddot{T} = 2 f^3 \tilde{T}$ such that $\hk_i \hk_j \hk_k \dddot{W}_{ijk} = \dddot{V}\mu + \dddot{T} \mu^3$ and $\hk_i \delta_{jk} \dddot{W}_{ijk} = \frac{5}{3} \dddot{V} + \dddot{T}.$

The fourth moment $\kappa_{ijkl} = \avg{(1+\delta_1)(1+\delta_2)\dDelta_i \dDelta_j \dDelta_k \dDelta_l}$ has only one contribution at one loop order; in Fourier space this is
\begin{equation}
     \kappa_{ijkl} = \dtq  e^{i\bk \cdot \bq -\frac{1}{2}k_ik_j A_{ij}} \avg{\dDelta_i \dDelta_j \dDelta_k \dDelta_l} = \dtq  e^{i\bk \cdot \bq -\frac{1}{2}k_ik_j A_{ij}} \ddot{A}_{\{ij} \ddot{A}_{kl\}}
\end{equation}
where the distinct unordered indices are now $\{ijkl\} = (ij)(kl) + (ik)(jl) + (il)(jk)$. This can be similarly decomposed as $\kappa_{ijkl} = \third \kappa_0 \delta_{\{ij}\delta_{kl\}} + \frac{1}{6} \kappa_2 \hk_{\{i} \hk_j \delta_{kl\}} + \kappa_4 \hk_i \hk_j \hk_k \hk_l$, for which the following linear equations hold
\begin{align}
    5 \kappa_0 + \frac{5}{3}\kappa_2 + \kappa_4 &= \dtq  e^{i\bk \cdot \bq -\frac{1}{2}k_ik_j A^{\rm lin}_{ij}} \Big\{15 \ddot{X}_{\rm lin}^2 + 10 \ddot{X}_{\rm lin}\ddot{Y}_{\rm lin} + 3 \ddot{Y}_{\rm lin}^2 \Big\} \nonumber \\
    \frac{5}{3}\kappa_0 + \frac{4}{3}\kappa_2 + \kappa_4 &= \dtq  e^{i\bk \cdot \bq -\frac{1}{2}k_ik_j A^{\rm lin}_{ij}} \Big\{5 \ddot{X}_{\rm lin}^2 + (1+7\mu^2) \ddot{X}_{\rm lin}\ddot{Y}_{\rm lin} + 3 \mu^2 \ddot{Y}_{\rm lin}^2 \Big\} \nonumber \\
    \kappa_0 + \kappa_2 + \kappa_4 &= \dtq  e^{i\bk \cdot \bq -\frac{1}{2}k_ik_j A^{\rm lin}_{ij}} \Big\{3 \ddot{X}_{\rm lin}^2 + 6\mu^2 \ddot{X}_{\rm lin}\ddot{Y}_{\rm lin} + 3 \mu^4 \ddot{Y}_{\rm lin}^2 \Big\}.
\end{align}

\subsection{EPT}

As described in Section~\ref{sec:comparison}, EPT is equivalent to LPT when the exponential of
\begin{equation}
    -\half k_i k_j A^{\rm lin}_{ij} = -\half k^2 (X^{\rm lin} + Y^{\rm lin}\mu^2) \nonumber
\end{equation} 
is expanded as its Taylor series. Doing so reduces Equation~\ref{eqn:lpt_hankel} to the simpler form
\begin{equation}
    \dtq\ e^{i\bk \cdot \bq}\ \mu^n \ f(q) = \sum_{\ell = 0}^n 4 \pi \int dq\ q^2\ f(q)\ \alpha_{\ell}^{(n)}\  j_\ell(kq),
\end{equation}
where Hankel transform no longer has $k$-dependence beyond the spherical bessel function $j_\ell(kq)$ and the coefficients $\alpha_\ell^{(n)}$ are defined such that $\mu^n = \sum_\ell \alpha_{\ell}^{(n)} \mathcal{L}_\ell(\mu)$. The fact that the $k$ dependence is isolated to the Bessel function in EPT means that bias contributions at each $k$ can be calculated all at once, instead of requiring one set of FFTLogs per $k$ point as in LPT. Transforming these expanded LPT integrands into the EPT bias basis using Equation~\ref{eqn:bias_map} yields Hankel transform expressions for all one-loop contributions to the EPT redshift-space power spectrum at one loop. An especially convenient feature of computing EPT integrals in the LPT basis is that the IR cancellations in the small $k$ limit are explicitly satisfied in each expression.

Since the expressions required in the calculation outlined above are essentially identical to those in the previous section for LPT, we have chosen not to explicitly enumerate them. However, let us briefly comment on two particular numerical choices that both simplify the calculation and improve stability. Firstly, a subset of the terms involved, due to ``connected'' correlators in Fourier space, can be Fourier-transformed explicitly as the kernels ($R_n$, $Q_n$) involved were themselves already computed using FFTLogs of products of generalized linear correlation functions (App.~\ref{app:fftpt}). For example the matter power spectrum contains both the connected terms
\begin{align*}
    P_{1}(k) \ni \dtq e^{i \bk \cdot \bq} \Big\{ 1 &- \half k_i k_j \Big(A^{\rm lin}_{ij} + A^{\rm loop}_{ij}\Big) - \frac{i}{6} k_i k_j k_k W_{ijk} \Big\} \\
    &= P_{\rm lin}(k) + \frac{9}{98} Q_1(k) + \frac{10}{21} R_1(k) + \frac{6}{7} R_2(k) + \frac{3}{7} Q_2(k) 
\end{align*}
and a disconnected contribution equal to the Fourier transform of $k_i k_j k_k k_l A^{\rm lin}_{ij} A^{\rm lin}_{kl}/8$ --- the former need not be Fourier transformed a second time. Note that since all displacement correlators appear as pairwise displacements $\Delta$, this split does not break Galilean invariance. The connected components for each of the velocity moments are given in Tables \ref{tab:pk_connected}, \ref{tab:vk_connected}, \ref{tab:sk_connected} and \ref{tab:gk_connected}.

\begin{table}
\begin{center}
\begin{tabular}{c | c | c}
$\tilde{\Xi}^{(0)}$ & $F(q)$  & $\tilde{F}(k)$ \\ \hline
$1$ & $- \half k_i k_j \Big(A^{\rm lin}_{ij} + A^{\rm loop}_{ij}\Big) - \frac{i}{6} k_i k_j k_k W_{ijk}$ & $\PL + \frac{9}{98} Q_1 + \frac{10}{21} R_1 + \frac{6}{7} R_2 + \frac{3}{7} Q_2$ \\
$b_1$ & $2 i k_i U_i - k_i k_j A^{10}_{ij}$ & $2\PL + \frac{10}{21}R_1 + \frac{1}{7}(6 R_1 + 12 R_2 + 6 Q_5)$ \\
$b_1^2$ & $\xi_{\rm lin} + ik_i U^{11}_i$ & $\PL + \frac{6}{7} (R_1 + R_2)$ \\
$b_2$ & $i k_i U^{20}_i$ & $\frac{3}{7} Q_8$ \\
$b_s$ & $2 i k_i \dot{V}^{10}_i$ & $\frac{2}{7} Q_{s^2}$ \\
$b_3$ & $2 i k_i U_{b_3}$ & $2 R_{b_3}$ \\
$b_1 b_3$ & $2 \theta$ & $2 R_{b_3}$
\end{tabular}
\caption{Contributions to the real-space power spectrum from ``connected'' cumulants in LPT.}
\label{tab:pk_connected}
\end{center}

\end{table}

\begin{table}[]
    \centering
    \begin{tabular}{c | c | c}
    $\tilde{\Xi}^{(1)}_i $ & $F(q)$  & $k^2\tilde{F}(k)$ \\ \hline
    $1$ & $i k_j \Big(\dot{A}^{\rm lin}_{ij} + \dot{A}^{\rm loop}_{ij}\Big) - \half k_j k_k \dot{W}_{ijk}$ & $ -i k_i ( 2 \PL + \frac{18}{49} Q_1 + \frac{40}{21} R_1 + \frac{12}{7} Q_2 + \frac{24}{7} R_2)$ \\
    $b_1$ & $2 \dot{U}_i + 2 i k_j \dot{A}^{10}_{ij}$ & $ -i k_i (2\PL + 4 R_1 + \frac{36}{7}R_2 + \frac{18}{7} Q_5)$ \\
    $b_1^2$ & $\dot{U}^{11}_i$ & $ -i k_i (12/7) (R_1 + R_2)$ \\
    $b_2$ & $\dot{U}^{20}_i$ & $ -i k_i (6/7) Q_8$ \\
    $b_s$ & $2 \dot{V}^{10}_i$ & $-i k_i (4/7) Q_{s^2}$ \\
    $b_3$ & $2 \dot{U}_{b_3,i}$ & $-i k_i 2 R_{b_3}$
\end{tabular}
    \caption{Contributions to the pairwise velocity spectrum from ``connected'' cumulants in LPT.}
    \label{tab:vk_connected}
\end{table}

\begin{table}[]
    \centering
    \begin{tabular}{c | c | c}
    $\tilde{\Xi}^{(2)}_{\hk \hk} $ & $F(q)$  & $k^2\tilde{F}(k)$ \\ \hline
    $1$ & $\ddot{A}_{\hk \hk} + i k_n \ddot{W}_{\hk \hk n}$ & $ -( 2 \PL + \frac{36}{49} Q_1 + \frac{20}{7} R_1 + \frac{30}{7} Q_2 + \frac{60}{7} R_2)$ \\
    $b_1$ & $2 \ddot{A}^{10}_{\hk \hk}$ & $-(24/7)(R_1 + 2 R_2 + Q_5)$ \\
    \hline
    $\tilde{\Xi}^{(2)}_{ii} $ & & \\
    \hline
    $1$ & $\ddot{A}_{ii} + i k_n \ddot{W}_{iin}$ & $-( 2 \PL - \frac{4}{7} R_1 - \frac{6}{49} Q_1 + \frac{60}{7} R_2 + \frac{30}{7} Q_2$    )\\
    $b_1$ & $2 \ddot{A}^{10}_{ii}$ & $ -(24/7)  (2R_2 + Q_5)$
\end{tabular}
    \caption{Contributions to the pairwise velocity dispersion from ``connected'' cumulants in LPT, decomposed into its trace $\sigma^2_{12,ii}$ and $\hk$ component $\sigma^2_{12,\hk \hk}$.}
    \label{tab:sk_connected}
\end{table}

\begin{table}[]
    \centering
    \begin{tabular}{c | c }
    $\tilde{\Xi}^{(3)}_{ijk} $ &  $k^3\tilde{F}(k)$ \\ \hline
    $-i \hk_i \hk_j \hk_k \gamma_{ijk}$  & $ (36/7) (2R_2 + Q_2 )$ \\
    $-i \hk_i \delta_{jk} \gamma_{ijk}$ & $-(12/7) (2R_1 -6R_2 + Q_1 -3Q_2)$
\end{tabular}
    \caption{Contributions to the third pairwise velocity moment from ``connected'' cumulants in LPT, decomposed into its contractions with the unit vector $\hk$ and $\delta_{ij}$. At one-loop order, all such contributions are due to matter velocities in the form of $\dddot{W}_{ijk}$ and therefore aren't multiplied by any bias paremeters.}
    \label{tab:gk_connected}
\end{table}

Secondly, the coefficients $\alpha^{(n)}_\ell$ are not unique and can be expressed in a number of ways by utilizing the recurrence relations of spherical Bessel functions. Perhaps the most obvious in the context of LPT corresponds taking the $B \rightarrow 0$ limit of Equation~\ref{eqn:IAB}, in which case for example
\begin{equation*}
    \dtq e^{i \bk \cdot \bq}\ \mu^2 f(q) = 4 \pi \int dq \ q^2 \ f(q)\ \Big( j_0(kq) - \frac{2 j_1(kq) }{kq} \Big).
\end{equation*}
This choice, however, leads to extra factors of $kq$ multiplying $j_\ell$ that make the separation of $k$-dependences messier. Thus, we have chosen in our calculations to use, e.g.
\begin{equation}
    \dtq e^{i \bk \cdot \bq}\ \mu^2 f(q) = 4 \pi \int dq \ q^2 \ f(q)\ \left( \third j_0(kq) - \frac{2}{3} j_2(kq) \right).
\end{equation}

\section{Useful Mathematical Identies}
\label{app:bessel}

To evaluate the power spectrum we make use of the angular integrals of the form
\begin{equation}
    I_{2m(+1)}(A,B) = \half \int d\mu \: \mu^{2m(+1)} e^{iA\mu - \frac{B\mu^2}{2}} = i^{0(+1)}e^{-B/2} \sum_{n=0}^\infty f^{2m}_n(B) \Big(\frac{B}{A}\Big)^n j_{n(+1)}(A)
    \label{eqn:IAB}
\end{equation}
where the series coefficients $f^{2m}_n(B)$ can be explicitly written using confluent hypergeometric functions of the second kind
\begin{equation}
    f^{2m}_n(B) = \Big( \frac{2}{B} \Big)^m\ U(-m,n-m+1,\frac{B}{2})
\end{equation}.
For convenience we list the first few such integrals (see also Refs.~\cite{VlaCasWhi16,VlaWhi19}):
\begin{align}
&\frac{1}{2} \int d\mu \: \mu^0 e^{iA\mu - \frac{B\mu^2}{2}} =e^{-B/2} \sum_{n=0}^{\infty} \Big(\frac{B}{A}\Big)^n j_n(A) \\
&\frac{1}{2} \int d\mu \: \mu^1 e^{iA\mu - \frac{B\mu^2}{2}} = i e^{-B/2} \sum_{n=0}^{\infty} \Big(\frac{B}{A}\Big)^n j_{n+1}(A) \\
&\frac{1}{2} \int d\mu \: \mu^2 e^{iA\mu - \frac{B\mu^2}{2}} =e^{-B/2} \sum_{n=0}^{\infty} \Big( 1 - \frac{2n}{B} \Big) \Big(\frac{B}{A}\Big)^n j_n(A) \\
&\frac{1}{2} \int d\mu \: \mu^3 e^{iA\mu - \frac{B\mu^2}{2}} =i e^{-B/2} \sum_{n=0}^{\infty} \Big( 1 - \frac{2n}{B} \Big)\Big(\frac{B}{A}\Big)^n j_{n+1}(A)
 \\
&\frac{1}{2} \int d\mu \: \mu^4 e^{iA\mu - \frac{B\mu^2}{2}} =e^{-B/2} \sum_{n=0}^{\infty} \Big( 1 - \frac{4n}{B} + \frac{4n(n-1)}{B^2}\Big) \Big(\frac{B}{A}\Big)^n j_n(A)
\end{align}
The integrals that begin with $j_{n+1}$ can be merged with those that do not by shifting indices, e.g.
\begin{align}
    \frac{1}{2} \int d\mu \: \mu^3 e^{iA\mu - \frac{B\mu^2}{2}} =i e^{-B/2} \sum_{n=0}^{\infty} \Big(\frac{A \Theta_{n}}{B} \Big) \Big( 1 - \frac{2(n-1)}{B} \Big)\Big(\frac{B}{A}\Big)^n j_{n}(A),
\end{align}
where $\Theta_{n} = 1$ for integers $n$ greater than zero and is zero for $n = 0$, such that we can write
\begin{equation}
    I_m(A,B) = \half \int d\mu \: \mu^{m} e^{iA\mu - \frac{B\mu^2}{2}} = e^{-B/2} \sum_{n=0}^\infty c^m_n(B) \Big(\frac{B}{A}\Big)^n j_{n}(A).
    \label{eqn:hankel_translate}
\end{equation}

\section{Implementation in Python}
\label{app:python}

Our {\sc python} code to calculate the velocity components and combine them into redshift-space power spectra, {\tt velocileptors}, is publicly available\footnote{https://github.com/sfschen/velocileptors} and includes example Jupyter notebooks and scripts introducing the main modules.

The library is split into two main subdirectories, {\tt LPT} and {\tt EPT} which house the calculations performed in LPT and EPT, respectively. The main workhorse module in each is called \textbf{moment\_expansion\_fftw.py}, which produces the IR-resummed velocity moments and contains functions to combine them into redshift-space power spectra. This is supplemented by \textbf{fourier\_streaming\_model\_fftw.py} and \textbf{gaussian\_streaming\_model\_fftw.py} in {\tt LPT} and \newline \textbf{ept\_fullresum\_fftw.py} in {\tt EPT}, the latter of which calculates the one-loop EPT redshift-space power spectrum directly.

In addtion to these the folder {\tt Utils} contains various useful functions necessary for the above calculations, the most important of which is \textbf{qfuncfft.py}, which comptues various one-loop PT kernels and correlators using the FFTLog formalism described in ref.~\cite{Schmittfull2016a} and expressions derived in Appendix~\ref{app:fftpt}.

The structure of the basic LPT, and by extension EPT, class \textbf{cleft\_fftw.py} is based on earlier code\footnote{C.~Modi. https://github.com/modichirag/CLEFT}, with a few modifications. Most importantly, the FFTLogs are evaluated using \textbf{spherical\_bessel\_transform\_fftw.py}, a custom FFTLog module based on {\tt mcfit}\footnote{Y.~Li. https://github.com/eelregit/mcfit} that saves time on the Hankel transforms used to compute the various LPT spectra by storing the FFTLog kernels\footnote{https://jila.colorado.edu/~ajsh/FFTLog/index.html}, whose evaluations were the slowest steps of previous LPT codes, rather than computing them on the fly. To further speed up these Hankel transforms we use a multi-threadable python wrapper for FFTW\footnote{http://www.fftw.org/}, pyFFTW\footnote{https://hgomersall.github.io/pyFFTW}, which can be installed via \textbf{pip}. Our LPT code takes less than one and a half seconds to generate power spectra at 50 wavenumbers running on one thread on a Macbook Pro purchased in 2013 and summing over spherical Bessel functions up to $\ell = 5$, generating all the bias contributions independently such that power spectra within the same cosmology (but potentially different $f$) can be re-computed essentially instantly, as can power spectra at different LOS angles $\mu$. We found that this setting was sufficient to produce $<0.5 \%$ errors out to $k = 0.25\kMpc$ on all relevant spectra.  Results at an arbitrary number of $k$s can then be provided via cubic spline interpolation with no loss of accuracy. The EPT code is slightly faster still and takes less than a second to run (independent of the number of $k$ points). For completeness, we include the capability to set 1-loop terms to zero (for Zeldovich calculations) as well as a module to compute correlation functions in redshift-space via the Gaussian streaming model.

\bibliographystyle{JHEP}
\bibliography{main}

\providecommand{\href}[2]{#2}\begingroup\raggedright\begin{thebibliography}{100}

\bibitem{Wei13}
D.~H. {Weinberg}, M.~J. {Mortonson}, D.~J. {Eisenstein}, C.~{Hirata}, A.~G.
  {Riess}, and E.~{Rozo}, {\it {Observational probes of cosmic acceleration}},
  {\em \physrep} {\bf 530} (Sept., 2013) 87--255,
  [\href{http://arxiv.org/abs/1201.2434}{{\tt arXiv:1201.2434}}].

\bibitem{PDG18}
M.~{Tanabashi}, K.~{Hagiwara}, K.~{Hikasa}, K.~{Nakamura}, Y.~{Sumino},
  F.~{Takahashi}, J.~{Tanaka}, K.~{Agashe}, G.~{Aielli}, C.~{Amsler},
  M.~{Antonelli}, D.~M. {Asner}, H.~{Baer}, S.~{Banerjee}, R.~M. {Barnett},
  T.~{Basaglia}, C.~W. {Bauer}, J.~J. {Beatty}, V.~I. {Belousov},
  J.~{Beringer}, S.~{Bethke}, A.~{Bettini}, H.~{Bichsel}, O.~{Biebel}, K.~M.
  {Black}, E.~{Blucher}, O.~{Buchmuller}, V.~{Burkert}, M.~A. {Bychkov}, R.~N.
  {Cahn}, M.~{Carena}, A.~{Ceccucci}, A.~{Cerri}, D.~{Chakraborty}, M.~C.
  {Chen}, R.~S. {Chivukula}, G.~{Cowan}, O.~{Dahl}, G.~{D'Ambrosio},
  T.~{Damour}, D.~{de Florian}, A.~{de Gouv{\^e}a}, T.~{DeGrand}, P.~{de Jong},
  G.~{Dissertori}, B.~A. {Dobrescu}, M.~{D'Onofrio}, M.~{Doser}, M.~{Drees},
  H.~K. {Dreiner}, D.~A. {Dwyer}, P.~{Eerola}, S.~{Eidelman}, J.~{Ellis},
  J.~{Erler}, V.~V. {Ezhela}, W.~{Fetscher}, B.~D. {Fields}, R.~{Firestone},
  B.~{Foster}, A.~{Freitas}, H.~{Gallagher}, L.~{Garren}, H.~J. {Gerber},
  G.~{Gerbier}, T.~{Gershon}, Y.~{Gershtein}, T.~{Gherghetta}, A.~A. {Godizov},
  M.~{Goodman}, C.~{Grab}, A.~V. {Gritsan}, C.~{Grojean}, D.~E. {Groom},
  M.~{Gr{\"u}newald}, A.~{Gurtu}, T.~{Gutsche}, H.~E. {Haber}, C.~{Hanhart},
  S.~{Hashimoto}, Y.~{Hayato}, K.~G. {Hayes}, A.~{Hebecker}, S.~{Heinemeyer},
  B.~{Heltsley}, J.~J. {Hern{\'a}ndez-Rey}, J.~{Hisano}, A.~{H{\"o}cker},
  J.~{Holder}, A.~{Holtkamp}, T.~{Hyodo}, K.~D. {Irwin}, K.~F. {Johnson},
  M.~{Kado}, M.~{Karliner}, U.~F. {Katz}, S.~R. {Klein}, E.~{Klempt}, R.~V.
  {Kowalewski}, F.~{Krauss}, M.~{Kreps}, B.~{Krusche}, Y.~V. {Kuyanov},
  Y.~{Kwon}, O.~{Lahav}, J.~{Laiho}, J.~{Lesgourgues}, A.~{Liddle},
  Z.~{Ligeti}, C.~J. {Lin}, C.~{Lippmann}, T.~M. {Liss}, L.~{Littenberg}, K.~S.
  {Lugovsky}, S.~B. {Lugovsky}, A.~{Lusiani}, Y.~{Makida}, F.~{Maltoni},
  T.~{Mannel}, A.~V. {Manohar}, W.~J. {Marciano}, A.~D. {Martin}, A.~{Masoni},
  J.~{Matthews}, U.~G. {Mei{\ss}ner}, D.~{Milstead}, R.~E. {Mitchell},
  K.~{M{\"o}nig}, P.~{Molaro}, F.~{Moortgat}, M.~{Moskovic}, H.~{Murayama},
  M.~{Narain}, P.~{Nason}, S.~{Navas}, M.~{Neubert}, P.~{Nevski}, Y.~{Nir},
  K.~A. {Olive}, S.~{Pagan Griso}, J.~{Parsons}, C.~{Patrignani}, J.~A.
  {Peacock}, M.~{Pennington}, S.~T. {Petcov}, V.~A. {Petrov}, E.~{Pianori},
  A.~{Piepke}, A.~{Pomarol}, A.~{Quadt}, J.~{Rademacker}, G.~{Raffelt}, B.~N.
  {Ratcliff}, P.~{Richardson}, A.~{Ringwald}, S.~{Roesler}, S.~{Rolli},
  A.~{Romaniouk}, L.~J. {Rosenberg}, J.~L. {Rosner}, G.~{Rybka}, R.~A.
  {Ryutin}, C.~T. {Sachrajda}, Y.~{Sakai}, G.~P. {Salam}, S.~{Sarkar},
  F.~{Sauli}, O.~{Schneider}, K.~{Scholberg}, A.~J. {Schwartz}, D.~{Scott},
  V.~{Sharma}, S.~R. {Sharpe}, T.~{Shutt}, M.~{Silari}, T.~{Sj{\"o}strand },
  P.~{Skands}, T.~{Skwarnicki}, J.~G. {Smith}, G.~F. {Smoot}, S.~{Spanier},
  H.~{Spieler}, C.~{Spiering}, A.~{Stahl}, S.~L. {Stone}, T.~{Sumiyoshi}, M.~J.
  {Syphers}, K.~{Terashi}, J.~{Terning}, U.~{Thoma}, R.~S. {Thorne},
  L.~{Tiator}, M.~{Titov}, N.~P. {Tkachenko}, N.~A. {T{\"o}rnqvist}, D.~R.
  {Tovey}, G.~{Valencia}, R.~{Van de Water}, N.~{Varelas}, G.~{Venanzoni},
  L.~{Verde}, M.~G. {Vincter}, P.~{Vogel}, A.~{Vogt}, S.~P. {Wakely},
  W.~{Walkowiak}, C.~W. {Walter}, D.~{Wands}, D.~R. {Ward}, M.~O. {Wascko},
  G.~{Weiglein}, D.~H. {Weinberg}, E.~J. {Weinberg}, M.~{White}, L.~R.
  {Wiencke}, S.~{Willocq}, C.~G. {Wohl}, J.~{Womersley}, C.~L. {Woody}, R.~L.
  {Workman}, W.~M. {Yao}, G.~P. {Zeller}, O.~V. {Zenin}, R.~Y. {Zhu}, S.~L.
  {Zhu}, F.~{Zimmermann}, P.~A. {Zyla}, J.~{Anderson}, L.~{Fuller}, V.~S.
  {Lugovsky}, P.~{Schaffner}, and {Particle Data Group}, {\it {Review of
  Particle Physics$^{*}$}},  {\em \prd} {\bf 98} (Aug, 2018) 030001.

\bibitem{Ame18}
L.~{Amendola}, S.~{Appleby}, A.~{Avgoustidis}, D.~{Bacon}, T.~{Baker},
  M.~{Baldi}, N.~{Bartolo}, A.~{Blanchard}, C.~{Bonvin}, S.~{Borgani},
  E.~{Branchini}, C.~{Burrage}, S.~{Camera}, C.~{Carbone}, L.~{Casarini},
  M.~{Cropper}, C.~{de Rham}, J.~P. {Dietrich}, C.~{Di Porto}, R.~{Durrer},
  A.~{Ealet}, P.~G. {Ferreira}, F.~{Finelli}, J.~{Garc{\'\i}a-Bellido},
  T.~{Giannantonio}, L.~{Guzzo}, A.~{Heavens}, L.~{Heisenberg}, C.~{Heymans},
  H.~{Hoekstra}, L.~{Hollenstein}, R.~{Holmes}, Z.~{Hwang}, K.~{Jahnke}, T.~D.
  {Kitching}, T.~{Koivisto}, M.~{Kunz}, G.~{La Vacca}, E.~{Linder}, M.~{March},
  V.~{Marra}, C.~{Martins}, E.~{Majerotto}, D.~{Markovic}, D.~{Marsh},
  F.~{Marulli}, R.~{Massey}, Y.~{Mellier}, F.~{Montanari}, D.~F. {Mota}, N.~J.
  {Nunes}, W.~{Percival}, V.~{Pettorino}, C.~{Porciani}, C.~{Quercellini},
  J.~{Read}, M.~{Rinaldi}, D.~{Sapone}, I.~{Sawicki}, R.~{Scaramella},
  C.~{Skordis}, F.~{Simpson}, A.~{Taylor}, S.~{Thomas}, R.~{Trotta},
  L.~{Verde}, F.~{Vernizzi}, A.~{Vollmer}, Y.~{Wang}, J.~{Weller}, and
  T.~{Zlosnik}, {\it {Cosmology and fundamental physics with the Euclid
  satellite}},  {\em Living Reviews in Relativity} {\bf 21} (Apr, 2018) 2,
  [\href{http://arxiv.org/abs/1606.00180}{{\tt arXiv:1606.00180}}].

\bibitem{Kai87}
N.~{Kaiser}, {\it {Clustering in real space and in redshift space}},  {\em
  \mnras} {\bf 227} (July, 1987) 1--21.

\bibitem{Ham92}
A.~J.~S. {Hamilton}, {\it {Measuring Omega and the real correlation function
  from the redshift correlation function}},  {\em \apjl} {\bf 385} (Jan., 1992)
  L5--L8.

\bibitem{DESI}
{DESI Collaboration}, A.~{Aghamousa}, J.~{Aguilar}, S.~{Ahlen}, S.~{Alam},
  L.~E. {Allen}, C.~{Allende Prieto}, J.~{Annis}, S.~{Bailey}, C.~{Balland},
  and et~al., {\it {The DESI Experiment Part I: Science,Targeting, and Survey
  Design}},  {\em ArXiv e-prints} (Oct., 2016)
  [\href{http://arxiv.org/abs/1611.00036}{{\tt arXiv:1611.00036}}].

\bibitem{EUCLID18}
L.~{Amendola}, S.~{Appleby}, A.~{Avgoustidis}, D.~{Bacon}, T.~{Baker},
  M.~{Baldi}, N.~{Bartolo}, A.~{Blanchard}, C.~{Bonvin}, S.~{Borgani},
  E.~{Branchini}, C.~{Burrage}, S.~{Camera}, C.~{Carbone}, L.~{Casarini},
  M.~{Cropper}, C.~{de Rham}, J.~P. {Dietrich}, C.~{Di Porto}, R.~{Durrer},
  A.~{Ealet}, P.~G. {Ferreira}, F.~{Finelli}, J.~{Garc{\'{\i}}a-Bellido},
  T.~{Giannantonio}, L.~{Guzzo}, A.~{Heavens}, L.~{Heisenberg}, C.~{Heymans},
  H.~{Hoekstra}, L.~{Hollenstein}, R.~{Holmes}, Z.~{Hwang}, K.~{Jahnke}, T.~D.
  {Kitching}, T.~{Koivisto}, M.~{Kunz}, G.~{La Vacca}, E.~{Linder}, M.~{March},
  V.~{Marra}, C.~{Martins}, E.~{Majerotto}, D.~{Markovic}, D.~{Marsh},
  F.~{Marulli}, R.~{Massey}, Y.~{Mellier}, F.~{Montanari}, D.~F. {Mota}, N.~J.
  {Nunes}, W.~{Percival}, V.~{Pettorino}, C.~{Porciani}, C.~{Quercellini},
  J.~{Read}, M.~{Rinaldi}, D.~{Sapone}, I.~{Sawicki}, R.~{Scaramella},
  C.~{Skordis}, F.~{Simpson}, A.~{Taylor}, S.~{Thomas}, R.~{Trotta},
  L.~{Verde}, F.~{Vernizzi}, A.~{Vollmer}, Y.~{Wang}, J.~{Weller}, and
  T.~{Zlosnik}, {\it {Cosmology and fundamental physics with the Euclid
  satellite}},  {\em Living Reviews in Relativity} {\bf 21} (Apr., 2018) 2,
  [\href{http://arxiv.org/abs/1606.00180}{{\tt arXiv:1606.00180}}].

\bibitem{SimonsObs}
N.~{Galitzki}, A.~{Ali}, K.~S. {Arnold}, P.~C. {Ashton}, J.~E. {Austermann},
  C.~{Baccigalupi}, T.~{Baildon}, D.~{Barron}, J.~A. {Beall}, S.~{Beckman},
  S.~M.~M. {Bruno}, S.~{Bryan}, P.~G. {Calisse}, G.~E. {Chesmore},
  Y.~{Chinone}, S.~K. {Choi}, G.~{Coppi}, K.~D. {Crowley}, K.~T. {Crowley},
  A.~{Cukierman}, M.~J. {Devlin}, S.~{Dicker}, B.~{Dober}, S.~M. {Duff},
  J.~{Dunkley}, G.~{Fabbian}, P.~A. {Gallardo}, M.~{Gerbino},
  N.~{Goeckner-Wald}, J.~E. {Golec}, J.~E. {Gudmundsson}, E.~E. {Healy},
  S.~{Henderson}, C.~A. {Hill}, G.~C. {Hilton}, S.-P.~P. {Ho}, L.~A. {Howe},
  J.~{Hubmayr}, O.~{Jeong}, B.~{Keating}, B.~J. {Koopman}, K.~{Kiuchi},
  A.~{Kusaka}, J.~{Lashner}, A.~T. {Lee}, Y.~{Li}, M.~{Limon}, M.~{Lungu},
  F.~{Matsuda}, P.~D. {Mauskopf}, A.~J. {May}, N.~{McCallum}, J.~{McMahon},
  F.~{Nati}, M.~D. {Niemack}, J.~L. {Orlowski-Scherer}, S.~C. {Parshley},
  L.~{Piccirillo}, M.~{Sathyanarayana Rao}, C.~{Raum}, M.~{Salatino}, J.~S.
  {Seibert}, C.~{Sierra}, M.~{Silva-Feaver}, S.~M. {Simon}, S.~T. {Staggs},
  J.~R. {Stevens}, A.~{Suzuki}, G.~{Teply}, R.~{Thornton}, C.~{Tsai}, J.~N.
  {Ullom}, E.~M. {Vavagiakis}, M.~R. {Vissers}, B.~{Westbrook}, E.~J.
  {Wollack}, Z.~{Xu}, and N.~{Zhu}, {\it {The Simons Observatory: instrument
  overview}},  in {\em Millimeter, Submillimeter, and Far-Infrared Detectors
  and Instrumentation for Astronomy IX}, vol.~10708 of {\em Society of
  Photo-Optical Instrumentation Engineers (SPIE) Conference Series},
  p.~1070804, July, 2018.
\newblock \href{http://arxiv.org/abs/1808.04493}{{\tt arXiv:1808.04493}}.

\bibitem{CMBS4}
K.~N. {Abazajian}, P.~{Adshead}, Z.~{Ahmed}, S.~W. {Allen}, D.~{Alonso}, K.~S.
  {Arnold}, C.~{Baccigalupi}, J.~G. {Bartlett}, N.~{Battaglia}, B.~A. {Benson},
  C.~A. {Bischoff}, J.~{Borrill}, V.~{Buza}, E.~{Calabrese}, R.~{Caldwell},
  J.~E. {Carlstrom}, C.~L. {Chang}, T.~M. {Crawford}, F.-Y. {Cyr-Racine},
  F.~{De Bernardis}, T.~{de Haan}, S.~{di Serego Alighieri}, J.~{Dunkley},
  C.~{Dvorkin}, J.~{Errard}, G.~{Fabbian}, S.~{Feeney}, S.~{Ferraro}, J.~P.
  {Filippini}, R.~{Flauger}, G.~M. {Fuller}, V.~{Gluscevic}, D.~{Green},
  D.~{Grin}, E.~{Grohs}, J.~W. {Henning}, J.~C. {Hill}, R.~{Hlozek},
  G.~{Holder}, W.~{Holzapfel}, W.~{Hu}, K.~M. {Huffenberger}, R.~{Keskitalo},
  L.~{Knox}, A.~{Kosowsky}, J.~{Kovac}, E.~D. {Kovetz}, C.-L. {Kuo},
  A.~{Kusaka}, M.~{Le Jeune}, A.~T. {Lee}, M.~{Lilley}, M.~{Loverde}, M.~S.
  {Madhavacheril}, A.~{Mantz}, D.~J.~E. {Marsh}, J.~{McMahon}, P.~D.
  {Meerburg}, J.~{Meyers}, A.~D. {Miller}, J.~B. {Munoz}, H.~N. {Nguyen}, M.~D.
  {Niemack}, M.~{Peloso}, J.~{Peloton}, L.~{Pogosian}, C.~{Pryke}, M.~{Raveri},
  C.~L. {Reichardt}, G.~{Rocha}, A.~{Rotti}, E.~{Schaan}, M.~M. {Schmittfull},
  D.~{Scott}, N.~{Sehgal}, S.~{Shandera}, B.~D. {Sherwin}, T.~L. {Smith},
  L.~{Sorbo}, G.~D. {Starkman}, K.~T. {Story}, A.~{van Engelen}, J.~D.
  {Vieira}, S.~{Watson}, N.~{Whitehorn}, and W.~L. {Kimmy Wu}, {\it {CMB-S4
  Science Book, First Edition}},  {\em ArXiv e-prints} (Oct., 2016)
  [\href{http://arxiv.org/abs/1610.02743}{{\tt arXiv:1610.02743}}].

\bibitem{Howlett17}
C.~{Howlett}, L.~{Staveley-Smith}, and C.~{Blake}, {\it {Cosmological forecasts
  for combined and next-generation peculiar velocity surveys}},  {\em \mnras}
  {\bf 464} (Jan, 2017) 2517--2544,
  [\href{http://arxiv.org/abs/1609.08247}{{\tt arXiv:1609.08247}}].

\bibitem{Kim19}
A.~{Kim}, G.~{Aldering}, P.~{Antilogus}, A.~{Bahmanyar}, S.~{BenZvi},
  H.~{Courtois}, T.~{Davis}, S.~{Ferraro}, S.~G.~A. {Gontcho}, O.~{Graur},
  R.~{Graziani}, J.~{Guy}, C.~{Harper}, R.~{Hlozek}, C.~{Howlett},
  D.~{Huterer}, C.~{Ju}, P.~F. {Leget}, E.~V. {Linder}, P.~{McDonald},
  J.~{Nordin}, P.~{Nugent}, S.~{Perlmutter}, N.~{Regnault}, M.~{Rigault},
  A.~{Shafieloo}, A.~{Slosar}, R.~B. {Tully}, M.~{White}, and M.~{Wood-Vasey},
  {\it {Testing Gravity Using Type Ia Supernovae Discovered by Next-Generation
  Wide-Field Imaging Surveys}},  {\em \baas} {\bf 51} (May, 2019) 140,
  [\href{http://arxiv.org/abs/1903.07652}{{\tt arXiv:1903.07652}}].

\bibitem{Graziani20}
R.~{Graziani}, M.~{Rigault}, N.~{Regnault}, P.~{Gris}, A.~{M{\"o}ller},
  P.~{Antilogus}, P.~{Astier}, M.~{Betoule}, S.~{Bongard}, M.~{Briday},
  J.~{Cohen-Tanugi}, Y.~{Copin}, H.~M. {Courtois}, D.~{Fouchez}, E.~{Gangler},
  D.~{Guinet}, A.~J. {Hawken}, Y.~L. {Kim}, P.~F. {L{\'e}get}, J.~{Neveu},
  P.~{Ntelis}, P.~{Rosnet}, and E.~{Nuss}, {\it {Peculiar velocity cosmology
  with type Ia supernovae}},  {\em arXiv e-prints} (Jan, 2020)
  arXiv:2001.09095, [\href{http://arxiv.org/abs/2001.09095}{{\tt
  arXiv:2001.09095}}].

\bibitem{Pee80}
P.~J.~E. {Peebles}, {\em {The large-scale structure of the universe}}.
\newblock 1980.

\bibitem{Pea99}
J.~A. {Peacock}, {\em {Cosmological Physics}}.
\newblock Jan., 1999.

\bibitem{Dod03}
S.~{Dodelson}, {\em {Modern cosmology}}.
\newblock 2003.

\bibitem{Ber02}
F.~{Bernardeau}, S.~{Colombi}, E.~{Gazta{\~n}aga}, and R.~{Scoccimarro}, {\it
  {Large-scale structure of the Universe and cosmological perturbation
  theory}},  {\em \physrep} {\bf 367} (Sept., 2002) 1--248,
  [\href{http://xxx.lanl.gov/abs/astro-ph/0112551}{{\tt astro-ph/0112551}}].

\bibitem{Juszkiewicz81}
R.~{Juszkiewicz}, {\it {On the evolution of cosmological adiabatic
  perturbations in the weakly non-linear regime}},  {\em \mnras} {\bf 197}
  (Dec., 1981) 931--940.

\bibitem{Vishniac83}
E.~T. {Vishniac}, {\it {Why weakly non-linear effects are small in a
  zero-pressure cosmology}},  {\em \mnras} {\bf 203} (Apr., 1983) 345--349.

\bibitem{Goroff86}
M.~H. {Goroff}, B.~{Grinstein}, S.~J. {Rey}, and M.~B. {Wise}, {\it {Coupling
  of modes of cosmological mass density fluctuations}},  {\em \apj} {\bf 311}
  (Dec., 1986) 6--14.

\bibitem{Makino92}
N.~{Makino}, M.~{Sasaki}, and Y.~{Suto}, {\it {Analytic approach to the
  perturbative expansion of nonlinear gravitational fluctuations in
  cosmological density and velocity fields}},  {\em \prd} {\bf 46} (July, 1992)
  585--602.

\bibitem{Jain94}
B.~{Jain} and E.~{Bertschinger}, {\it {Second-Order Power Spectrum and
  Nonlinear Evolution at High Redshift}},  {\em \apj} {\bf 431} (Aug., 1994)
  495, [\href{http://xxx.lanl.gov/abs/astro-ph/9311070}{{\tt
  astro-ph/9311070}}].

\bibitem{BNSZ12}
D.~{Baumann}, A.~{Nicolis}, L.~{Senatore}, and M.~{Zaldarriaga}, {\it
  {Cosmological non-linearities as an effective fluid}},  {\em \jcap} {\bf 7}
  (July, 2012) 51, [\href{http://arxiv.org/abs/1004.2488}{{\tt
  arXiv:1004.2488}}].

\bibitem{CHS12}
J.~J.~M. {Carrasco}, M.~P. {Hertzberg}, and L.~{Senatore}, {\it {The effective
  field theory of cosmological large scale structures}},  {\em Journal of High
  Energy Physics} {\bf 9} (Sept., 2012) 82,
  [\href{http://arxiv.org/abs/1206.2926}{{\tt arXiv:1206.2926}}].

\bibitem{McDRoy09}
P.~{McDonald} and A.~{Roy}, {\it {Clustering of dark matter tracers:
  generalizing bias for the coming era of precision LSS}},  {\em \jcap} {\bf 8}
  (Aug., 2009) 020, [\href{http://arxiv.org/abs/0902.0991}{{\tt
  arXiv:0902.0991}}].

\bibitem{Perko16}
A.~{Perko}, L.~{Senatore}, E.~{Jennings}, and R.~H. {Wechsler}, {\it {Biased
  Tracers in Redshift Space in the EFT of Large-Scale Structure}},  {\em arXiv
  e-prints} (Oct, 2016) arXiv:1610.09321,
  [\href{http://arxiv.org/abs/1610.09321}{{\tt arXiv:1610.09321}}].

\bibitem{Des16}
V.~{Desjacques}, D.~{Jeong}, and F.~{Schmidt}, {\it {Large-Scale Galaxy Bias}},
   {\em ArXiv e-prints} (Nov., 2016)
  [\href{http://arxiv.org/abs/1611.09787}{{\tt arXiv:1611.09787}}].

\bibitem{MerPaj14}
L.~{Mercolli} and E.~{Pajer}, {\it {On the velocity in the Effective Field
  Theory of Large Scale Structures}},  {\em \jcap} {\bf 3} (Mar., 2014) 6,
  [\href{http://arxiv.org/abs/1307.3220}{{\tt arXiv:1307.3220}}].

\bibitem{Buc89}
T.~{Buchert}, {\it {A class of solutions in Newtonian cosmology and the pancake
  theory}},  {\em \aap} {\bf 223} (Oct., 1989) 9--24.

\bibitem{Mou91}
F.~{Moutarde}, J.-M. {Alimi}, F.~R. {Bouchet}, R.~{Pellat}, and A.~{Ramani},
  {\it {Precollapse scale invariance in gravitational instability}},  {\em
  \apj} {\bf 382} (Dec., 1991) 377--381.

\bibitem{Hiv95}
E.~{Hivon}, F.~R. {Bouchet}, S.~{Colombi}, and R.~{Juszkiewicz}, {\it {Redshift
  distortions of clustering: a Lagrangian approach.}},  {\em \aap} {\bf 298}
  (June, 1995) 643, [\href{http://xxx.lanl.gov/abs/astro-ph/9407049}{{\tt
  astro-ph/9407049}}].

\bibitem{TayHam96}
A.~N. {Taylor} and A.~J.~S. {Hamilton}, {\it {Non-linear cosmological power
  spectra in real and redshift space}},  {\em \mnras} {\bf 282} (Oct., 1996)
  767--778, [\href{http://xxx.lanl.gov/abs/astro-ph/9604020}{{\tt
  astro-ph/9604020}}].

\bibitem{Mat08a}
T.~{Matsubara}, {\it {Resumming cosmological perturbations via the Lagrangian
  picture: One-loop results in real space and in redshift space}},  {\em \prd}
  {\bf 77} (Mar., 2008) 063530, [\href{http://arxiv.org/abs/0711.2521}{{\tt
  arXiv:0711.2521}}].

\bibitem{Mat08b}
T.~{Matsubara}, {\it {Nonlinear perturbation theory with halo bias and
  redshift-space distortions via the Lagrangian picture}},  {\em \prd} {\bf 78}
  (Oct., 2008) 083519, [\href{http://arxiv.org/abs/0807.1733}{{\tt
  arXiv:0807.1733}}].

\bibitem{CLPT}
J.~{Carlson}, B.~{Reid}, and M.~{White}, {\it {Convolution Lagrangian
  perturbation theory for biased tracers}},  {\em \mnras} {\bf 429} (Feb.,
  2013) 1674--1685, [\href{http://arxiv.org/abs/1209.0780}{{\tt
  arXiv:1209.0780}}].

\bibitem{Whi14}
M.~{White}, {\it {The Zel'dovich approximation}},  {\em \mnras} {\bf 439}
  (Apr., 2014) 3630--3640, [\href{http://arxiv.org/abs/1401.5466}{{\tt
  arXiv:1401.5466}}].

\bibitem{ZheFri14}
V.~{Zheligovsky} and U.~{Frisch}, {\it {Time-analyticity of Lagrangian particle
  trajectories in ideal fluid flow}},  {\em Journal of Fluid Mechanics} {\bf
  749} (June, 2014) 404--430, [\href{http://arxiv.org/abs/1312.6320}{{\tt
  arXiv:1312.6320}}].

\bibitem{Mat15}
T.~{Matsubara}, {\it {Recursive solutions of Lagrangian perturbation theory}},
  {\em \prd} {\bf 92} (July, 2015) 023534,
  [\href{http://arxiv.org/abs/1505.01481}{{\tt arXiv:1505.01481}}].

\bibitem{VlaSelBal15}
Z.~{Vlah}, U.~{Seljak}, and T.~{Baldauf}, {\it {Lagrangian perturbation theory
  at one loop order: Successes, failu res, and improvements}},  {\em \prd} {\bf
  91} (Jan., 2015) 023508, [\href{http://arxiv.org/abs/1410.1617}{{\tt
  arXiv:1410.1617}}].

\bibitem{VlaWhiAvi15}
Z.~{Vlah}, M.~{White}, and A.~{Aviles}, {\it {A Lagrangian effective field
  theory}},  {\em \jcap} {\bf 9} (Sept., 2015) 014,
  [\href{http://arxiv.org/abs/1506.05264}{{\tt arXiv:1506.05264}}].

\bibitem{PorSenZal14}
R.~A. {Porto}, L.~{Senatore}, and M.~{Zaldarriaga}, {\it {The Lagrangian-space
  Effective Field Theory of large scale structures}},  {\em \jcap} {\bf 5}
  (May, 2014) 022, [\href{http://arxiv.org/abs/1311.2168}{{\tt
  arXiv:1311.2168}}].

\bibitem{McQWhi16}
M.~{McQuinn} and M.~{White}, {\it {Cosmological perturbation theory in 1+1
  dimensions}},  {\em \jcap} {\bf 1} (Jan., 2016) 043,
  [\href{http://arxiv.org/abs/1502.07389}{{\tt arXiv:1502.07389}}].

\bibitem{DAmico19}
G.~{d'Amico}, J.~{Gleyzes}, N.~{Kokron}, K.~{Markovic}, L.~{Senatore},
  P.~{Zhang}, F.~{Beutler}, and H.~{Gil-Mar{\'\i}n}, {\it {The cosmological
  analysis of the SDSS/BOSS data from the Effective Field Theory of Large-Scale
  Structure}},  {\em \jcap} {\bf 2020} (May, 2020) 005,
  [\href{http://arxiv.org/abs/1909.05271}{{\tt arXiv:1909.05271}}].

\bibitem{Ivanov19}
M.~M. {Ivanov}, M.~{Simonovi{\'c}}, and M.~{Zaldarriaga}, {\it {Cosmological
  parameters from the BOSS galaxy power spectrum}},  {\em \jcap} {\bf 2020}
  (May, 2020) 042, [\href{http://arxiv.org/abs/1909.05277}{{\tt
  arXiv:1909.05277}}].

\bibitem{Colas19}
T.~{Colas}, G.~{D'Amico}, L.~{Senatore}, P.~{Zhang}, and F.~{Beutler}, {\it
  {Efficient Cosmological Analysis of the SDSS/BOSS data from the Effective
  Field Theory of Large-Scale Structure}},  {\em arXiv e-prints} (Sep, 2019)
  arXiv:1909.07951, [\href{http://arxiv.org/abs/1909.07951}{{\tt
  arXiv:1909.07951}}].

\bibitem{Jackson72}
J.~C. {Jackson}, {\it {A critique of Rees's theory of primordial gravitational
  radiation}},  {\em \mnras} {\bf 156} (Jan, 1972) 1P,
  [\href{http://arxiv.org/abs/0810.3908}{{\tt arXiv:0810.3908}}].

\bibitem{Vlah16}
Z.~{Vlah}, U.~{Seljak}, M.~{Yat Chu}, and Y.~{Feng}, {\it {Perturbation theory,
  effective field theory, and oscillations in the power spectrum}},  {\em
  \jcap} {\bf 2016} (Mar, 2016) 057,
  [\href{http://arxiv.org/abs/1509.02120}{{\tt arXiv:1509.02120}}].

\bibitem{Okumura12}
T.~{Okumura}, U.~{Seljak}, P.~{McDonald}, and V.~{Desjacques}, {\it
  {Distribution function approach to redshift space distortions. Part II:
  N-body simulations}},  {\em \jcap} {\bf 2012} (Feb, 2012) 010,
  [\href{http://arxiv.org/abs/1109.1609}{{\tt arXiv:1109.1609}}].

\bibitem{Okumura12II}
T.~{Okumura}, U.~{Seljak}, and V.~{Desjacques}, {\it {Distribution function
  approach to redshift space distortions. Part III: halos and galaxies}},  {\em
  \jcap} {\bf 2012} (Nov., 2012) 014,
  [\href{http://arxiv.org/abs/1206.4070}{{\tt arXiv:1206.4070}}].

\bibitem{ReiWhi11}
B.~A. {Reid} and M.~{White}, {\it {Towards an accurate model of the
  redshift-space clustering of haloes in the quasi-linear regime}},  {\em
  \mnras} {\bf 417} (Nov., 2011) 1913--1927,
  [\href{http://arxiv.org/abs/1105.4165}{{\tt arXiv:1105.4165}}].

\bibitem{Wan14}
L.~{Wang}, B.~{Reid}, and M.~{White}, {\it {An analytic model for
  redshift-space distortions}},  {\em \mnras} {\bf 437} (Jan., 2014) 588--599,
  [\href{http://arxiv.org/abs/1306.1804}{{\tt arXiv:1306.1804}}].

\bibitem{VlaCasWhi16}
Z.~{Vlah}, E.~{Castorina}, and M.~{White}, {\it {The Gaussian streaming model
  and convolution Lagrangian effective field theory}},  {\em \jcap} {\bf 12}
  (Dec., 2016) 007, [\href{http://arxiv.org/abs/1609.02908}{{\tt
  arXiv:1609.02908}}].

\bibitem{Uhlemann15}
C.~{Uhlemann}, M.~{Kopp}, and T.~{Haugg}, {\it {Edgeworth streaming model for
  redshift space distortions}},  {\em \prd} {\bf 92} (Sep, 2015) 063004,
  [\href{http://arxiv.org/abs/1503.08837}{{\tt arXiv:1503.08837}}].

\bibitem{Fis95}
K.~B. {Fisher}, {\it {On the Validity of the Streaming Model for the
  Redshift-Space Correlation Function in the Linear Regime}},  {\em \apj} {\bf
  448} (Aug., 1995) 494, [\href{http://xxx.lanl.gov/abs/astro-ph/9412081}{{\tt
  astro-ph/9412081}}].

\bibitem{Rei12}
B.~A. {Reid}, L.~{Samushia}, M.~{White}, W.~J. {Percival}, M.~{Manera},
  N.~{Padmanabhan}, A.~J. {Ross}, A.~G. {S{\'a}nchez}, S.~{Bailey},
  D.~{Bizyaev}, A.~S. {Bolton}, H.~{Brewington}, J.~{Brinkmann}, J.~R.
  {Brownstein}, A.~J. {Cuesta}, D.~J. {Eisenstein}, J.~E. {Gunn},
  K.~{Honscheid}, E.~{Malanushenko}, V.~{Malanushenko}, C.~{Maraston}, C.~K.
  {McBride}, D.~{Muna}, R.~C. {Nichol}, D.~{Oravetz}, K.~{Pan}, R.~{de Putter},
  N.~A. {Roe}, N.~P. {Ross}, D.~J. {Schlegel}, D.~P. {Schneider}, H.-J. {Seo},
  A.~{Shelden}, E.~S. {Sheldon}, A.~{Simmons}, R.~A. {Skibba}, S.~{Snedden},
  M.~E.~C. {Swanson}, D.~{Thomas}, J.~{Tinker}, R.~{Tojeiro}, L.~{Verde}, D.~A.
  {Wake}, B.~A. {Weaver}, D.~H. {Weinberg}, I.~{Zehavi}, and G.-B. {Zhao}, {\it
  {The clustering of galaxies in the SDSS-III Baryon Oscillation Spectroscopic
  Survey: measurements of the growth of structure and expansion rate at z =
  0.57 from anisotropic clustering}},  {\em \mnras} {\bf 426} (Nov., 2012)
  2719--2737, [\href{http://arxiv.org/abs/1203.6641}{{\tt arXiv:1203.6641}}].

\bibitem{Sunayama16}
T.~{Sunayama}, N.~{Padmanabhan}, K.~{Heitmann}, S.~{Habib}, and E.~{Rangel},
  {\it {Efficient construction of mock catalogs for baryon acoustic oscillation
  surveys}},  {\em \jcap} {\bf 5} (May, 2016) 051,
  [\href{http://arxiv.org/abs/1510.06665}{{\tt arXiv:1510.06665}}].

\bibitem{nbdkit}
N.~{Hand}, Y.~{Feng}, F.~{Beutler}, Y.~{Li}, C.~{Modi}, U.~{Seljak}, and
  Z.~{Slepian}, {\it {nbodykit: An Open-source, Massively Parallel Toolkit for
  Large-scale Structure}},  {\em \aj} {\bf 156} (Oct, 2018) 160,
  [\href{http://arxiv.org/abs/1712.05834}{{\tt arXiv:1712.05834}}].

\bibitem{NFW}
J.~F. {Navarro}, C.~S. {Frenk}, and S.~D.~M. {White}, {\it {A Universal Density
  Profile from Hierarchical Clustering}},  {\em \apj} {\bf 490} (Dec, 1997)
  493--508, [\href{http://xxx.lanl.gov/abs/astro-ph/9611107}{{\tt
  astro-ph/9611107}}].

\bibitem{VlaWhi19}
Z.~{Vlah} and M.~{White}, {\it {Exploring redshift-space distortions in
  large-scale structure}},  {\em \jcap} {\bf 2019} (Mar, 2019) 007,
  [\href{http://arxiv.org/abs/1812.02775}{{\tt arXiv:1812.02775}}].

\bibitem{White15}
M.~{White}, B.~{Reid}, C.-H. {Chuang}, J.~L. {Tinker}, C.~K. {McBride},
  F.~{Prada}, and L.~{Samushia}, {\it {Tests of redshift-space distortions
  models in configuration space for the analysis of the BOSS final data
  release}},  {\em \mnras} {\bf 447} (Feb, 2015) 234--245,
  [\href{http://arxiv.org/abs/1408.5435}{{\tt arXiv:1408.5435}}].

\bibitem{GF19}
J.~E. {Garc{\'\i}a-Farieta}, F.~{Marulli}, L.~{Moscardini}, A.~{Veropalumbo},
  and R.~A. {Casas-Mirand a}, {\it {Validating the methodology for constraining
  the linear growth rate from clustering anisotropies}},  {\em \mnras} {\bf
  494} (Mar., 2020) 1658--1674, [\href{http://arxiv.org/abs/1909.08016}{{\tt
  arXiv:1909.08016}}].

\bibitem{Cuesta20}
C.~{Cuesta-Lazaro}, B.~{Li}, A.~{Eggemeier}, P.~{Zarrouk}, C.~M. {Baugh},
  T.~{Nishimichi}, and M.~{Takada}, {\it {Towards a non-Gaussian model of
  redshift space distortions}},  {\em arXiv e-prints} (Feb, 2020)
  arXiv:2002.02683, [\href{http://arxiv.org/abs/2002.02683}{{\tt
  arXiv:2002.02683}}].

\bibitem{SelMcD11}
U.~{Seljak} and P.~{McDonald}, {\it {Distribution function approach to redshift
  space distortions}},  {\em \jcap} {\bf 11} (Nov., 2011) 039,
  [\href{http://arxiv.org/abs/1109.1888}{{\tt arXiv:1109.1888}}].

\bibitem{Ferraro19}
S.~{Ferraro} and M.~J. {Wilson}, {\it {Inflation and Dark Energy from
  spectroscopy at z \&gt; 2}},  {\em \baas} {\bf 51} (May, 2019) 72,
  [\href{http://arxiv.org/abs/1903.09208}{{\tt arXiv:1903.09208}}].

\bibitem{Schlegel19}
D.~{Schlegel}, J.~A. {Kollmeier}, and S.~{Ferraro}, {\it {The MegaMapper: a
  z\&gt;2 spectroscopic instrument for the study of Inflation and Dark
  Energy}},  in {\em \baas}, vol.~51, p.~229, Sep, 2019.
\newblock \href{http://arxiv.org/abs/1907.11171}{{\tt arXiv:1907.11171}}.

\bibitem{MSE19}
{The MSE Science Team}, C.~{Babusiaux}, M.~{Bergemann}, A.~{Burgasser},
  S.~{Ellison}, D.~{Haggard}, D.~{Huber}, M.~{Kaplinghat}, T.~{Li},
  J.~{Marshall}, S.~{Martell}, A.~{McConnachie}, W.~{Percival}, A.~{Robotham},
  Y.~{Shen}, S.~{Thirupathi}, K.-V. {Tran}, C.~{Yeche}, D.~{Yong},
  V.~{Adibekyan}, V.~{Silva Aguirre}, G.~{Angelou}, M.~{Asplund}, M.~{Balogh},
  P.~{Banerjee}, M.~{Bannister}, D.~{Barr{\'\i}a}, G.~{Battaglia}, A.~{Bayo},
  K.~{Bechtol}, P.~G. {Beck}, T.~C. {Beers}, E.~P. {Bellinger}, T.~{Berg},
  J.~M. {Bestenlehner}, M.~{Bilicki}, B.~{Bitsch}, J.~{Bland-Hawthorn}, A.~S.
  {Bolton}, A.~{Boselli}, J.~{Bovy}, A.~{Bragaglia}, D.~{Buzasi}, E.~{Caffau},
  J.~{Cami}, T.~{Carleton}, L.~{Casagrande}, S.~{Cassisi}, M.~{Catelan},
  C.~{Chang}, L.~{Cortese}, I.~{Damjanov}, L.~J.~M. {Davies}, R.~{de Grijs},
  G.~{de Rosa}, A.~{Deason}, P.~{di Matteo}, A.~{Drlica-Wagner}, D.~{Erkal},
  A.~{Escorza}, L.~{Ferrarese}, S.~W. {Fleming}, A.~{Font-Ribera},
  K.~{Freeman}, B.~T. {G{\"a}nsicke}, M.~{Gabdeev}, S.~{Gallagher},
  D.~{Gandolfi}, R.~A. {Garc{\'\i}a}, P.~{Gaulme}, M.~{Geha}, M.~{Gennaro},
  M.~{Gieles}, K.~{Gilbert}, Y.~{Gordon}, A.~{Goswami}, J.~P. {Greco},
  C.~{Grillmair}, G.~{Guiglion}, V.~{H{\'e}nault-Brunet}, P.~{Hall}, G.~{Hand
  ler}, T.~{Hansen}, N.~{Hathi}, D.~{Hatzidimitriou}, M.~{Haywood}, J.~V.
  {Hern{\'a}ndez Santisteban}, L.~{Hillenbrand}, A.~M. {Hopkins}, C.~{Howlett},
  M.~J. {Hudson}, R.~{Ibata}, D.~{Ili{\'c}}, P.~{Jablonka}, A.~{Ji},
  L.~{Jiang}, S.~{Juneau}, A.~{Karakas}, D.~{Karinkuzhi}, S.~Y. {Kim},
  X.~{Kong}, I.~{Konstantopoulos}, J.-K. {Krogager}, C.~{Lagos},
  R.~{Lallement}, C.~{Laporte}, Y.~{Lebreton}, K.-G. {Lee}, G.~F. {Lewis},
  S.~{Lianou}, X.~{Liu}, N.~{Lodieu}, J.~{Loveday}, S.~{M{\'e}sz{\'a}ros},
  M.~{Makler}, Y.-Y. {Mao}, D.~{Marchesini}, N.~{Martin}, M.~{Mateo},
  C.~{Melis}, T.~{Merle}, A.~{Miglio}, F.~{Gohar Mohammad},
  K.~{Molaverdikhani}, R.~{Monier}, T.~{Morel}, B.~{Mosser}, D.~{Nataf},
  L.~{Necib}, H.~R. {Neilson}, J.~A. {Newman}, A.~M. {Nierenberg}, B.~{Nord},
  P.~{Noterdaeme}, C.~{O'Dea}, M.~{Oshagh}, A.~B. {Pace},
  N.~{Palanque-Delabrouille}, G.~{Pandey}, L.~C. {Parker}, M.~S. {Pawlowski},
  A.~H.~G. {Peter}, P.~{Petitjean}, A.~{Petric}, V.~{Placco}, L.~{\v{C}}.
  {Popovi{\'c}}, A.~M. {Price-Whelan}, A.~{Prsa}, S.~{Ravindranath}, R.~M.
  {Rich}, J.~{Ruan}, J.~{Rybizki}, C.~{Sakari}, R.~E. {Sanderson},
  R.~{Schiavon}, C.~{Schimd}, A.~{Serenelli}, A.~{Siebert}, M.~{Siudek},
  R.~{Smiljanic}, D.~{Smith}, J.~{Sobeck}, E.~{Starkenburg}, D.~{Stello}, G.~M.
  {Szab{\'o}}, R.~{Szabo}, M.~A. {Taylor}, K.~{Thanjavur}, G.~{Thomas},
  E.~{Tollerud}, S.~{Toonen}, P.-E. {Tremblay}, L.~{Tresse}, M.~{Tsantaki},
  M.~{Valentini}, S.~{Van Eck}, A.~{Variu}, K.~{Venn}, E.~{Villaver}, M.~G.
  {Walker}, Y.~{Wang}, Y.~{Wang}, M.~J. {Wilson}, N.~{Wright}, S.~{Xu},
  M.~{Yildiz}, H.~{Zhang}, K.~{Zwintz}, B.~{Anguiano}, M.~{Bedell},
  W.~{Chaplin}, R.~{Collet}, J.-C. {Cuillandre}, P.-A. {Duc}, N.~{Flagey},
  J.~{Hermes}, A.~{Hill}, D.~{Kamath}, M.~B. {Laychak}, K.~{Ma{\l}ek},
  M.~{Marley}, A.~{Sheinis}, D.~{Simons}, S.~G. {Sousa}, K.~{Szeto}, Y.-S.
  {Ting}, S.~{Vegetti}, L.~{Wells}, F.~{Babas}, S.~{Bauman}, A.~{Bosselli},
  P.~{C{\^o}t{\'e}}, M.~{Colless}, J.~{Comparat}, H.~{Courtois}, D.~{Crampton},
  S.~{Croom}, L.~{Davies}, R.~{de Grijs}, K.~{Denny}, D.~{Devost}, P.~{di
  Matteo}, S.~{Driver}, M.~{Fernandez-Lorenzo}, R.~{Guhathakurta}, Z.~{Han},
  C.~{Higgs}, V.~{Hill}, K.~{Ho}, A.~{Hopkins}, M.~{Hudson}, R.~{Ibata},
  S.~{Isani}, M.~{Jarvis}, A.~{Johnson}, E.~{Jullo}, N.~{Kaiser}, J.-P.
  {Kneib}, J.~{Koda}, G.~{Koshy}, S.~{Mignot}, R.~{Murowinski}, J.~{Newman},
  A.~{Nusser}, A.~{Pancoast}, E.~{Peng}, C.~{Peroux}, C.~{Pichon},
  B.~{Poggianti}, J.~{Richard}, D.~{Salmon}, A.~{Seibert}, P.~{Shastri},
  D.~{Smith}, F.~{Sutaria}, C.~{Tao}, E.~{Taylor}, B.~{Tully}, L.~{van
  Waerbeke}, T.~{Vermeulen}, M.~{Walker}, J.~{Willis}, C.~{Willot}, and
  K.~{Withington}, {\it {The Detailed Science Case for the Maunakea
  Spectroscopic Explorer, 2019 edition}},  {\em arXiv e-prints} (Apr, 2019)
  arXiv:1904.04907, [\href{http://arxiv.org/abs/1904.04907}{{\tt
  arXiv:1904.04907}}].

\bibitem{Ellis19}
R.~{Ellis} and K.~{Dawson}, {\it {SpecTel: A 10-12 meter class Spectroscopic
  Survey Telescope}},  in {\em \baas}, vol.~51, p.~45, Sep, 2019.
\newblock \href{http://arxiv.org/abs/1907.06797}{{\tt arXiv:1907.06797}}.

\bibitem{Slosar19}
A.~{Slosar}, Z.~{Ahmed}, D.~{Alonso}, M.~A. {Amin}, E.~J. {Arena},
  K.~{Bandura}, N.~{Battaglia}, J.~{Blazek}, P.~{Bull}, E.~{Castorina}, T.-C.
  {Chang}, L.~{Connor}, R.~{Dav{\'e}}, C.~{Dvorkin}, A.~{van Engelen},
  S.~{Ferraro}, R.~{Flauger}, S.~{Foreman}, J.~{Frisch}, D.~{Green},
  G.~{Holder}, D.~{Jacobs}, M.~C. {Johnson}, J.~S. {Dillon}, D.~{Karagiannis},
  A.~A. {Kaurov}, L.~{Knox}, A.~{Liu}, M.~{Loverde}, Y.-Z. {Ma}, K.~W. {Masui},
  T.~{McClintock}, K.~{Moodley}, M.~{Munchmeyer}, L.~B. {Newburgh}, C.~{Ng},
  A.~{Nomerotski}, P.~{O'Connor}, A.~{Obuljen}, H.~{Padmanabhan},
  D.~{Parkinson}, J.~X. {Prochaska}, S.~{Rajendran}, D.~{Rapetti},
  B.~{Saliwanchik}, E.~{Schaan}, N.~{Sehgal}, J.~R. {Shaw}, C.~{Sheehy},
  E.~{Sheldon}, R.~{Shirley}, E.~{Silverstein}, T.~{Slatyer}, A.~{Slosar},
  P.~{Stankus}, A.~{Stebbins}, P.~T. {Timbie}, G.~S. {Tucker}, W.~{Tyndall},
  F.~{Villaescusa Navarro}, B.~{Wallisch}, and M.~{White}, {\it {Packed
  Ultra-wideband Mapping Array (PUMA): A Radio Telescope for Cosmology and
  Transients}},  in {\em \baas}, vol.~51, p.~53, Sep, 2019.
\newblock \href{http://arxiv.org/abs/1907.12559}{{\tt arXiv:1907.12559}}.

\bibitem{Vla12}
Z.~{Vlah}, U.~{Seljak}, P.~{McDonald}, T.~{Okumura}, and T.~{Baldauf}, {\it
  {Distribution function approach to redshift space distortions. Part IV:
  perturbation theory applied to dark matter}},  {\em \jcap} {\bf 2012} (Nov,
  2012) 009, [\href{http://arxiv.org/abs/1207.0839}{{\tt arXiv:1207.0839}}].

\bibitem{Vla13}
Z.~{Vlah}, U.~{Seljak}, T.~{Okumura}, and V.~{Desjacques}, {\it {Distribution
  function approach to redshift space distortions. Part V: perturbation theory
  applied to dark matter halos}},  {\em \jcap} {\bf 10} (Oct., 2013) 053,
  [\href{http://arxiv.org/abs/1308.6294}{{\tt arXiv:1308.6294}}].

\bibitem{Sug16}
N.~S. {Sugiyama}, T.~{Okumura}, and D.~N. {Spergel}, {\it {Understanding
  redshift space distortions in density-weighted peculiar velocity}},  {\em
  \jcap} {\bf 7} (July, 2016) 001, [\href{http://arxiv.org/abs/1509.08232}{{\tt
  arXiv:1509.08232}}].

\bibitem{Ham00}
A.~J.~S. {Hamilton}, {\it {Uncorrelated modes of the non-linear power
  spectrum}},  {\em \mnras} {\bf 312} (Feb., 2000) 257--284,
  [\href{http://xxx.lanl.gov/abs/astro-ph/9905191}{{\tt astro-ph/9905191}}].

\bibitem{SenZal15}
L.~{Senatore} and M.~{Zaldarriaga}, {\it {The IR-resummed Effective Field
  Theory of Large Scale Structures}},  {\em \jcap} {\bf 2} (Feb., 2015) 13,
  [\href{http://arxiv.org/abs/1404.5954}{{\tt arXiv:1404.5954}}].

\bibitem{Fujita+:2020}
T.~Fujita and Z.~Vlah, {\it {Perturbative description of bias tracers using
  consistency relations of LSS}},  \href{http://arxiv.org/abs/2003.10114}{{\tt
  arXiv:2003.10114}}.

\bibitem{Dalal08}
N.~{Dalal}, M.~{White}, J.~R. {Bond}, and A.~{Shirokov}, {\it {Halo Assembly
  Bias in Hierarchical Structure Formation}},  {\em \apj} {\bf 687} (Nov.,
  2008) 12--21, [\href{http://arxiv.org/abs/0803.3453}{{\tt arXiv:0803.3453}}].

\bibitem{Cha12}
K.~C. {Chan}, R.~{Scoccimarro}, and R.~K. {Sheth}, {\it {Gravity and
  large-scale nonlocal bias}},  {\em \prd} {\bf 85} (Apr., 2012) 083509,
  [\href{http://arxiv.org/abs/1201.3614}{{\tt arXiv:1201.3614}}].

\bibitem{Saito14}
S.~{Saito}, T.~{Baldauf}, Z.~{Vlah}, U.~{Seljak}, T.~{Okumura}, and
  P.~{McDonald}, {\it {Understanding higher-order nonlocal halo bias at large
  scales by combining the power spectrum with the bispectrum}},  {\em \prd}
  {\bf 90} (Dec., 2014) 123522, [\href{http://arxiv.org/abs/1405.1447}{{\tt
  arXiv:1405.1447}}].

\bibitem{Sen14}
L.~{Senatore}, {\it {Bias in the effective field theory of large scale
  structures}},  {\em \jcap} {\bf 11} (Nov., 2015) 007,
  [\href{http://arxiv.org/abs/1406.7843}{{\tt arXiv:1406.7843}}].

\bibitem{Abidi18}
M.~M. {Abidi} and T.~{Baldauf}, {\it {Cubic halo bias in Eulerian and
  Lagrangian space}},  {\em \jcap} {\bf 2018} (July, 2018) 029,
  [\href{http://arxiv.org/abs/1802.07622}{{\tt arXiv:1802.07622}}].

\bibitem{Schmittfull19}
M.~{Schmittfull}, M.~{Simonovi{\'c}}, V.~{Assassi}, and M.~{Zaldarriaga}, {\it
  {Modeling biased tracers at the field level}},  {\em \prd} {\bf 100} (Aug.,
  2019) 043514, [\href{http://arxiv.org/abs/1811.10640}{{\tt
  arXiv:1811.10640}}].

\bibitem{Modi19}
C.~{Modi}, M.~{White}, A.~{Slosar}, and E.~{Castorina}, {\it {Reconstructing
  large-scale structure with neutral hydrogen surveys}},  {\em \jcap} {\bf
  2019} (Nov., 2019) 023, [\href{http://arxiv.org/abs/1907.02330}{{\tt
  arXiv:1907.02330}}].

\bibitem{Zel70}
Y.~B. {Zel'dovich}, {\it {Gravitational instability: An approximate theory for
  large density perturbations.}},  {\em \aap} {\bf 5} (Mar., 1970) 84--89.

\bibitem{Baldauf2015}
T.~{Baldauf}, M.~{Mirbabayi}, M.~{Simonovi{\'c}}, and M.~{Zaldarriaga}, {\it
  {Equivalence principle and the baryon acoustic peak}},  {\em \prd} {\bf 92}
  (Aug., 2015) 043514, [\href{http://arxiv.org/abs/1504.04366}{{\tt
  arXiv:1504.04366}}].

\bibitem{Blas2016}
D.~{Blas}, M.~{Garny}, M.~M. {Ivanov}, and S.~{Sibiryakov}, {\it {Time-sliced
  perturbation theory II: baryon acoustic oscillations and infrared
  resummation}},  {\em \jcap} {\bf 2016} (July, 2016) 028,
  [\href{http://arxiv.org/abs/1605.02149}{{\tt arXiv:1605.02149}}].

\bibitem{Peloso2017}
M.~{Peloso} and M.~{Pietroni}, {\it {Galilean invariant resummation schemes of
  cosmological perturbations}},  {\em \jcap} {\bf 2017} (Jan., 2017) 056,
  [\href{http://arxiv.org/abs/1609.06624}{{\tt arXiv:1609.06624}}].

\bibitem{Ivanov2018}
M.~M. {Ivanov} and S.~{Sibiryakov}, {\it {Infrared resummation for biased
  tracers in redshift space}},  {\em \jcap} {\bf 2018} (July, 2018) 053,
  [\href{http://arxiv.org/abs/1804.05080}{{\tt arXiv:1804.05080}}].

\bibitem{Peacock92}
J.~A. {Peacock}, {\it {Errors on the measurement of omega via cosmological
  dipoles.}},  {\em \mnras} {\bf 258} (Oct, 1992) 581--586.

\bibitem{Park94}
C.~{Park}, M.~S. {Vogeley}, M.~J. {Geller}, and J.~P. {Huchra}, {\it {Power
  Spectrum, Correlation Function, and Tests for Luminosity Bias in the CfA
  Redshift Survey}},  {\em \apj} {\bf 431} (Aug, 1994) 569.

\bibitem{Peacock94}
J.~A. {Peacock} and S.~J. {Dodds}, {\it {Reconstructing the Linear Power
  Spectrum of Cosmological Mass Fluctuations}},  {\em \mnras} {\bf 267} (Apr,
  1994) 1020, [\href{http://xxx.lanl.gov/abs/astro-ph/9311057}{{\tt
  astro-ph/9311057}}].

\bibitem{TNS10}
A.~{Taruya}, T.~{Nishimichi}, and S.~{Saito}, {\it {Baryon acoustic
  oscillations in 2D: Modeling redshift-space power spectrum from perturbation
  theory}},  {\em \prd} {\bf 82} (Sept., 2010) 063522,
  [\href{http://arxiv.org/abs/1006.0699}{{\tt arXiv:1006.0699}}].

\bibitem{RSDmock}
M.~{White}, B.~{Reid}, C.-H. {Chuang}, J.~L. {Tinker}, C.~K. {McBride},
  F.~{Prada}, and L.~{Samushia}, {\it {Tests of redshift-space distortions
  models in configuration space for the analysis of the BOSS final data
  release}},  {\em \mnras} {\bf 447} (Feb., 2015) 234--245,
  [\href{http://arxiv.org/abs/1408.5435}{{\tt arXiv:1408.5435}}].

\bibitem{Bella18}
L.~{Fonseca de la Bella}, D.~{Regan}, D.~{Seery}, and D.~{Parkinson}, {\it
  {Impact of bias and redshift-space modelling for the halo power spectrum:
  Testing the effective field theory of large-scale structure}},  {\em ArXiv
  e-prints} (May, 2018) [\href{http://arxiv.org/abs/1805.12394}{{\tt
  arXiv:1805.12394}}].

\bibitem{Hand17}
N.~{Hand}, Y.~{Li}, Z.~{Slepian}, and U.~{Seljak}, {\it {An optimal FFT-based
  anisotropic power spectrum estimator}},  {\em \jcap} {\bf 2017} (Jul, 2017)
  002, [\href{http://arxiv.org/abs/1704.02357}{{\tt arXiv:1704.02357}}].

\bibitem{Ding2017}
Z.~{Ding}, H.-J. {Seo}, Z.~{Vlah}, Y.~{Feng}, M.~{Schmittfull}, and
  F.~{Beutler}, {\it {Theoretical systematics of Future Baryon Acoustic
  Oscillation Surveys}},  {\em \mnras} {\bf 479} (Sept., 2018) 1021--1054,
  [\href{http://arxiv.org/abs/1708.01297}{{\tt arXiv:1708.01297}}].

\bibitem{Beutler19}
F.~{Beutler}, M.~{Biagetti}, D.~{Green}, A.~{Slosar}, and B.~{Wallisch}, {\it
  {Primordial features from linear to nonlinear scales}},  {\em Physical Review
  Research} {\bf 1} (Dec., 2019) 033209,
  [\href{http://arxiv.org/abs/1906.08758}{{\tt arXiv:1906.08758}}].

\bibitem{Vasudevan19}
A.~{Vasudevan}, M.~M. {Ivanov}, S.~{Sibiryakov}, and J.~{Lesgourgues}, {\it
  {Time-sliced perturbation theory with primordial non-Gaussianity and effects
  of large bulk flows on inflationary oscillating features}},  {\em \jcap} {\bf
  2019} (Sept., 2019) 037, [\href{http://arxiv.org/abs/1906.08697}{{\tt
  arXiv:1906.08697}}].

\bibitem{ESW07}
D.~J. {Eisenstein}, H.-J. {Seo}, and M.~{White}, {\it {On the Robustness of the
  Acoustic Scale in the Low-Redshift Clustering of Matter}},  {\em \apj} {\bf
  664} (Aug., 2007) 660--674,
  [\href{http://xxx.lanl.gov/abs/astro-ph/0604361}{{\tt astro-ph/0604361}}].

\bibitem{Xu12}
X.~{Xu}, N.~{Padmanabhan}, D.~J. {Eisenstein}, K.~T. {Mehta}, and A.~J.
  {Cuesta}, {\it {A 2 per cent distance to z = 0.35 by reconstructing baryon
  acoustic oscillations - II. Fitting techniques}},  {\em \mnras} {\bf 427}
  (Dec., 2012) 2146--2167, [\href{http://arxiv.org/abs/1202.0091}{{\tt
  arXiv:1202.0091}}].

\bibitem{Chudaykin++:2020}
A.~Chudaykin, M.~M. Ivanov, and M.~Simonovi\'c, {\it {CLASS-PT: non-linear
  perturbation theory extension of the Boltzmann code CLASS}},
  \href{http://arxiv.org/abs/2004.10607}{{\tt arXiv:2004.10607}}.

\bibitem{DAmico20}
G.~{D'Amico}, L.~{Senatore}, and P.~{Zhang}, {\it {Limits on $w$CDM from the
  EFTofLSS with the PyBird code}},  {\em arXiv e-prints} (Mar., 2020)
  arXiv:2003.07956, [\href{http://arxiv.org/abs/2003.07956}{{\tt
  arXiv:2003.07956}}].

\bibitem{Nishimichi20}
T.~{Nishimichi}, G.~{D'Amico}, M.~M. {Ivanov}, L.~{Senatore},
  M.~{Simonovi{\'c}}, M.~{Takada}, M.~{Zaldarriaga}, and P.~{Zhang}, {\it
  {Blinded challenge for precision cosmology with large-scale structure:
  results from effective field theory for the redshift-space galaxy power
  spectrum}},  {\em arXiv e-prints} (Mar., 2020) arXiv:2003.08277,
  [\href{http://arxiv.org/abs/2003.08277}{{\tt arXiv:2003.08277}}].

\bibitem{Hetdex}
G.~J. {Hill}, K.~{Gebhardt}, E.~{Komatsu}, N.~{Drory}, P.~J. {MacQueen},
  J.~{Adams}, G.~A. {Blanc}, R.~{Koehler}, M.~{Rafal}, M.~M. {Roth}, A.~{Kelz},
  C.~{Gronwall}, R.~{Ciardullo}, and D.~P. {Schneider}, {\em {The Hobby-Eberly
  Telescope Dark Energy Experiment (HETDEX): Description and Early Pilot Survey
  Results}}, vol.~399 of {\em Astronomical Society of the Pacific Conference
  Series}, p.~115.
\newblock 2008.

\bibitem{VN18}
F.~{Villaescusa-Navarro}, S.~{Genel}, E.~{Castorina}, A.~{Obuljen}, D.~N.
  {Spergel}, L.~{Hernquist}, D.~{Nelson}, I.~P. {Carucci}, A.~{Pillepich},
  F.~{Marinacci}, B.~{Diemer}, M.~{Vogelsberger}, R.~{Weinberger}, and
  R.~{Pakmor}, {\it {Ingredients for 21 cm Intensity Mapping}},  {\em \apj}
  {\bf 866} (Oct, 2018) 135, [\href{http://arxiv.org/abs/1804.09180}{{\tt
  arXiv:1804.09180}}].

\bibitem{Jain10}
B.~{Jain} and J.~{Khoury}, {\it {Cosmological tests of gravity}},  {\em Annals
  of Physics} {\bf 325} (Jul, 2010) 1479--1516,
  [\href{http://arxiv.org/abs/1004.3294}{{\tt arXiv:1004.3294}}].

\bibitem{Joyce15}
A.~{Joyce}, B.~{Jain}, J.~{Khoury}, and M.~{Trodden}, {\it {Beyond the
  cosmological standard model}},  {\em \physrep} {\bf 568} (Mar, 2015) 1--98,
  [\href{http://arxiv.org/abs/1407.0059}{{\tt arXiv:1407.0059}}].

\bibitem{Yamamoto06}
K.~{Yamamoto}, M.~{Nakamichi}, A.~{Kamino}, B.~A. {Bassett}, and H.~{Nishioka},
  {\it {A Measurement of the Quadrupole Power Spectrum in the Clustering of the
  2dF QSO Survey}},  {\em \pasj} {\bf 58} (Feb, 2006) 93--102,
  [\href{http://xxx.lanl.gov/abs/astro-ph/0505115}{{\tt astro-ph/0505115}}].

\bibitem{Scoccimarro15}
R.~{Scoccimarro}, {\it {Fast estimators for redshift-space clustering}},  {\em
  \prd} {\bf 92} (Oct, 2015) 083532,
  [\href{http://arxiv.org/abs/1506.02729}{{\tt arXiv:1506.02729}}].

\bibitem{Bianchi15}
D.~{Bianchi}, H.~{Gil-Mar{\'\i}n}, R.~{Ruggeri}, and W.~J. {Percival}, {\it
  {Measuring line-of-sight-dependent Fourier-space clustering using FFTs}},
  {\em \mnras} {\bf 453} (Oct, 2015) L11--L15,
  [\href{http://arxiv.org/abs/1505.05341}{{\tt arXiv:1505.05341}}].

\bibitem{Chudaykin19}
A.~{Chudaykin} and M.~M. {Ivanov}, {\it {Measuring neutrino masses with
  large-scale structure: Euclid forecast with controlled theoretical error}},
  {\em \jcap} {\bf 2019} (Nov, 2019) 034,
  [\href{http://arxiv.org/abs/1907.06666}{{\tt arXiv:1907.06666}}].

\bibitem{Ross15}
A.~J. {Ross}, W.~J. {Percival}, and M.~{Manera}, {\it {The information content
  of anisotropic Baryon Acoustic Oscillation scale measurements}},  {\em
  \mnras} {\bf 451} (Aug, 2015) 1331--1340,
  [\href{http://arxiv.org/abs/1501.05571}{{\tt arXiv:1501.05571}}].

\bibitem{Rei14}
B.~A. {Reid}, H.-J. {Seo}, A.~{Leauthaud}, J.~L. {Tinker}, and M.~{White}, {\it
  {A 2.5 per cent measurement of the growth rate from small-scale redshift
  space clustering of SDSS-III CMASS galaxies}},  {\em \mnras} {\bf 444} (Oct.,
  2014) 476--502, [\href{http://arxiv.org/abs/1404.3742}{{\tt
  arXiv:1404.3742}}].

\bibitem{Moh16}
F.~G. {Mohammad}, S.~{de la Torre}, D.~{Bianchi}, L.~{Guzzo}, and J.~A.
  {Peacock}, {\it {Group-galaxy correlations in redshift space as a probe of
  the growth of structure}},  {\em \mnras} {\bf 458} (May, 2016) 1948--1963,
  [\href{http://arxiv.org/abs/1502.05045}{{\tt arXiv:1502.05045}}].

\bibitem{Schmittfull2016a}
M.~{Schmittfull}, Z.~{Vlah}, and P.~{McDonald}, {\it {Fast large scale
  structure perturbation theory using one-dimensional fast Fourier
  transforms}},  {\em \prd} {\bf 93} (May, 2016) 103528,
  [\href{http://arxiv.org/abs/1603.04405}{{\tt arXiv:1603.04405}}].

\bibitem{Schmittfull+:2016}
M.~Schmittfull and Z.~Vlah, {\it {FFT-PT: Reducing the two-loop large-scale
  structure power spectrum to low-dimensional radial integrals}},  {\em Phys.
  Rev. D} {\bf 94} (2016), no.~10 103530,
  [\href{http://arxiv.org/abs/1609.00349}{{\tt arXiv:1609.00349}}].

\bibitem{Fang17}
X.~{Fang}, J.~A. {Blazek}, J.~E. {McEwen}, and C.~M. {Hirata}, {\it {FAST-PT
  II: an algorithm to calculate convolution integrals of general tensor
  quantities in cosmological perturbation theory}},  {\em \jcap} {\bf 2017}
  (Feb., 2017) 030, [\href{http://arxiv.org/abs/1609.05978}{{\tt
  arXiv:1609.05978}}].

\bibitem{Simonovic18}
M.~{Simonovi{\'c}}, T.~{Baldauf}, M.~{Zaldarriaga}, J.~J. {Carrasco}, and J.~A.
  {Kollmeier}, {\it {Cosmological perturbation theory using the FFTLog:
  formalism and connection to QFT loop integrals}},  {\em \jcap} {\bf 2018}
  (Apr., 2018) 030, [\href{http://arxiv.org/abs/1708.08130}{{\tt
  arXiv:1708.08130}}].

\bibitem{Tomlinson20}
J.~{Tomlinson}, H.~S. {Grasshorn Gebhardt}, and D.~{Jeong}, {\it {Fast
  Calculation of Nonlinear Redshift-space Galaxy Power Spectrum Including
  Selection Bias}},  {\em arXiv e-prints} (Apr., 2020) arXiv:2004.03629,
  [\href{http://arxiv.org/abs/2004.03629}{{\tt arXiv:2004.03629}}].

\end{thebibliography}\endgroup
\end{document}